\documentclass[12pt]{elsarticle}

\usepackage{amssymb}
\usepackage{amsmath}
\usepackage{subfigure}
\usepackage{siunitx}
\usepackage{lmodern}
\usepackage{anyfontsize}

\usepackage{dcolumn}
\usepackage{makecell}
\usepackage{lipsum}
\usepackage{bm}
\usepackage{adjustbox}
\usepackage{booktabs,makecell}
\usepackage{nomencl}
\usepackage[nameinlink,capitalize]{cleveref}
\usepackage{booktabs}
\usepackage{array}
\biboptions{sort&compress}

\usepackage{graphicx}% Include figure files
\usepackage{epstopdf}
\usepackage{dcolumn}% Align table columns on decimal point
\usepackage{bm}% bold math
\usepackage[utf8]{inputenc}
\usepackage[T1]{fontenc}
\usepackage{mathptmx}
\usepackage{etoolbox}
\makeatletter
\def\ps@pprintTitle{%
 \let\@oddhead\@empty
 \let\@evenhead\@empty
 \def\@oddfoot{}%
 \let\@evenfoot\@oddfoot}
\makeatother

\usepackage{gensymb} %for degree sign
\usepackage[numbers]{natbib}

\usepackage[symbol]{footmisc}

\begin{document}

\begin{frontmatter}



\title{Impact of a Cold Control Plate on Fluid Flow and Heat Transfer across an Isothermally Heated Rotary Oscillating Circular Cylinder}


\author[inst1]{Amarjit Haty}

%\author[inst1,inst2]{Author Three}

%\affiliation[inst2]{organization={Department Two},%Department and Organization
%            addressline={Address Two}, 
%            city={City Two},
%            postcode={22222}, 
%            state={State Two},
%            country={Country Two}}

\author[inst1]{Rajendra K. Ray \corref{cor1}}
\ead{rajendra@iitmandi.ac.in}
\cortext[cor1]{Corresponding author}

\affiliation[inst1]{organization={School of Mathematical and Statistical Sciences, Indian Institute of Technology Mandi},%Department and Organization
        %    addressline={Address One}, 
            city={Mandi},
            postcode={175005}, 
            state={Himachal Pradesh},
            country={India.}}

\author[inst2]{H.V.R. Mittal}
\affiliation[inst2]{organization={Department of Mathematics, Indian Institute of Technology Palakkad},%Department and Organization
        %    addressline={Address One}, 
            city={Palakkad},
            postcode={678557}, 
            state={Kerala},
            country={India.}}

\begin{abstract}
The main objective of this paper is to study the effect of a cold, vertical, arc-shaped control plate on the flow characteristics and forced convective heat transfer mechanism across a rotary oscillating, isothermally heated circular cylinder. Two-dimensional, unsteady, incompressible, laminar, and viscous flow of a Newtonian, constant property fluid is considered across the cylinder. The simulations are performed with an in-house code for various gap ratios between the control plate and the cylinder ($0\leq d/R_0 \leq 3$), maximum angular velocity ($0.5\leq \alpha_m \leq 4$) and frequency ratio of oscillation ($f/f_0=0.5,\ 3$) at Prandtl number $0.7$ and Reynolds number $150$. Here, $d$ denotes the gap between the surface of the cylinder and the leading surface of the control plate, $R_0$ denotes the radius of the cylinder, $f$ is the frequency of oscillation and $f_0$ is the frequency of natural vortex shedding. $d/R_0=0$ corresponds to the no plate case. Heat transfer and vortex shedding phenomena are discussed in relation to one another. A significant increase in heat transmission is observed for all $\alpha_m$ with the gap ratio of $d/R_0=0.5$ and $f/f_0=0.5$. The heat absorption on the surface of the control plate decreases to zero with increasing gap ratio when $\alpha_m=0.5$ and $f/f_0=0.5$ but never becomes zero when $\alpha_m=4$ and $f/f_0=3$. Additionally, when compared to the no plate case with $(\alpha_m,\ f/f_0)=(0.5,\ 0.5)$, the maximum peak of the drag coefficient is decreased by $9.877\%$ for the gap ratio of $d/R_0=3$. For $\alpha_m=4$ and $f/f_0=3$, the smallest gap ratio of $d/R_0=0.5$ is found to significantly increase the lift coefficient relative to other cases.
\end{abstract}




\end{frontmatter}


\section{Introduction\protect} \label{Introduction}
Over the years, a topic of major significance has been the flow of fluid and the transmission of heat around bluff bodies like cylinders. Because of its theoretical underpinnings and practical applications, this subject is widely explored \cite{kalita2009transformation,ray2016higher,mittal2017numerical,
mittal2017locked}. Numerous industrial operations, including mooring lines \cite{gao2017experimental}, offshore oil platforms \cite{yawar2019transient,ganta2019analysis}, eolian tones \cite{strouhal1878besondere}, flow control \cite{kumar2013flow,lu1996numerical}, and tube-tank heat exchangers \cite{sellappan2014vortex}, are major applications. The cooling of electrical components and chips with different forms are other uses \cite{Zebib1989,yang2001thermal}. Various active and passive control techniques are adopted to reduce the vortex shedding as well as the aerodynamic forces, such as lift and drag. In the past, researchers have employed splitter plates \cite{roshko1955wake,Apelt1973,apelt1975effects,kwon1996} as passive control approaches and rotating control cylinders \cite{MITTAL2001291,asadullah2018counter} or rotary oscillation motion of cylinder \cite{tokumaru1991rotary,shiels2001investigation} as an active control strategy.\\ 

In 1991, Tokumaru and Dimotakis \cite{tokumaru1991rotary} experimentally studied the active control of flow across a circular cylinder by forced rotary oscillation. They were able to reduce the drag by $80\%$ for Reynolds number $15000$. With an active control by rotation of the cylinder, He et al. \cite{he2000active} were able to reduce the drag by $31\%$ for Reynolds number $200$ and by $61\%$ for Reynolds number $1000$. Reduction of drag force is very significant at a frequency of oscillation higher than the lock-in frequency \cite{cheng2001,cheng2001numerical}. Higher maximum angular velocity found to be more effective in flow control and aerodynamic forces reduction \cite{mittal2017locked}. As the cylinder oscillation frequency becomes closer to the vortex shedding frequency, Saxena and Laird \cite{saxena1978heat} found that the forced oscillation significantly improves heat transmission. According to Cheng et al. \cite{cheng1997experimental}, the turbulence and lock-on effects are crucial to the heat transmission mechanism. The forced convection from an isothermally heated cylinder with rotary oscillation was studied by Mahfouz and Badr \cite{mahfouz2000forced}. They found a significant improvement in heat transmission in the lock-on frequency band. Fu and Tong \cite{FU20023033} numerically examined the heat transfer and flow across an isothermally heated transversely oscillating cylinder. They discovered that the oscillating cylinder and vortex shedding interact to dominate the wake, causing the thermal fields to be periodic in the lock-on regime and greatly enhancing heat transfer. Ghazanfarian and Nobari \cite{ghazanfarian2009numerical,nobari2010convective} numerically studied the heat transmission from a rotating circular cylinder that oscillates in a cross and inline mode. Their research shows that the lock-on regime considerably increases heat transmission. They also concluded that beyond a threshold rotation speed, vortex shedding is minimised, and as rotational speed rises, the average Nusselt number and drag coefficient rapidly decrease. Al-Mdallal and Mahfouz \cite{al2017heat} conducted an investigation of the heat transfer phenomena from a stationary heated cylinder with circular motion, and they found that the rate of heat transfer significantly increased as the amplitude of the circular motion increased. In the existing literature, there are many studies on how to improve force convective heat transfer and reduce drag from circular cylinders by introducing a rotary oscillating motion. On wake structure, drag reduction, and forced convective heat transfer in channel flow, passive control with a splitter plate is investigated by very few researchers \cite{celik2008flow,ghiasi2018numerical} combined with active control of transverse oscillation of a circular cylinder.\\

In 1966, Gerrard \cite{gerrard1966mechanics} conducted an experimental study on the flow past bluff bodies along with flow past circular cylinder with splitter plates at high Reynolds number, $2\times10^4$. Besides the trailing edge of the plate, the only essential parameter is the gap ratio between the cylinder and a splitter plate parallel to the flow, and he found that the length of the effective vortex generation region was equal to the plate's distance from the domain boundary. The effect of a splitter plate placed downstream of a bluff body and parallel to the open stream was studied by Roshko \cite{roshko1955wake} at a high Reynolds number ($Re = 10^5$). The author found that the shedding frequency and base suction were decreased by moving the plate closer to the cylinder. According to Bearman \cite{bearman1965investigation}, spinning the circular cylinder with the end plate downstream at a constant speed forces the separate shear flow on the surface to reconnect. The outcome is a reduction in the impacts and vibrations brought on by boundary-layer growth, and a suppression of vortex formation. Apelt et al. \cite{Apelt1973} experimentally studied the effect of a horizontal splitter plate with varied length on the flow across a stationary circular cylinder at $10^4< Re<5\times10^4$. By utilising a splitter plate with a length equivalent to the cylinder diameter, they were able to lower the drag coefficient by $31\%$. Kwon and Choi \cite{kwon1996} concluded that there is a threshold length of the splitter plate that, in proportion to the Reynolds number, results in the complete disappearance of vortex shedding. The reduction in drag coefficient and total heat transfer from the stationary cylinder and splitter plate surface increment are caused by the splitter length increment, as demonstrated by Razavi et al. \cite{razavi2008impact}. Ghiasi et al. \cite{ghiasi2018numerical} found that the drag is decreased up to a critical length of the splitter plate, after which the enlargement of the lock-on zone causes the drag to be increased for the majority of transverse oscillation frequencies. They also came to the conclusion that as the length of splitter plate increases, the thermal boundary layer around the cylinder thickens and reduces heat transfer. Significant reduction in drag coefficient and in amplitude of lift coefficient on a stationary cylinder with an attached splitter plate at Reynolds number $100$, $125$ and $150$ is reported by Deep et al. \cite{deep2022pod}. More studies with control plate are found for circular cylinder \cite{liu2016experimental}, transversely oscillating cylinder \cite{celik2008flow}, parallel dual plate with circular cylinder \cite{bao2013passive} and multiple control rods with circular cylinder \cite{lu2014numerical}. A few researchers studied the effects of curved fins and plates on missiles  \cite{eastman1985aerodynamics} and formula $1$ cars \cite{martins2021influence}, and these are now being used in real life.\\

There is no published research that shows how rotary oscillation and a splitter plate can both be used to influence wake formation, reduce drag, and convect heat away under forced convection from the circular cylinder. The current investigation focuses on the importance of rotary oscillation and gap ratios of an arc-shaped vertical control plate on the flow and forced convective heat transfer across a circular cylinder. Vortex shedding modes are defined in terms of the number of vortices that are shed from either side of the cylinder during vortex shedding cycles in a lock-on period $p_l = nT$, where $n$ can be either an integer or a fraction. In this case, $T$ is the period of cylinder oscillation and $p_l$ is the lock-on period. For instance,  $2S(p_l)$ mode indicates that the downstream wake is fed with an alternating counter-rotating single vortex from each side of the cylinder over the time period $p_l$. When $p_l=T$, the mode $2S(p_l)$ designates the traditional K\'arm\'an vortex street. $2P(p_l)$ mode means that two pairs of counter-rotating vortices shed from each side of the cylinder over the time period $p_l$. On the other hand, the mode $P+S(p_l)$  indicates the shedding of two vortices from one side of the cylinder, followed by one vortex from the other side per $p_l$. Reynolds number and Prandtl number are kept constant at $150$ and $0.7$, respectively. The gap ratio ($d/R_0$) of the control plate varies between $0$ and $3$, while the maximum angular velocity ranges from $0.5$ to $4$ and the frequency ratio varies from $0.5$ to $3$. Note that, $d/R_0=0$ denotes the case without any plate. The Higher Order Compact (HOC) technique \cite{kalita2009transformation,ray2016higher,mittal2016class,mittal2017numerical} based on non-uniform polar grids is used to discretize the 2-D unsteady Navier-Stokes equations and the energy equation. The discretized system is solved by using the bi-conjugate gradient stabilized method with the help of an in-house code.\\

The following is the order in which the paper is organised: The governing equations, initial, and boundary conditions are covered in Section 2. Numerical solution techniques, finite difference discretizations, and numerical scheme validation are covered in Section 3. Section 4 discusses results and analysis. Finally, in Section 5, the conclusions summarise our observations.

\begin{table*}
%\raggedleft
%\begingroup
\setlength{\tabcolsep}{6pt} % Default value: 6pt
\begin{tabular}{|ll|}
\hline
\textbf{Nomenclature}&\\
&\\
$d$ (\si{mm})& \thead[l]{Gap between the control plate and the \\cylinder surface}\\

$\hat{f}$ (\si{Hz}), $f$ & \thead[l]{Frequency of oscillation in dimensional and \\nondimensional form ($f=\hat{f}R_0/U_\infty)$}\\

$h$ (\si{W/m^2K}), $h_{avg}$ (\si{W/m^2K}) & \thead[l]{Coefficients of heat transfer (local and average)}\\

$Nu$, $\overline{Nu}$, $\overline{Nu}_t$ & \thead[l]{Nusselt number (local, average, and time-averaged \\total)}\\

%$Nu_{max}$ & \thead{Peak value of local Nusselt number}\\

$Pr$ & \thead[l]{Prandtl number ($=\nu/\beta)$}\\

$Q''$ (\si{W/m^2}) & \thead[l]{Radial heat flux on the surface (Local)}\\

$Re$ & \thead[l]{Reynolds number ($=2R_0U_{\infty}/\nu$)}\\

$R_0$ (\si{mm}) & \thead[l]{Radius of the circular cylinder}\\

$R_{\infty}$ (\si{mm}) & \thead[l]{Radius of the far field boundary}\\

$\hat{r}$ (\si{mm}), $r$ & \thead[l]{Radius in dimensional and nondimensional form}\\

$T_{\infty}$ (\si{K}) & \thead[l]{Free-stream fluid temperature}\\

$\hat{t}$ (\si{s}), $t$ & \thead[l]{Time in dimensional and nondimensional form}\\

$T_s$ (\si{K}) & \thead[l]{Surface temperature of the cylinder in dimensional form}\\

$U_{\infty}$ (\si{m/s}) & \thead[l]{Free-stream fluid velocity}\\

$\hat{u}$ (\si{m/s}), $u$ & \thead[l]{Radial velocity in dimensional and nondimensional form}\\

$\hat{v}$ (\si{m/s}), $v$ & \thead[l]{Tangential velocity in dimensional and nondimensional \\form}\\

$\hat{\alpha}$ (\si{m/s}), $\alpha$ & \thead[l]{Rotational velocity in dimensional and \\nondimensional form ($\alpha=\hat{\alpha}R_0/U_\infty)$}\\

$\hat{\alpha_m}$ (\si{m/s}), $\alpha_m$ & \thead[l]{Maximum angular velocity in dimensional and \\nondimensional form ($\alpha_m=\hat{\alpha_m}R_0/U_\infty)$}\\

$\beta$ (\si{m^2/s}) & \thead[l]{The thermal diffusivity of the fluid}\\

$\kappa$ (\si{W/mK}) & \thead[l]{The thermal conductivity of the fluid}\\

$\nu$ (\si{m^2/s}) & \thead[l]{The kinematic viscosity of the fluid}\\

$\hat{T}$ (\si{K}), $\Phi$ & \thead[l]{Temperature in dimensional and nondimensional form}\\

$\hat{\Psi}$ (\si{kg/ms}), $\Psi$ & \thead[l]{Stream function in dimensional and nondimensional form}\\

$\hat{\Omega}$ (\si{rad/s}), $\Omega$ & \thead[l]{Vorticity in dimensional and nondimensional form}\\

\hline

\end{tabular}
%\endgroup
\end{table*}
 
\section{The governing equations and the problem\protect}\label{The governing equations and the problem}

\begin{figure}[!t]
       \centering
 \includegraphics[width=0.8\textwidth,trim={0.5cm 0.5cm 0.5cm 0.5cm},clip]{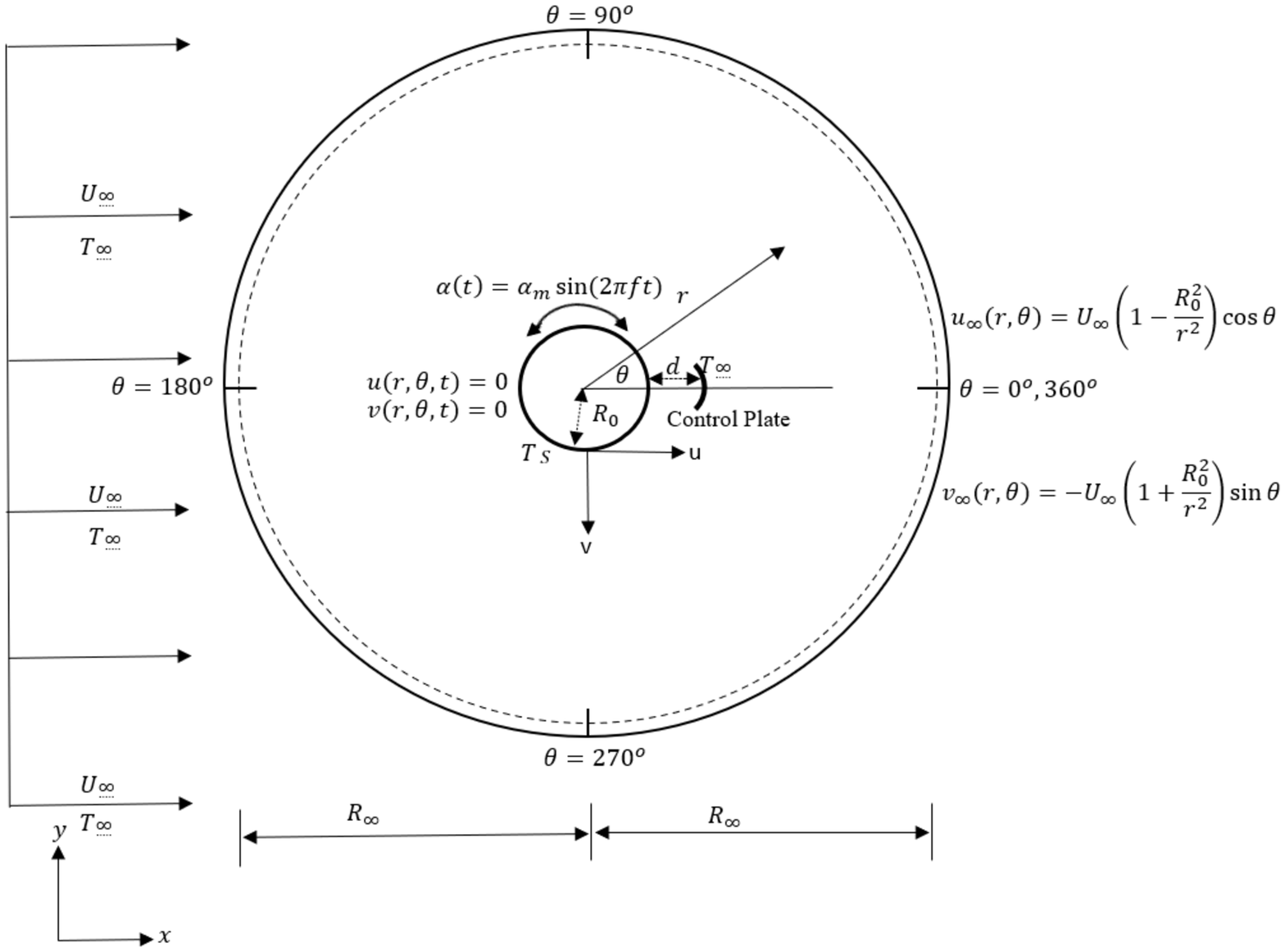}\\
\scriptsize{(a)}
\\
\vspace{0.5em}
 \includegraphics[width=0.55\textwidth,trim={0.1cm 0.1cm 0.1cm 0.1cm},clip]{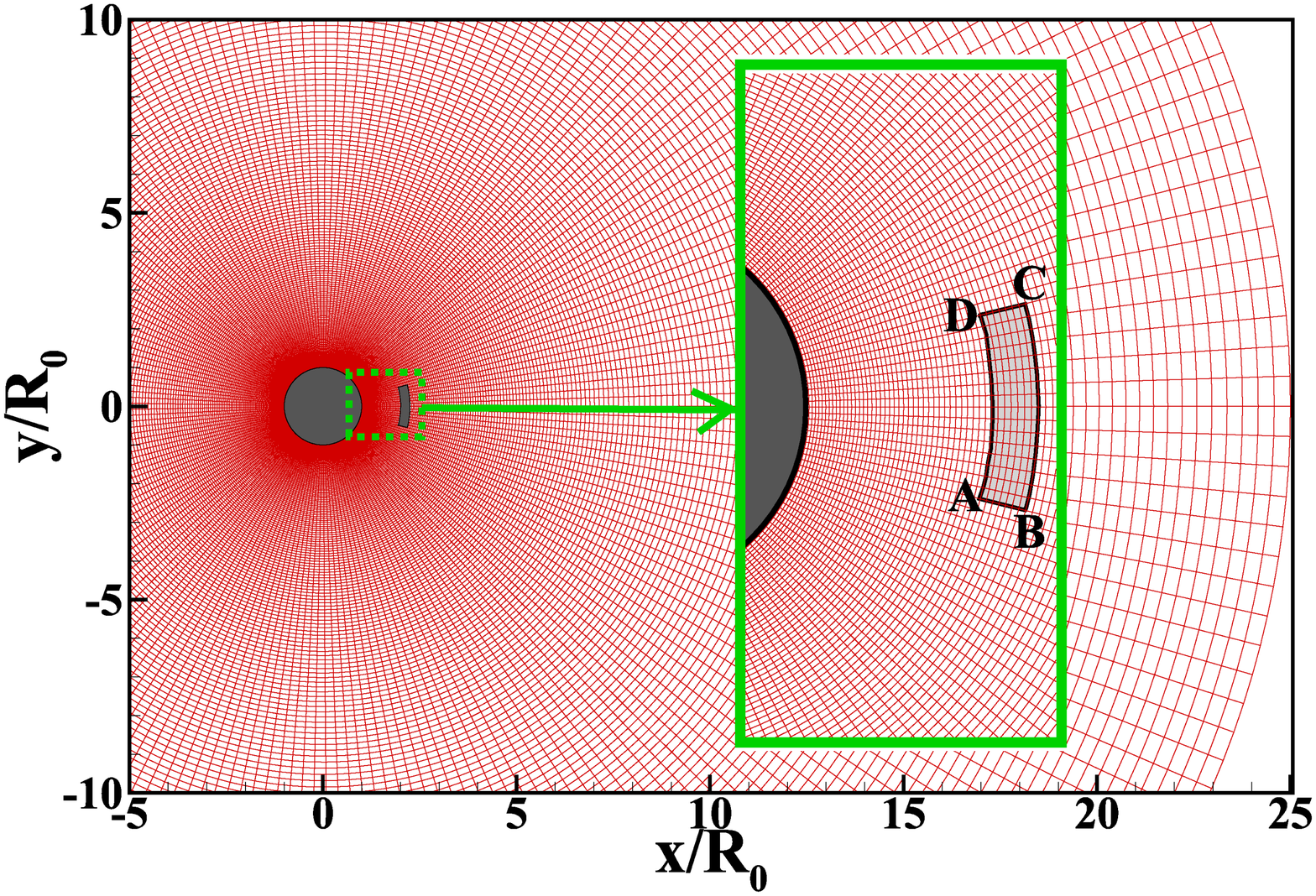}
 \includegraphics[width=0.25\textwidth,trim={0.9cm 0.9cm 0.9cm 0.9cm},clip]{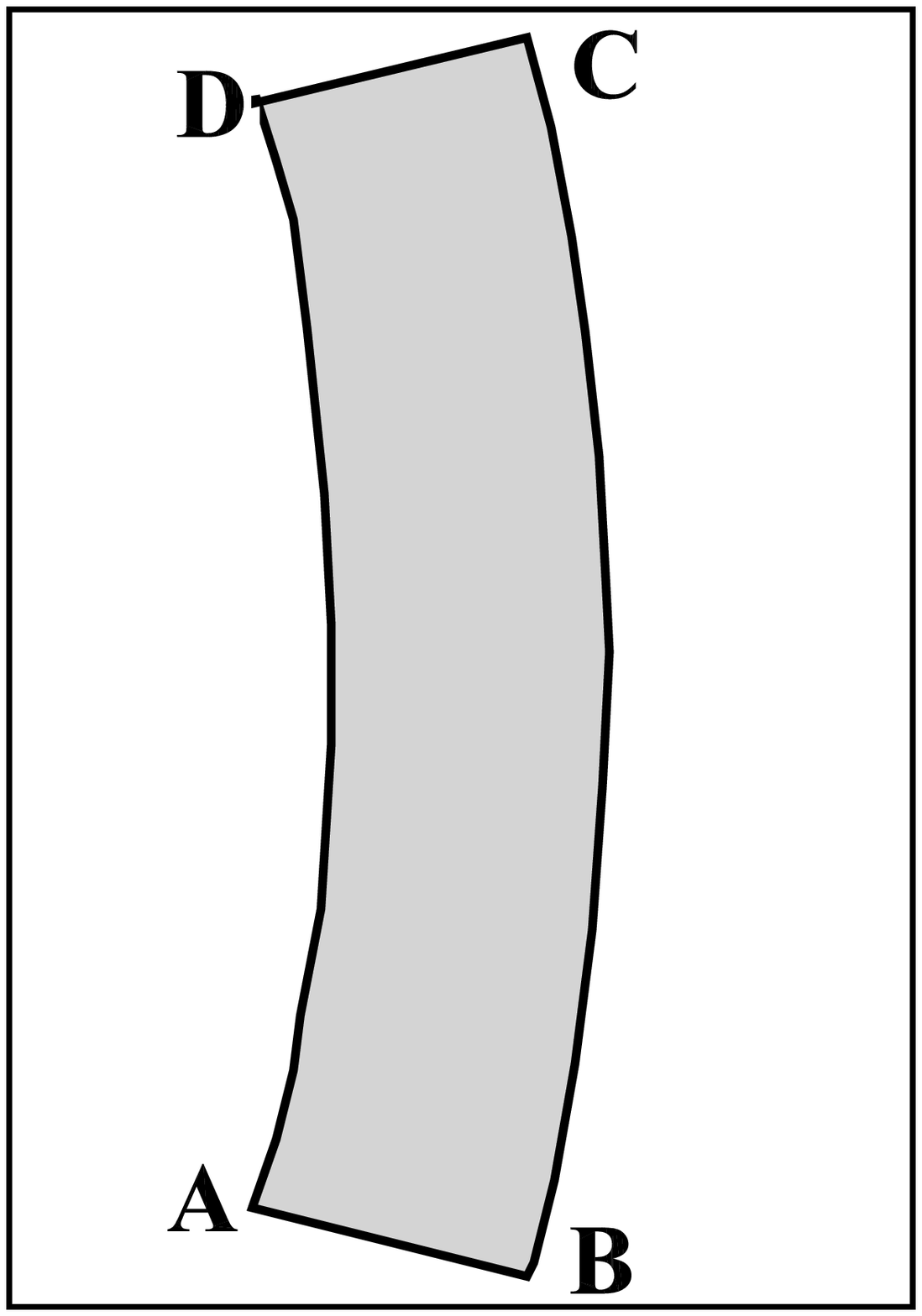} \\
\hspace{2cm} \scriptsize{(b)} \hspace{4cm} \scriptsize{(c)}
\\
       \caption{(a) The schematic illustration of the current problem, (b) an enlarged image of the control plate and the non-uniform polar mesh around the cylinder, and (c) the geometry of the control plate $ABCD$.}
       \label{fig:diagram}
    \end{figure}

An unsteady, 2-D, incompressible, laminar, and viscous flow of a Newtonian, constant property fluid across an isothermally heated, circular cylinder with a radius of $R_0$ is depicted in \cref{fig:diagram}(a). The non-uniform mesh structure around the cylinder and control plate is depicted in \cref{fig:diagram}(b). The geometry of the control plate is shown in \cref{fig:diagram}(c). The top ($CD$) and bottom ($AB$) surfaces of the plate are flat, but the front ($DA$) and rear ($BC$) surfaces are arc-shaped. The arc-length of the surface $DA$ is smaller than the arc-length of the surface $BC$ but the arc-lengths of the surfaces $AB$ and $CD$ are the same. The fluid approaches the cylinder with a uniform velocity $U_\infty$ and at a uniform temperature $T_\infty$. The cylinder impulsively obtains the surface temperature $T_s$ at $\hat{t}=0$. Dimensional parameters are converted to dimensionless form using the following formulas: $r=\frac{\hat{r}}{R_0}$, $t=\frac{\hat{t}U_\infty}{R_0}$, $v=\frac{\hat{v}}{U_\infty}$, $u=\frac{\hat{u}}{U_\infty}$, $\Phi=\frac{(\hat{T}-T_\infty)}{(T_s-T_\infty)}$, $\Psi=\frac{\hat{\Psi} U_\infty}{R_0}$, $\Omega=\frac{\hat{\Omega}R_0}{U_\infty}$. The control plate is placed at a gap $d$ from the cylinder surface and has a unit arc length and a constant thickness that is approximately equivalent to $0.18$ times the cylinder radius. Impermeability and no-slip boundary criteria are taken into account for the surface of the control plate. The control plate is maintained constant at the same temperature as the initial free stream fluid.\\

2-D, non-dimensional, stream function-vorticity formulation of the Navier-Stokes equations and the energy equation in polar coordinates $(r,\theta)$ are represented as,
\begin{equation}\label{eq:1}
\dfrac{\partial^2 \Omega}{\partial r^2}+\frac{1}{r}\dfrac{\partial \Omega}{\partial r}+\frac{1}{r^2}\dfrac{\partial^2 \Omega}{\partial \theta^2}=\frac{Re}{2}\left(u\dfrac{\partial \Omega}{\partial r}+\frac{v}{r}\dfrac{\partial \Omega}{\partial \theta}+\dfrac{\partial \Omega}{\partial t} \right),
\end{equation}
\begin{equation}\label{eq:2}
\dfrac{\partial^2 \Psi}{\partial r^2}+\frac{1}{r}\dfrac{\partial \Psi}{\partial r}+\frac{1}{r^2}\dfrac{\partial^2 \Psi}{\partial \theta^2}=-\Omega,
\end{equation}
\begin{equation}\label{eq:3}
\dfrac{\partial^2 \Phi}{\partial r^2}+\frac{1}{r}\dfrac{\partial \Phi}{\partial r}+\frac{1}{r^2}\dfrac{\partial^2 \Phi}{\partial \theta^2}=\frac{RePr}{2}\left(u\dfrac{\partial \Phi}{\partial r}+\frac{v}{r}\dfrac{\partial \Phi}{\partial \theta}+\dfrac{\partial \Phi}{\partial t} \right).
\end{equation}
The velocities, $v$ and $u$ are defined as 
\begin{equation}\label{eq:4}
v=-\dfrac{\partial \Psi}{\partial r} \hspace{0.5cm} \text{and} \hspace{0.5cm} u=\frac{1}{r}\dfrac{\partial \Psi}{\partial \theta},
\end{equation}
$\Omega$ is stated as
\begin{equation}\label{eq:5}
\Omega=\frac{1}{r}\left[\dfrac{\partial}{\partial r}(vr)-\dfrac{\partial u}{\partial \theta}\right].
\end{equation}

The non-dimensional expression for the rotary oscillating velocity, $\alpha(t)$, is given by,
\begin{equation}\label{eq:6}
\alpha(t)=\alpha_m sin(2\pi ft).
\end{equation}

On the surface of the cylinder, there are three boundary conditions: impermeability, no-slip, and constant temperature, i.e.
\begin{equation}\label{eq:7}
\Psi=0,\hspace{0.5cm}\frac{\partial \Psi}{\partial r}=-\alpha\hspace{0.5cm}\text{and}\hspace{0.5cm}\Phi=1.0\hspace{0.5cm}\text{when}\hspace{0.5cm}r=1.
\end{equation}

Boundary conditions on the surfaces $BC$ and $DA$ of the control plate are,
\begin{equation}\label{eq:8}
\Psi=0,\hspace{0.5cm}\frac{\partial \Psi}{\partial r}=0\hspace{0.5cm}\text{and}\hspace{0.5cm}\Phi=0.
\end{equation}

Boundary conditions on the surfaces $AB$ and $CD$ of the control plate are,
\begin{equation}\label{eq:9}
\Psi=0,\hspace{0.5cm}\frac{\partial \Psi}{\partial \theta}=0\hspace{0.5cm}\text{and}\hspace{0.5cm}\Phi=0.
\end{equation} 

The condition of vorticity on the surface of the cylinder is given by
\begin{equation}\label{eq:10}
\Omega=-\dfrac{\partial^2 \Psi}{\partial r^2} \hspace{0.5cm} \text{when} \hspace{0.5cm}r=1.
\end{equation}

In the distant field, $R_\infty$, the vorticity's resulting decay and the free-stream condition are taken to constitute the boundary conditions, i.e.
\begin{equation}\label{eq:11}
\begin{split}
\Psi\,\to\, \left(r-\frac{1}{r}\right)sin\ \theta,\ \ \ \dfrac{\partial \Psi}{\partial r}\,\to\,\left(1+\frac{1}{r^2}\right)sin\ \theta,\\ \text{and}\ \ \Phi\,\to\, 0 \ \ \ \text{as}\ \ \ r\,\to\,\frac{R_{\infty}}{R_0},
\end{split}
\end{equation}
\begin{equation}\label{eq:12}
\Omega\,\to\,0\ \ \ \text{as}\ \ \ r\,\to\,\frac{R_{\infty}}{R_0}.
\end{equation}

The initial conditions of the stream function and temperature are specified by \cref{eq:11}. An initial zero vorticity assumption is made for the far field in \cref{eq:12}. The following initial conditions are given for the velocities, derived from \cref{eq:4,eq:11}:
\begin{equation}\label{eq:13}
v=-\left(1+\frac{1}{r^2}\right)sin\ \theta\ \ \ \text{and}\ \ \ u=\left(1-\frac{1}{r^2}\right)cos\ \theta.
\end{equation}

\section{Numerical Scheme\protect}\label{Numerical Scheme}

The governing equations of motion and the energy equation are discretized on non-uniform polar grids using a higher order compact (HOC) finite difference technique \cite{kalita2009transformation,ray2016higher,mittal2017numerical} that is temporally second order accurate and spatially at least third order accurate. The non-uniform grid is concentrated around the cylinder which is created using the stretching function $r_i=exp\left(\dfrac{\lambda \pi i}{i_{max}}\right),\ \ 0\leq i\leq i_{max}$. The function $\theta_j$ is given by, $\theta_j=\dfrac{2\pi j}{j_{max}}$, $0\leq j\leq j_{max}$. The discretized equations can be expressed as follows \cite{kalita2009transformation,mittal2017numerical,mittal2018numerical}:
\begin{equation}
\begin{split}
[\mathbb{X}1_{ij}\delta^2_r+\mathbb{X}2_{ij}\delta^2_\theta+\mathbb{X}3_{ij}\delta_r+\mathbb{X}4_{ij}\delta_r\delta_\theta+\mathbb{X}5_{ij}\delta_r\delta^2_\theta\\+\mathbb{X}6_{ij}\delta^2_r\delta_\theta+\mathbb{X}7_{ij}\delta^2_r\delta^2_\theta]\Psi_{ij}=\mathbb{G}_{ij},
\end{split}
\end{equation}
\begin{equation}
\begin{split}
[\mathbb{Y}11_{ij}\delta^2_r+\mathbb{Y}12_{ij}\delta^2_\theta+\mathbb{Y}13_{ij}\delta_r+\mathbb{Y}14_{ij}\delta_\theta+\mathbb{Y}15_{ij}\delta_r\delta_\theta\\+\mathbb{Y}16_{ij}\delta_r\delta^2_\theta+\mathbb{Y}17_{ij}\delta^2_r\delta_\theta+\mathbb{Y}18_{ij}\delta^2_r\delta^2_\theta]\Omega^{n+1}_{ij}\\=[\mathbb{Y}21_{ij}\delta^2_r+\mathbb{Y}22_{ij}\delta^2_\theta+\mathbb{Y}23_{ij}\delta_r+\mathbb{Y}24_{ij}\delta_\theta+\mathbb{Y}25_{ij}\delta_r\delta_\theta\\+\mathbb{Y}26_{ij}\delta_r\delta^2_\theta+\mathbb{Y}27_{ij}\delta^2_r\delta_\theta+\mathbb{Y}28_{ij}\delta^2_r\delta^2_\theta]\Omega^{n}_{ij},
\end{split}
\end{equation}

\begin{equation}
\begin{split}
[\mathbb{Z}11_{ij}\delta^2_r+\mathbb{Z}12_{ij}\delta^2_\theta+\mathbb{Z}13_{ij}\delta_r+\mathbb{Z}14_{ij}\delta_\theta+\mathbb{Z}15_{ij}\delta_r\delta_\theta\\+\mathbb{Z}16_{ij}\delta_r\delta^2_\theta+\mathbb{Z}17_{ij}\delta^2_r\delta_\theta+\mathbb{Z}18_{ij}\delta^2_r\delta^2_\theta]\Phi^{n+1}_{ij}\\=[\mathbb{Z}21_{ij}\delta^2_r+\mathbb{Z}22_{ij}\delta^2_\theta+\mathbb{Z}23_{ij}\delta_r+\mathbb{Z}24_{ij}\delta_\theta+\mathbb{Z}25_{ij}\delta_r\delta_\theta\\+\mathbb{Z}26_{ij}\delta_r\delta^2_\theta+\mathbb{Z}27_{ij}\delta^2_r\delta_\theta+\mathbb{Z}28_{ij}\delta^2_r\delta^2_\theta]\Phi^{n}_{ij}.
\end{split}
\end{equation}
The coefficients $\mathbb{X}1_{ij}$, $\mathbb{X}2_{ij}$,$...$, $\mathbb{X}7_{ij}$; $\mathbb{G}_{ij}$;  $\mathbb{Y}11_{ij}$, $\mathbb{Y}12_{ij}$,$...$, $\mathbb{Y}18_{ij}$; $\mathbb{Y}21_{ij}$, $\mathbb{Y}22_{ij}$,$...$, $\mathbb{Y}28_{ij}$; $\mathbb{Z}11_{ij}$, $\mathbb{Z}12_{ij}$,$...$, $\mathbb{Z}18_{ij}$ and $\mathbb{Z}21_{ij}$, $\mathbb{Z}22_{ij}$,$...$, $\mathbb{Z}28_{ij}$  are the functions of $r$ and $\theta$. \cite{kalita2009transformation,mittal2017numerical,mittal2018numerical} provide the expressions for the central difference operators $\delta_{\theta}$, $\delta^2_{\theta}$, $\delta_r$ and $\delta^2_r$ which are non-uniform, as well as the notations $\theta_f$, $\theta_b$, $r_f$, $r_b$ and the coefficients.

\subsection{\textbf{Drag and lift coefficients}}

Surface friction and surface pressure distribution are the main factors responsible for the forces acting on a circular cylinder submerged in fluids of uniform flow. The drag ($C_D$) and lift ($C_L$) coefficient expressions are obtained from \cite{kalita2009transformation, mittal2017numerical}. These are the expressions:
\begin{equation}
C_D=\frac{1}{Re}\int^{2\pi}_0\left[ \left(\frac{\partial \Omega}{\partial r}\right)_{R_0}-\Omega_{R_0}\right]\cos{\theta}d\theta,
\end{equation}

\begin{equation}
C_L=\frac{1}{Re}\int^{2\pi}_0\left[ \left(\frac{\partial \Omega}{\partial r}\right)_{R_0}-\Omega_{R_0}\right]\sin{\theta}d\theta.
\end{equation}

The time-averaged drag, $\overline{C}_D$, is given as
\begin{equation}
\overline{C}_D=\frac{1}{t_1-t_2}\int_{t_1}^{t_2}{C}_Ddt.
\end{equation}
The time span between $t_1$ and $t_2$ is chosen when the flow enters a periodic phase or completes several cycles.

\subsection{\textbf{The parameters of heat transfer}}

Heat first transfers from the surface of the cylinder to the nearby fluid by conduction, and then It convects with the flow away. The heat conduction occurs in the radial direction from the surface of the cylinder. The dimensionless local heat flux in the radial direction is described as the local Nusselt number ($Nu$) as follows:
\begin{equation}
Nu=\frac{2hR_0}{\kappa}=\frac{Q''(2R_0)}{\kappa(T_s-T_\infty)},
\end{equation}
where $Q''$ is the local radial heat flux at the surface, $\kappa$ is the fluid thermal conductivity, and $h$ is the local heat transfer coefficient.\\

The formula for $Q''$ over the surface of the cylinder is $Q''=-\kappa\frac{\partial T}{\partial r}|_{r=R_0}$. Over the surfaces $BC$ and $DA$ of the control plate, $Q''$ is computed by, $Q''=-\kappa\frac{\partial T}{\partial r}$. $Q''$ is computed over the surfaces $AB$ and $CD$ of the control plate as, $Q''=-\kappa\frac{\partial T}{\partial \theta}$.\\

The average Nusselt number $\overline{Nu}$, which denotes the dimensionless heat transfer from the surface and it is defined on the cylinder surface as follows:
\begin{equation}
\overline{Nu}=\frac{2h_{avg}R_0}{\kappa}=\frac{1}{2\pi}\int_0^{2\pi}Nud\theta.
\end{equation}

The average heat transfer coefficient is expressed as $h_{avg}=$ $\frac{1}{2\pi}\int_0^{2\pi}hd\theta$. The time-averaged Nusselt number $\overline{Nu}_t$ is obtained as follows:
\begin{equation}
\overline{Nu}_t=\frac{1}{t_1-t_2}\int_{t_1}^{t_2}\overline{Nu}dt.
\end{equation}
The time span between $t_1$ and $t_2$ is chosen when the flow enters a periodic phase and completes several cycles.

\subsection{\textbf{Validation and independence studies for grid, time and domain}}
\begin{figure*}[!htbp]
\centering
\subfigure[]{
\includegraphics[width=0.3\textwidth,trim={0.1cm 0.1cm 0.1cm 0.1cm},clip]{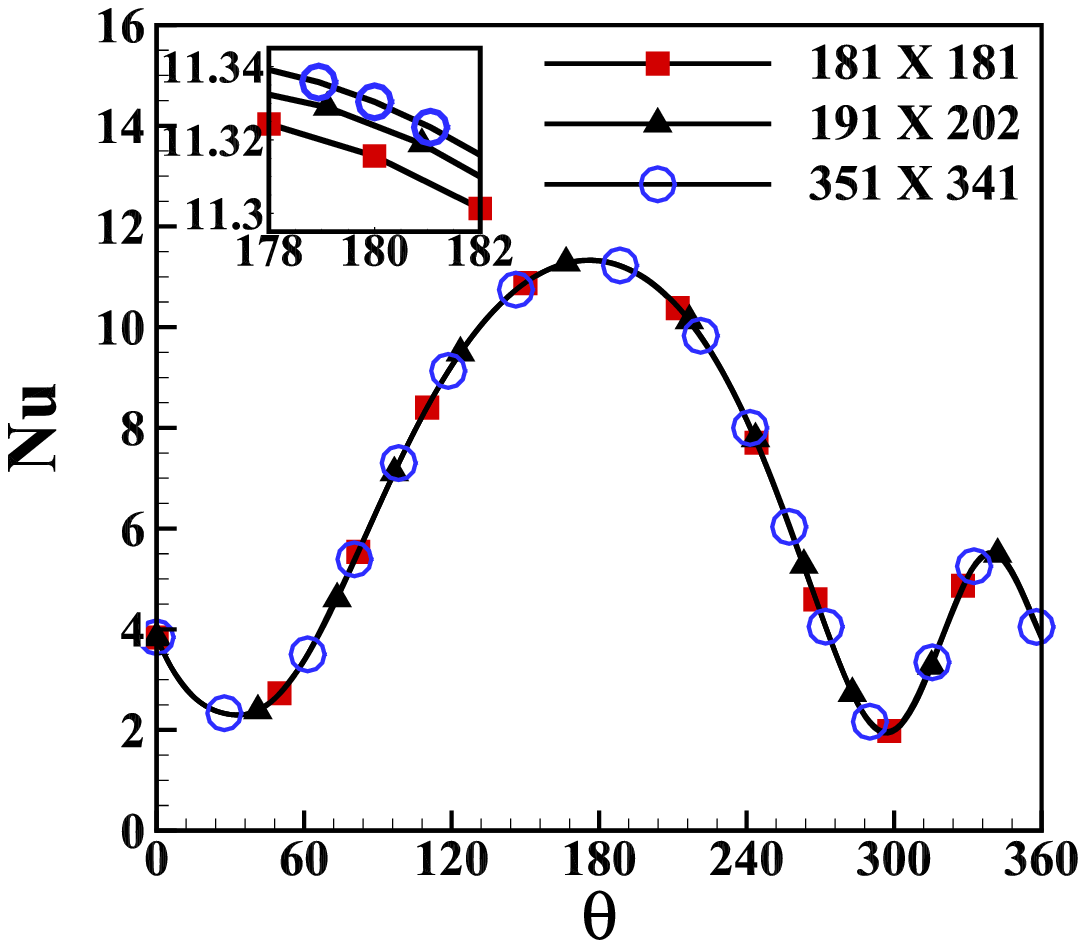}\label{fig:Grid_Independence}
}
\quad
\subfigure[]{\includegraphics[width=0.3\textwidth,trim={0.1cm 0.1cm 0.1cm 0.1cm},clip]{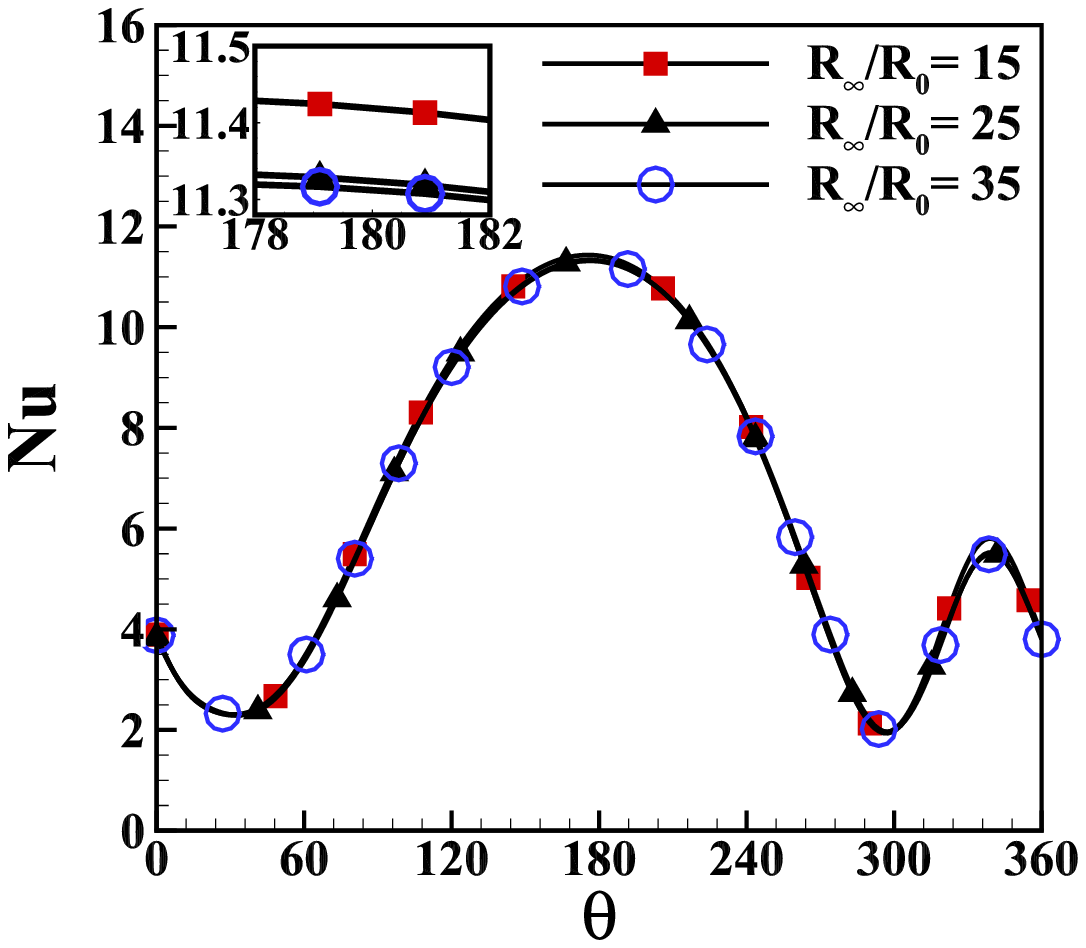}\label{fig:Space_Independence}}
\quad
\subfigure[]{\includegraphics[width=0.3\textwidth,trim={0.1cm 0.1cm 0.1cm 0.1cm},clip]{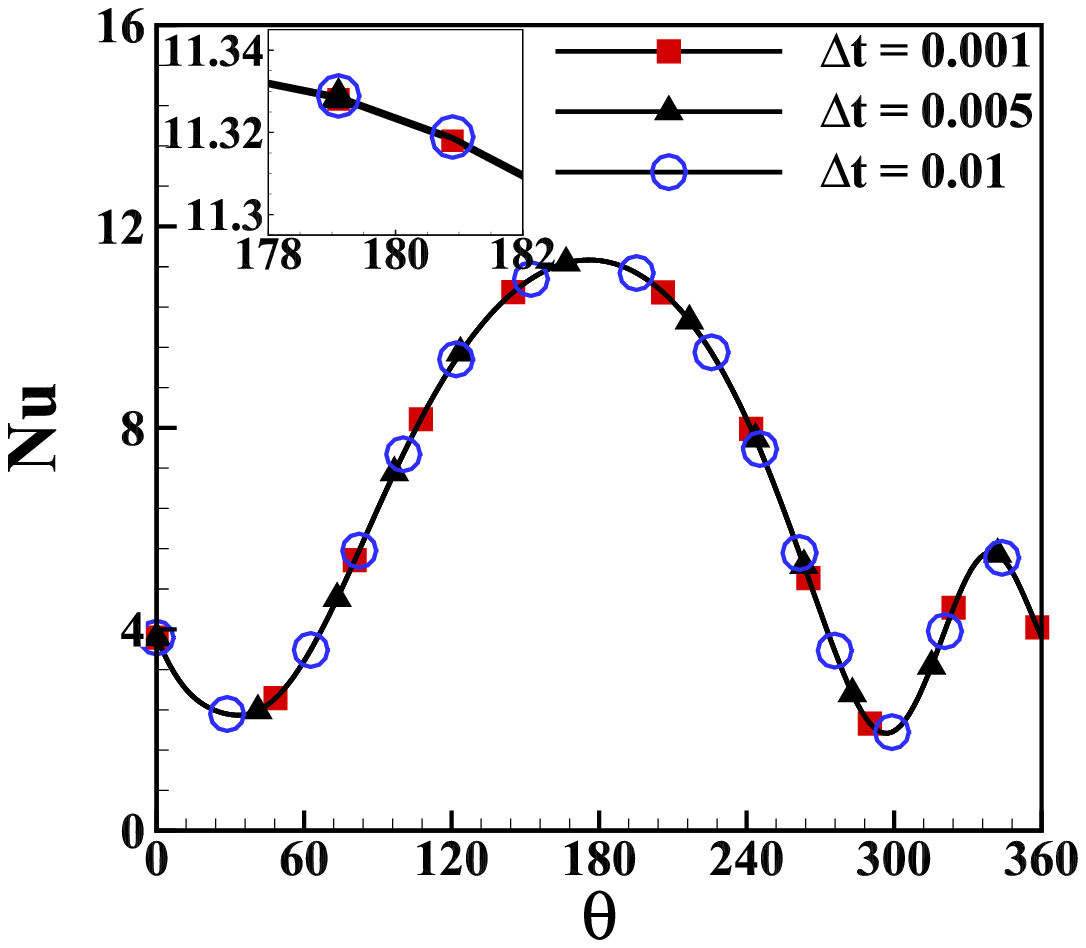}\label{fig:Time_Independence}}
\caption{Variation in the distribution of local Nusselt number $Nu$ (a) grid independence test with grid sizes $181\times181$, $191\times202$, $351\times341$, (b) domain independence test with outer boundary radius $15$, $25$, $35$ and (c) time independence test with time steps $0.001$, $0.005$, $0.01$ at $t = 100$ for $Re = 150$, $Pr=0.7$, $(\alpha_m,\ f/f_0)=(0.5,\ 0.5)$ and $d/R_0=1$.}
\label{fig:Independence}
\end{figure*}

The grid independence test is shown for three distinct grid sizes $(181\times181)$, $(191\times202)$  and $(351\times341)$ in \cref{fig:Independence}\subref{fig:Grid_Independence} with $Re=150$, $Pr=0.7$, $(\alpha_m,\ f/f_0)=(0.5,\ 0.5)$, $d/R_0=1$, fixed time step $\Delta t=0.01$ and a constant domain to cylinder radii ratio of $25$. All grid sizes appear to yield nearly identical outcomes. The grid size $191\times 202$ is chosen for future computations.  The independence test for the domain is conducted in \cref{fig:Independence}\subref{fig:Space_Independence} with three distinct radii of the outer boundary, $R_\infty/R_0=$ $15$, $25$ and $35$ , with grid size and time step, $(181\times181)$ and $\Delta t=0.01$, respectively. The values for the remaining parameters are left unchanged from those used in the grid independence test. Far field radius, $R_\infty=25R_0$ is chosen for further computations. The time independence test is then performed in \cref{fig:Independence}\subref{fig:Time_Independence} with time increments $\Delta t=0.001,\ 0.005,\ 0.01,\ 0.02$ with a predetermined grid size of $(181\times181)$ and $R_\infty=25R_0$. According to these test results, we chose $\frac{R \infty}{R_0}=25$ and $\Delta t=0.01$ for future computations.\\

\begin{figure}[!htbp]
\centering
\includegraphics[width=0.3\textwidth,trim={0.5cm 0.25cm 0.5cm 0.5cm},clip]{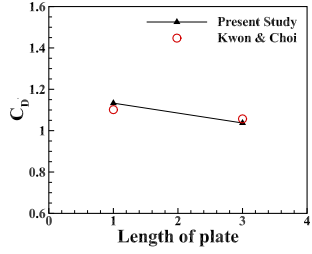}
\caption{Comparison of the time-averaged drag coefficient estimates from the current investigation with Kwon and Choi's work \cite{kwon1996} for $Re=160$.}
\label{tab:valid_kwon}
\end{figure}

To demonstrate the accuracy of our numerical scheme, we compare the results to those of flow past a circular cylinder with an attached splitter plate \cite{kwon1996}. \cref{tab:valid_kwon} indicates that the largest difference in time-averaged drag coefficients between the findings of the current study and those of the earlier works is $2.57\%$, which is likewise within a satisfactory range.

\section{Results and Discussions\protect}\label{Results and Discussions}
\begin{figure*}[!t]
\centering
\includegraphics[width=0.4\textwidth,trim={0.7cm 0.7cm 14cm 0.7cm},clip]{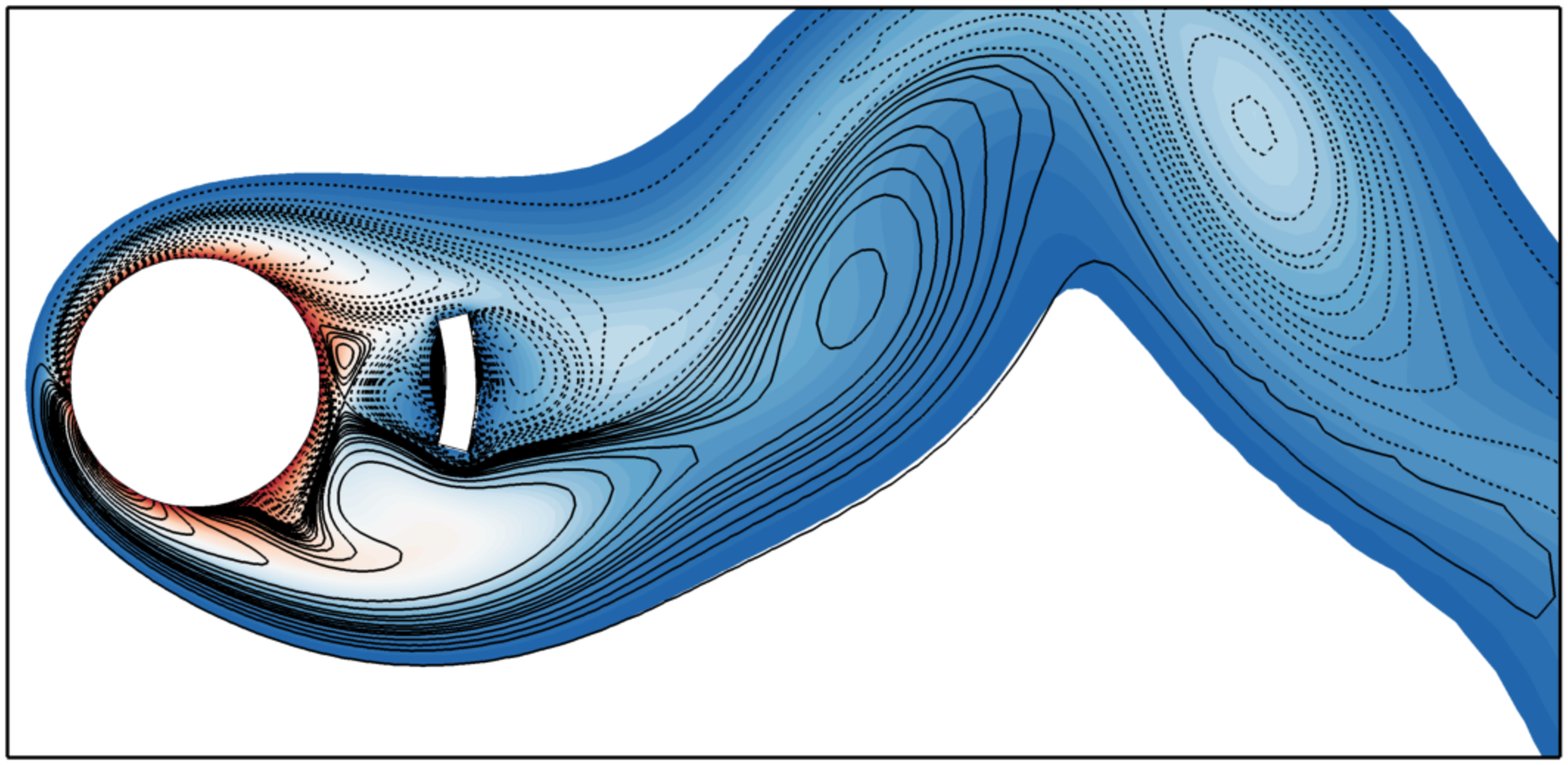}
\includegraphics[width=0.4\textwidth,trim={0.7cm 0.7cm 14cm 0.7cm},clip]{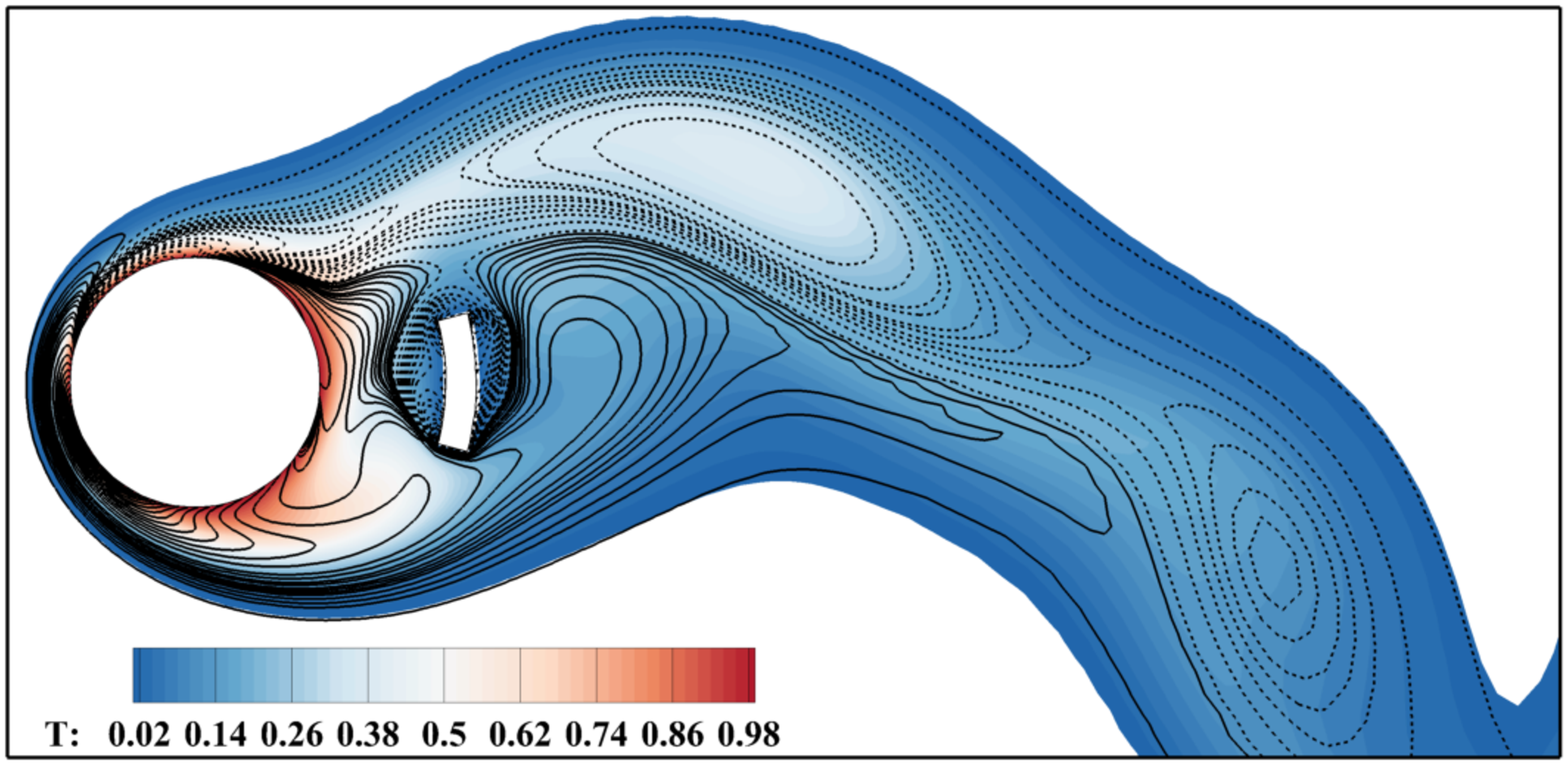}
\caption{Vorticity and isotherm contours are overlaid at periods $t=380$ (left) and $t=395$ (right) for $Re=150$, $\alpha_m=1$, $d/R_0=1$, and $f/f_0=0.5$. Black lines denote vorticity contours, whereas coloured contours indicate isotherm contours.}
 \label{fig:d_1_a_1_f_0-5}
\end{figure*}

The notable variables that affect heat and flow fields include the Reynolds number ($Re$), Prandtl number ($Pr$), maximum angular velocity ($\alpha_m$), frequency ratio ($f/f_0$), and the gap ratio of control plate ($d/R_0$) \cite{mittal2018numerical,gerrard1966mechanics}. The current investigations are conducted at $Re=150$, $\alpha_m\in[0.5,\ 4]$, $f/f_0=\{0.5,\ 3\}$, and $d/R _0\in[0,\ 3]$, with $Pr=0.7$. $d/R_0=0$ is assumed to be the case without the control plate. The values of $\alpha_m$ and $f/f_0$ are typically chosen in accordance with \cite{mittal2018numerical}. \cref{fig:d_1_a_1_f_0-5} illustrates the isotherm contours overlaid with vorticity contours at time steps $t=380,\ 395$ for $Re=150$, $\alpha_m=1$, $d/R_0=1$, and $f/f_0=0.5$. It is found that the vorticity contours nearly overlap with the isotherm contours. This suggests that the similar convection and diffusion phenomena are experienced by the heat transfer and vorticity. Heat advection occurs from the cylinder wall to the near wake in a mechanism similar to the advection of vorticity.\\

\begin{figure*}[!htbp]
\centering
\includegraphics[width=0.4\textwidth,trim={0cm 0cm 0cm 0cm},clip]{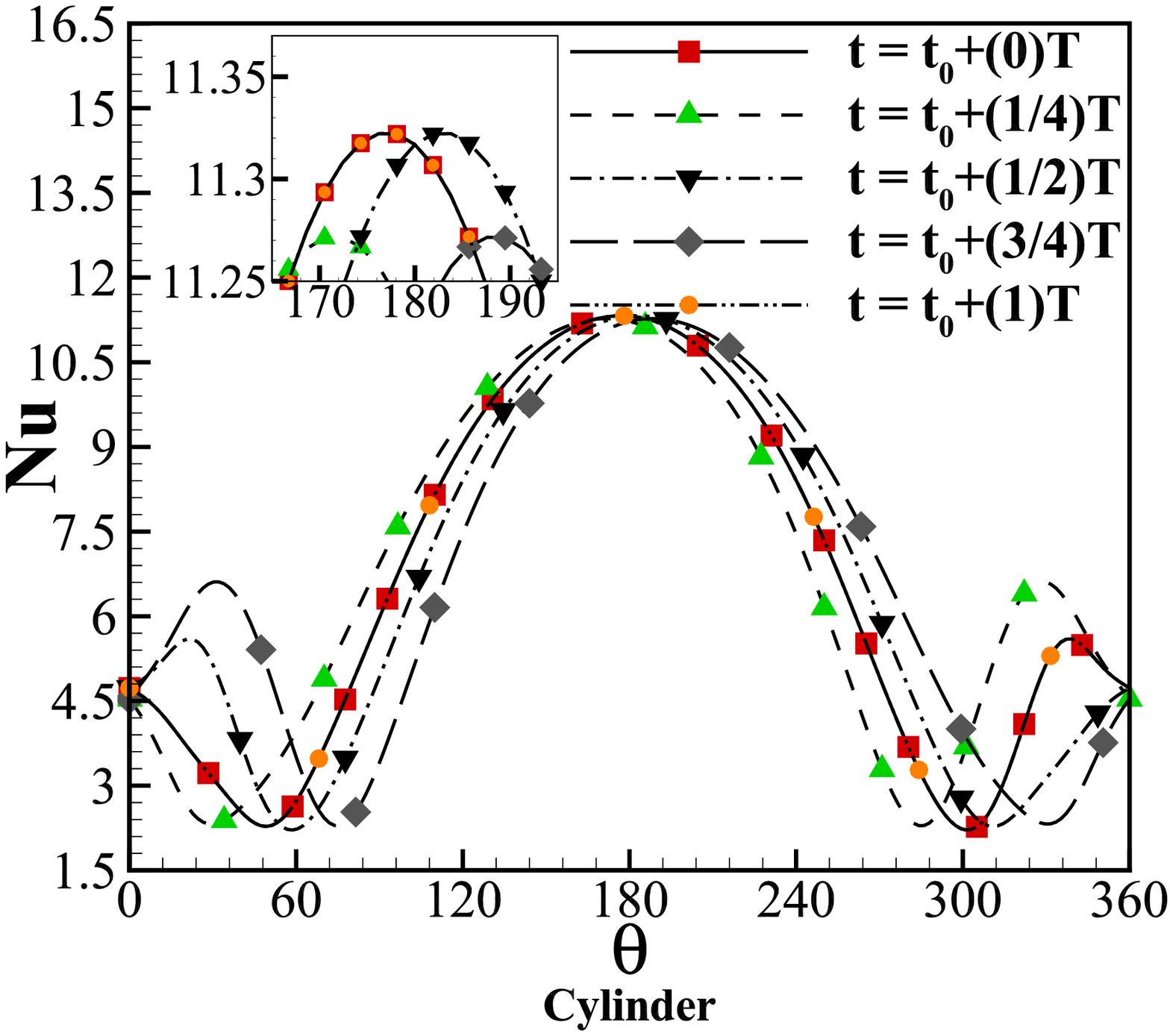}
\includegraphics[width=0.4\textwidth,trim={0cm 0cm 0cm 0cm},clip]{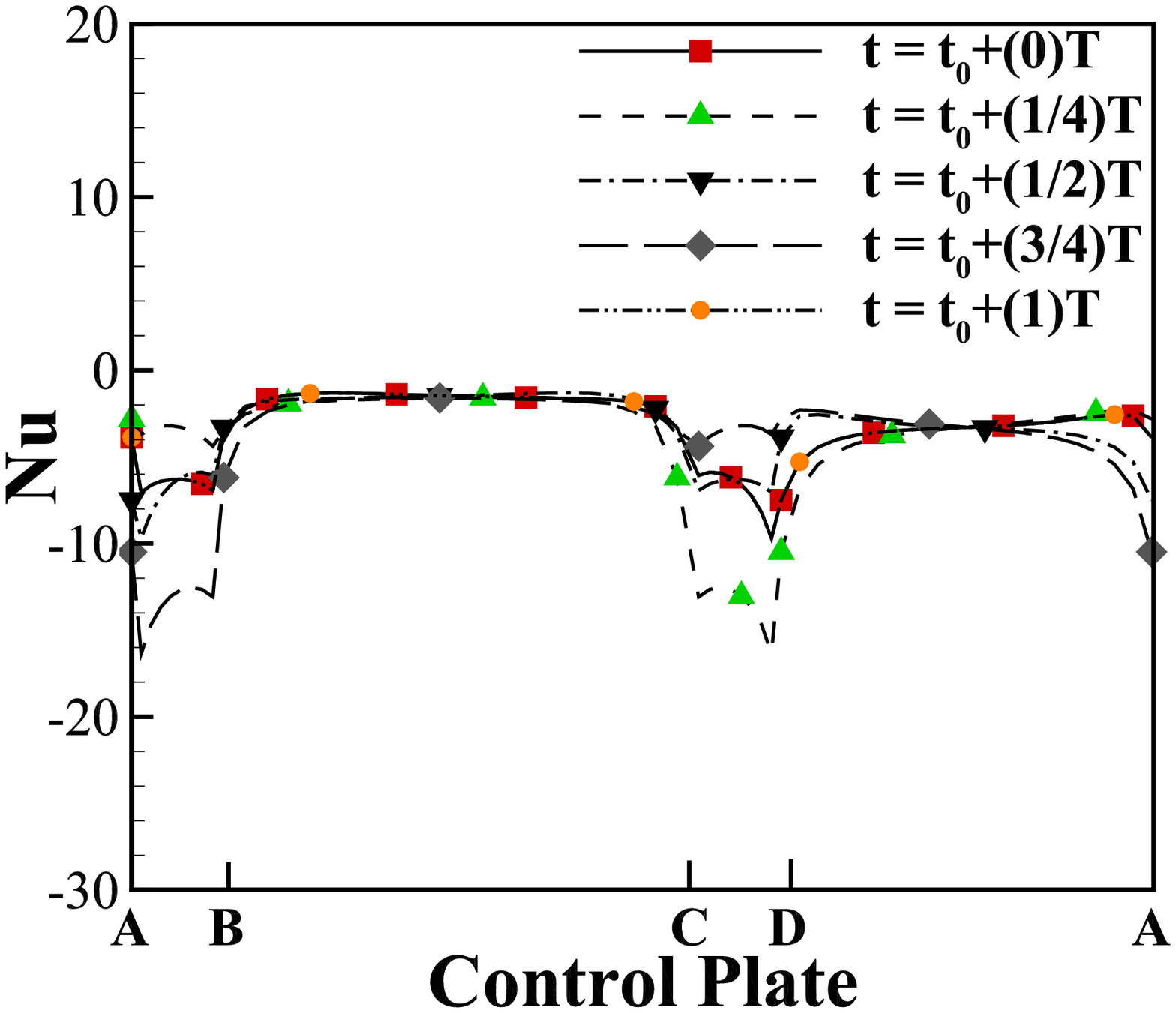}
\\
\scriptsize{$t=t_0+(0)T$}\hspace{4cm}\scriptsize{$t=t_0+(1/4)T$}
\\
\includegraphics[width=0.43\textwidth,trim={0.7cm 0.7cm 8cm 0.7cm},clip]{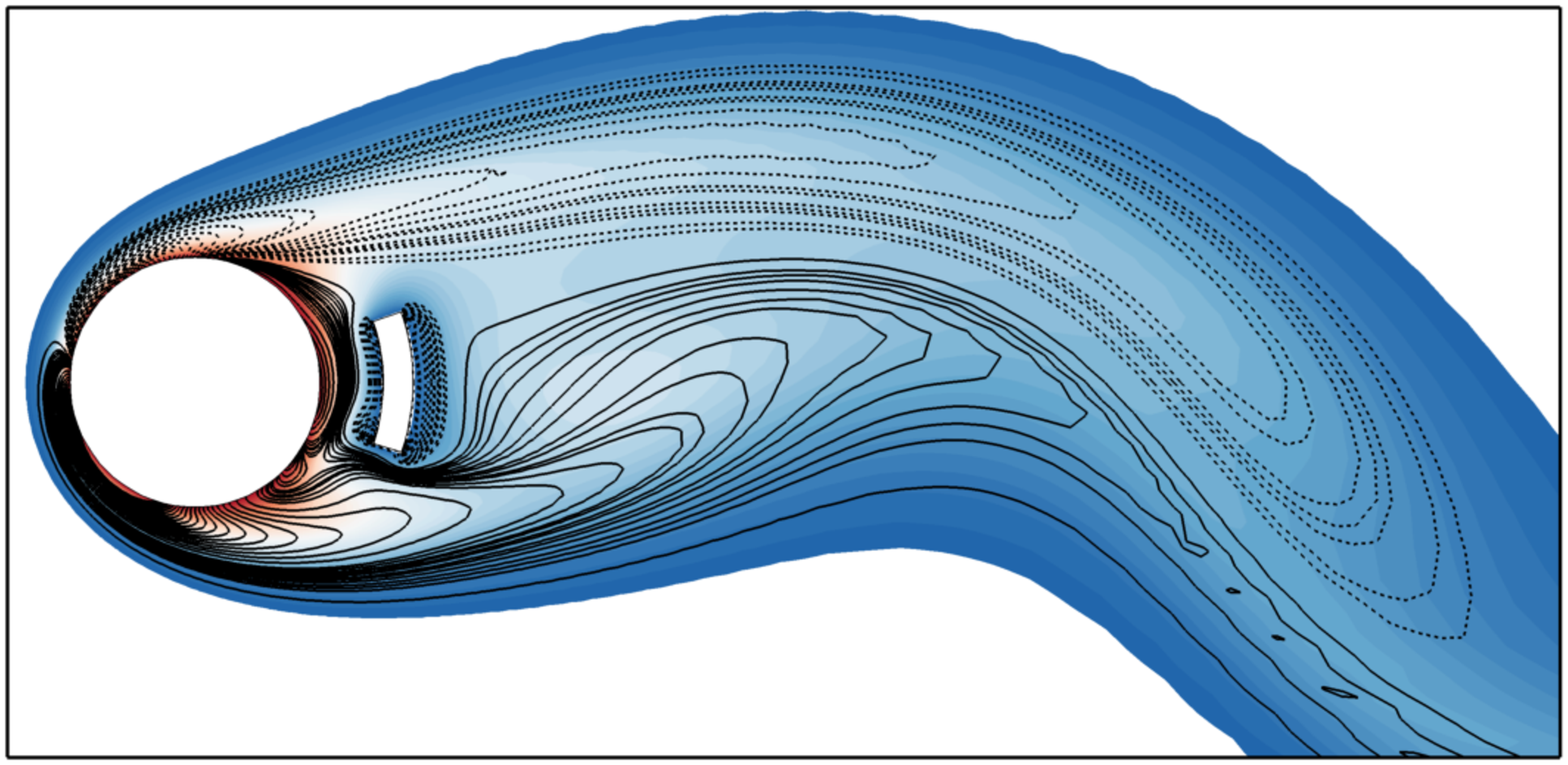}
\includegraphics[width=0.43\textwidth,trim={0.7cm 0.7cm 8cm 0.7cm},clip]{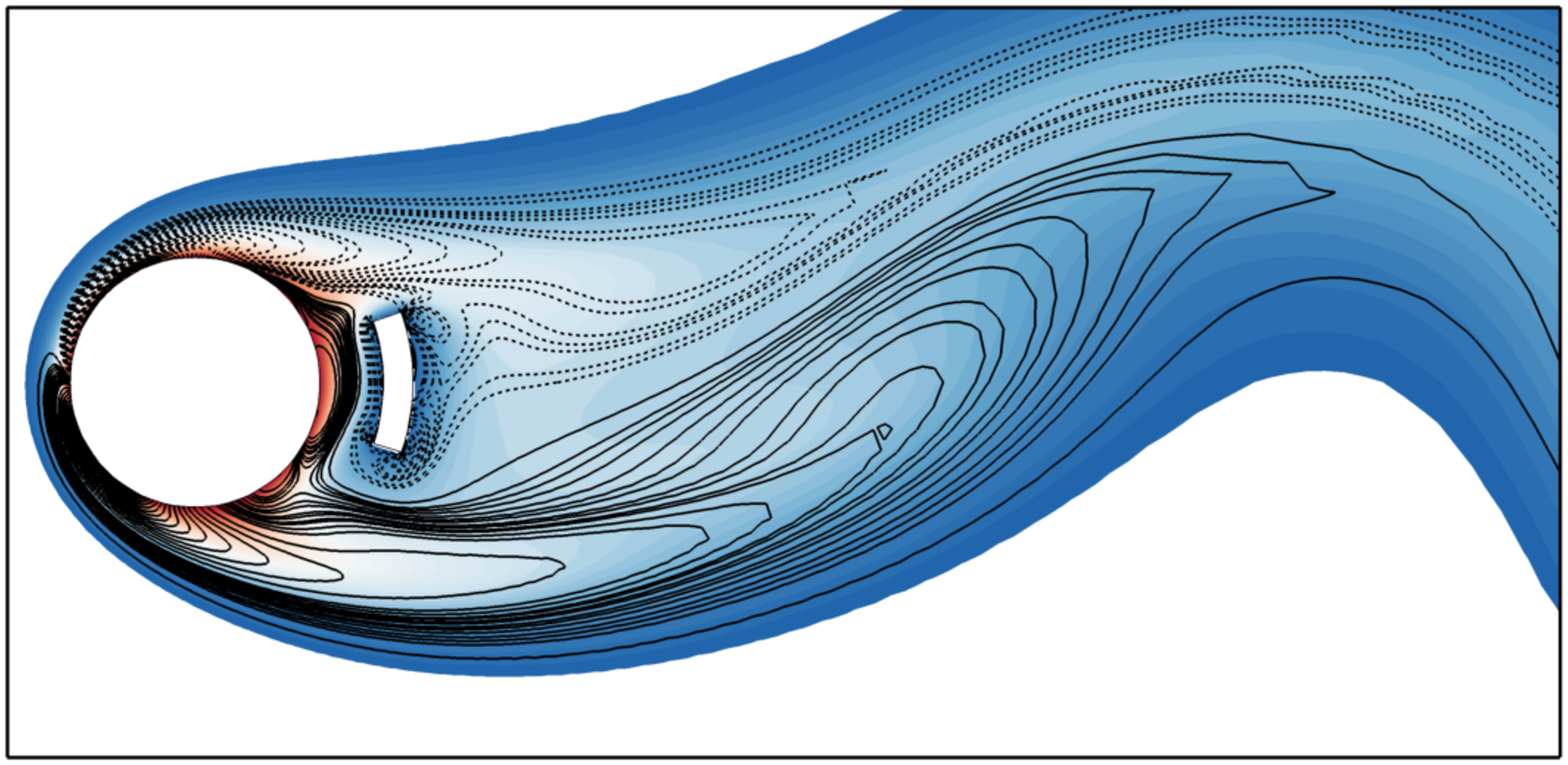}
\\
\scriptsize{$t=t_0+(1/2)T$}\hspace{4cm}\scriptsize{$t=t_0+(3/4)T$}
\\
\includegraphics[width=0.43\textwidth,trim={0.7cm 0.7cm 8cm 0.7cm},clip]{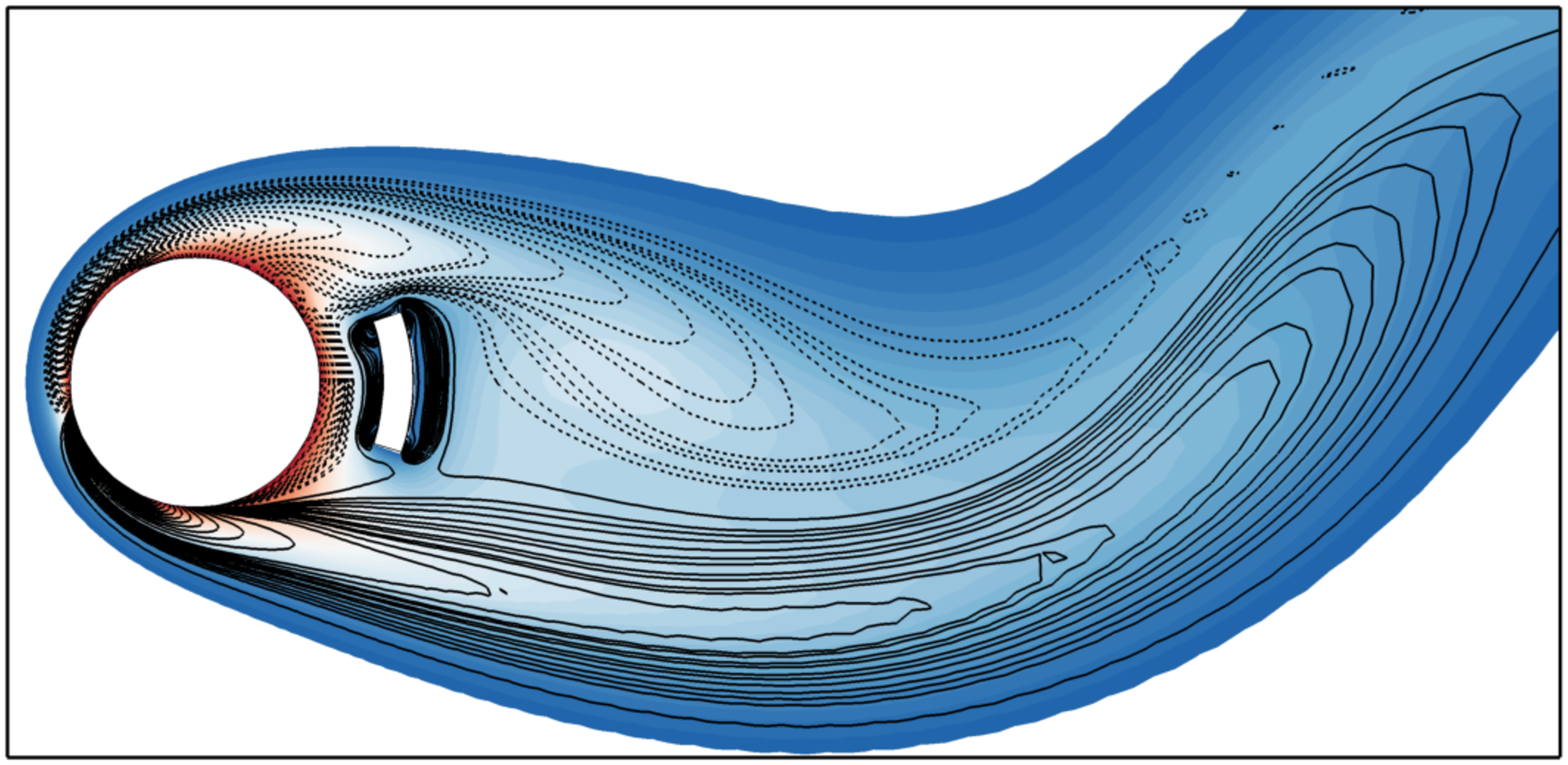}
\includegraphics[width=0.43\textwidth,trim={0.7cm 0.7cm 8cm 0.7cm},clip]{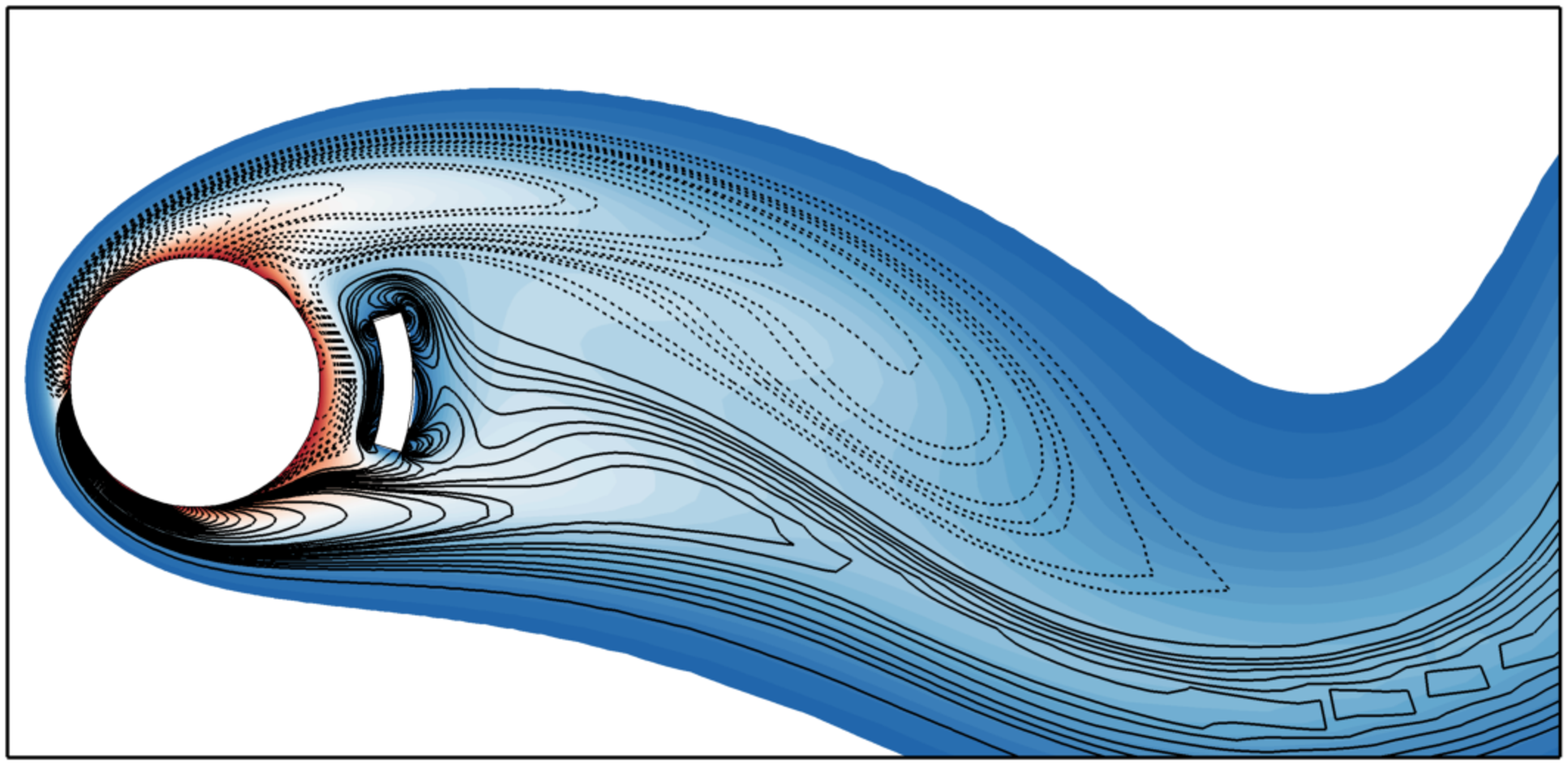}
\\
\scriptsize{$t=t_0+(1)T$}
\\
\includegraphics[width=0.43\textwidth,trim={0.7cm 0.7cm 8cm 0.7cm},clip]{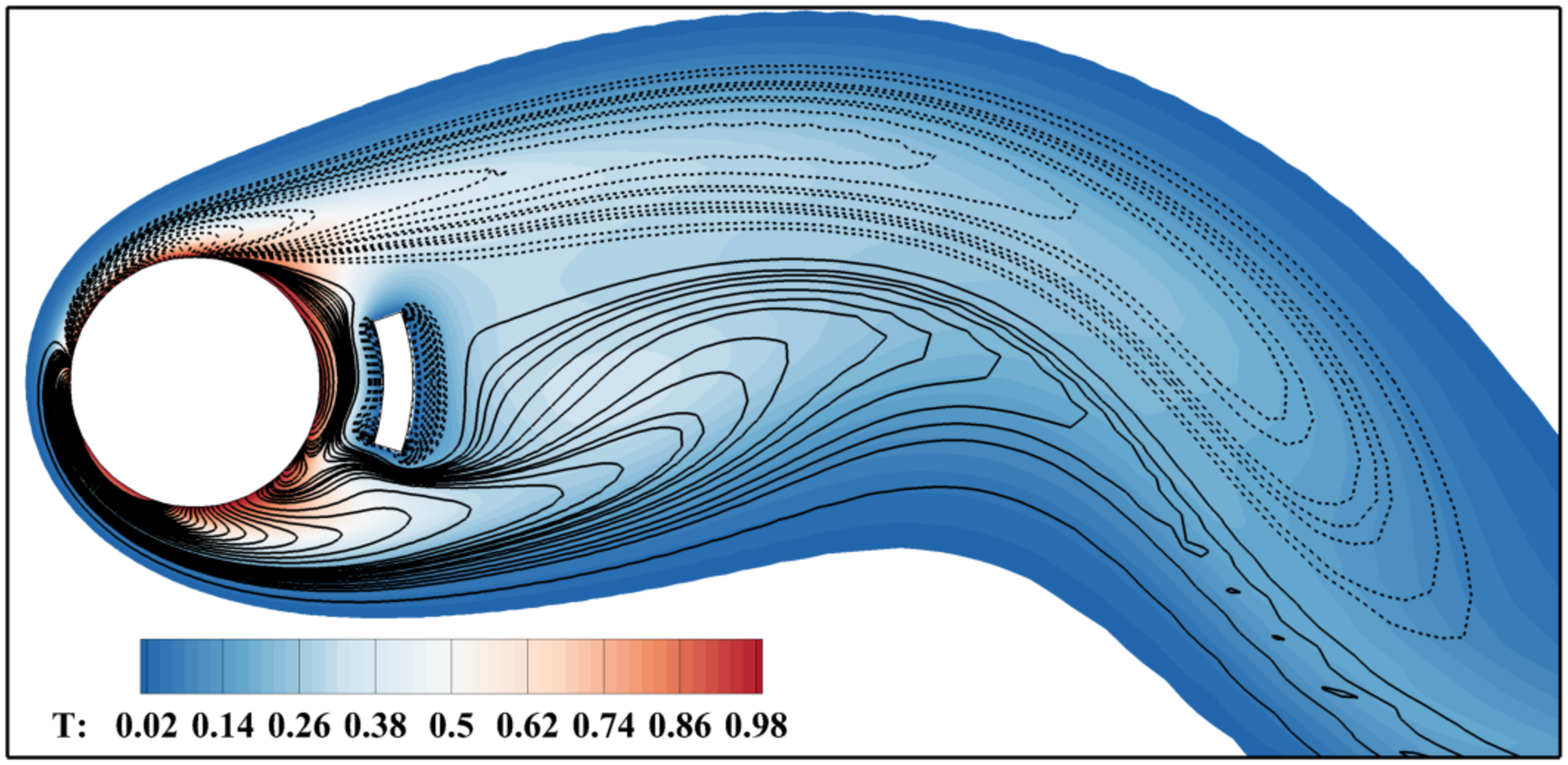}
\\
\caption{The distribution of the local Nusselt number $Nu$, over the surface of the cylinder and the control plate, and the isotherm contours during one oscillation period of the cylinder $T$, for $\alpha_m=0.5$ and $f/f_0=0.5$ with $d/R_0=0.5$ and $Re=150$.}
 \label{fig:d_0-5_a_0-5_f_0-5}
\end{figure*}

\cref{fig:d_0-5_a_0-5_f_0-5} illustrates the isotherm contours and the fluctuation of local Nusselt number during a single cylinder oscillation period $T$, given the following values: $Re=150$, $\alpha_m=0.5$, $f/f_0=0.5$ and $d/R_0=0.5$. In this case, the mode of vortex shedding is categorised as $2S(T)$ and is locked on for one oscillation cycle of the cylinder. A preliminary study of these isotherm contours reveals emerging vortices from either side of the cylinder as lumps of hot fluid convecting downstream. The vortex street is almost symmetric around the x-axis. Large temperature gradients are present close to the cylinder surface because of a very thin thermal boundary layer. High isothermal concentrations near the cylinder and the control plate and low isothermal concentrations away from them suggest that the heat convects away quickly to downstream.\\

\cref{fig:d_0-5_a_0-5_f_0-5} shows that the highest values of the local Nusselt number distribution on the cylinder are close to point, $\theta\approx\ang{180}$. There is a higher local heat flux near $\theta=\ang{180}$, which is supported by the fact that the locations of greatest values for local Nusselt number do not change much during the cylinder oscillation period. The fluid in $\theta\approx\ang{180}$ of the cylinder flows along the surface of the cylinder as it oscillates in the flow, thickening the thermal boundary layer and reducing the local heat transfer ($Nu$) in the wake. The control plate is placed behind the cylinder where the surface of the control plate is colder than the surface of the cylinder. As a result, some of the heat transferred from the cylinder surface will be absorbed on the control plate, which can be confirmed by the negative local Nusselt number on the surface of the control plate. This phenomenon is called ``backward heat transfer'' \cite{huang2015natural2}. A similar phenomenon is reported for heat transfer from a circular cylinder heated at a time-periodic pulsating temperature \cite{ray2022heat}. The lowest peaks of the local Nusselt number are observed at the top ($CD$) and bottom ($AB$) surfaces of the control plate, i.e., these parts of the surface will absorb the highest amount of heat from the fluid. The maximum peak of local Nusselt number is slightly reduced at $t=t_0+(1/4)T$ and $t_0+(3/4)T$. Additional local maximum peaks of local Nusselt number distribution are found between $\ang{0}$ and $\ang{60}$ and between $\ang{300}$ and $\ang{360}$. The local maximum peak values sharply rise at $t=t_0+(1/4)T$ and $t_0+(3/4)T$. This enhancement near $\theta\approx\ang{360}$ is caused by the vortex shedding phenomenon in the wake. The $Nu$ distribution demonstrates that the distribution curves at $t=t_0+(0)T$ and $t_0+(1/4)T$ are reflections of the distribution curves at $t=t_0+(1/2)T$ and $t_0+(3/4)T$, respectively. This suggests that the heat convection process that took place during the beginning half of the oscillation period will repeat itself during the second half on the opposite surface of the cylinder.\\

\begin{figure*}[!t]
\centering
\includegraphics[width=0.45\textwidth,trim={0cm 0cm 0cm 0cm},clip]{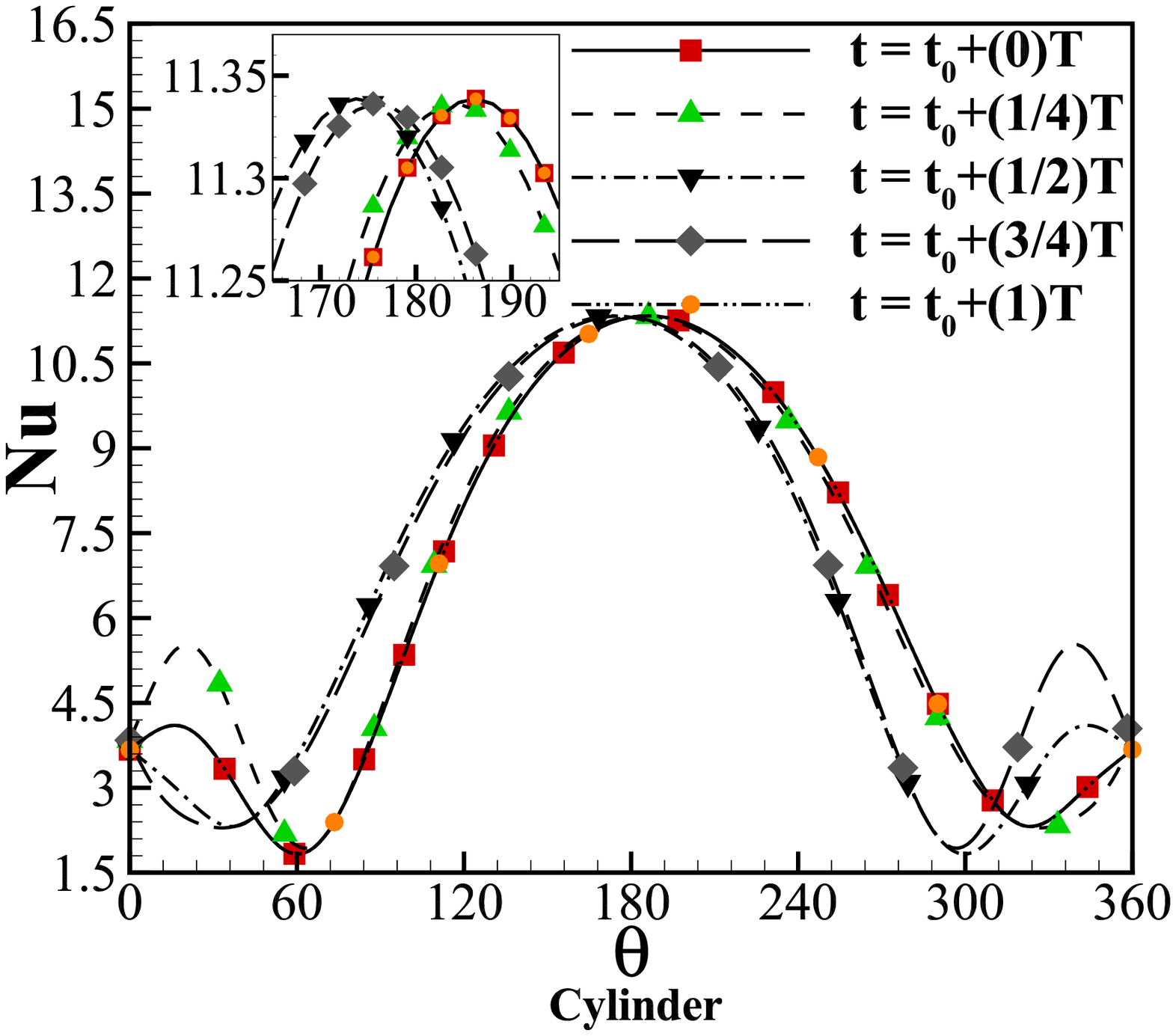}
\includegraphics[width=0.45\textwidth,trim={0cm 0cm 0cm 0cm},clip]{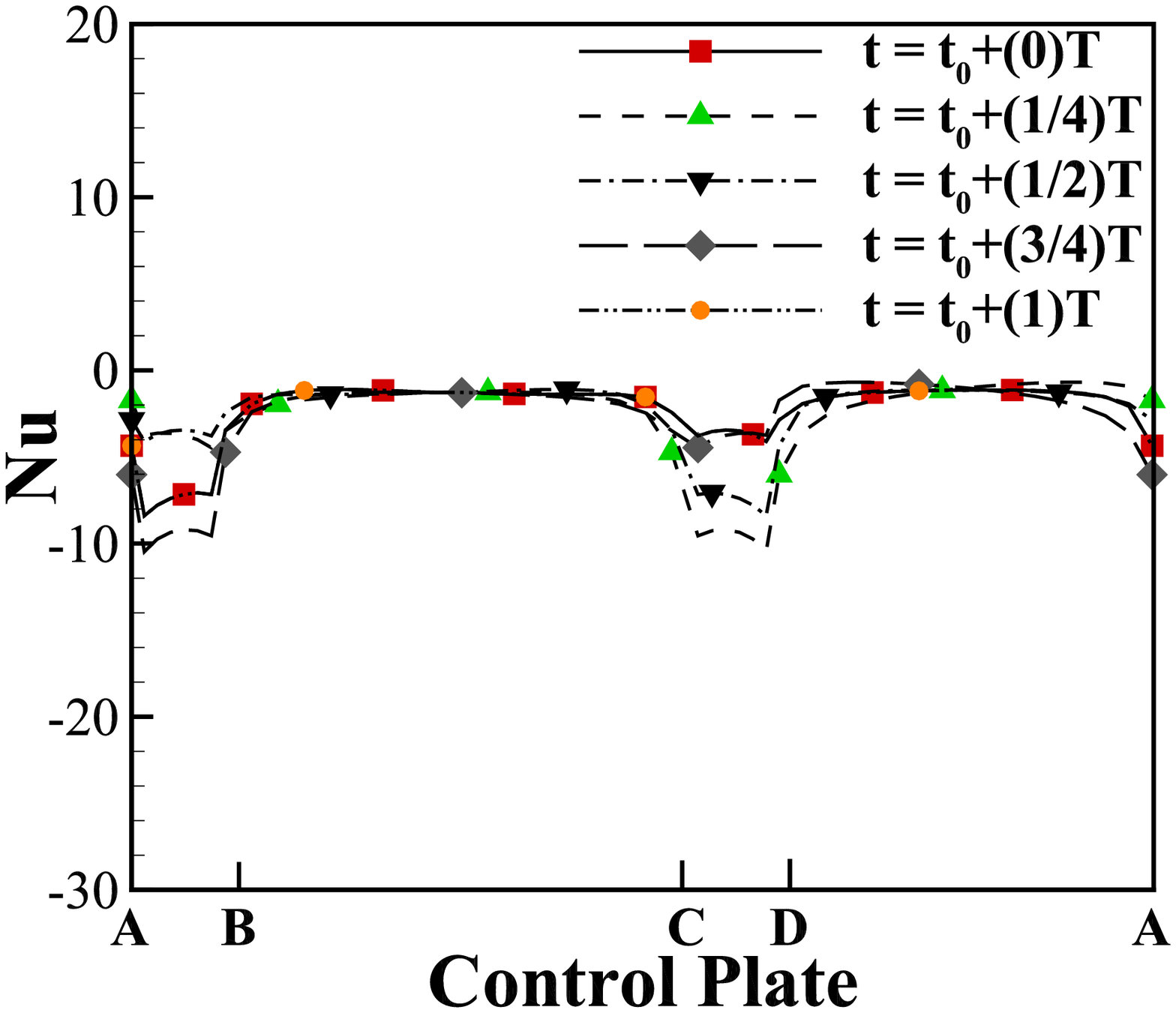}
\\
\scriptsize{$t=t_0+(0)T$}\hspace{4cm}\scriptsize{$t=t_0+(1/2)T$}
\\
\includegraphics[width=0.45\textwidth,trim={0.7cm 0.7cm 8cm 0.7cm},clip]{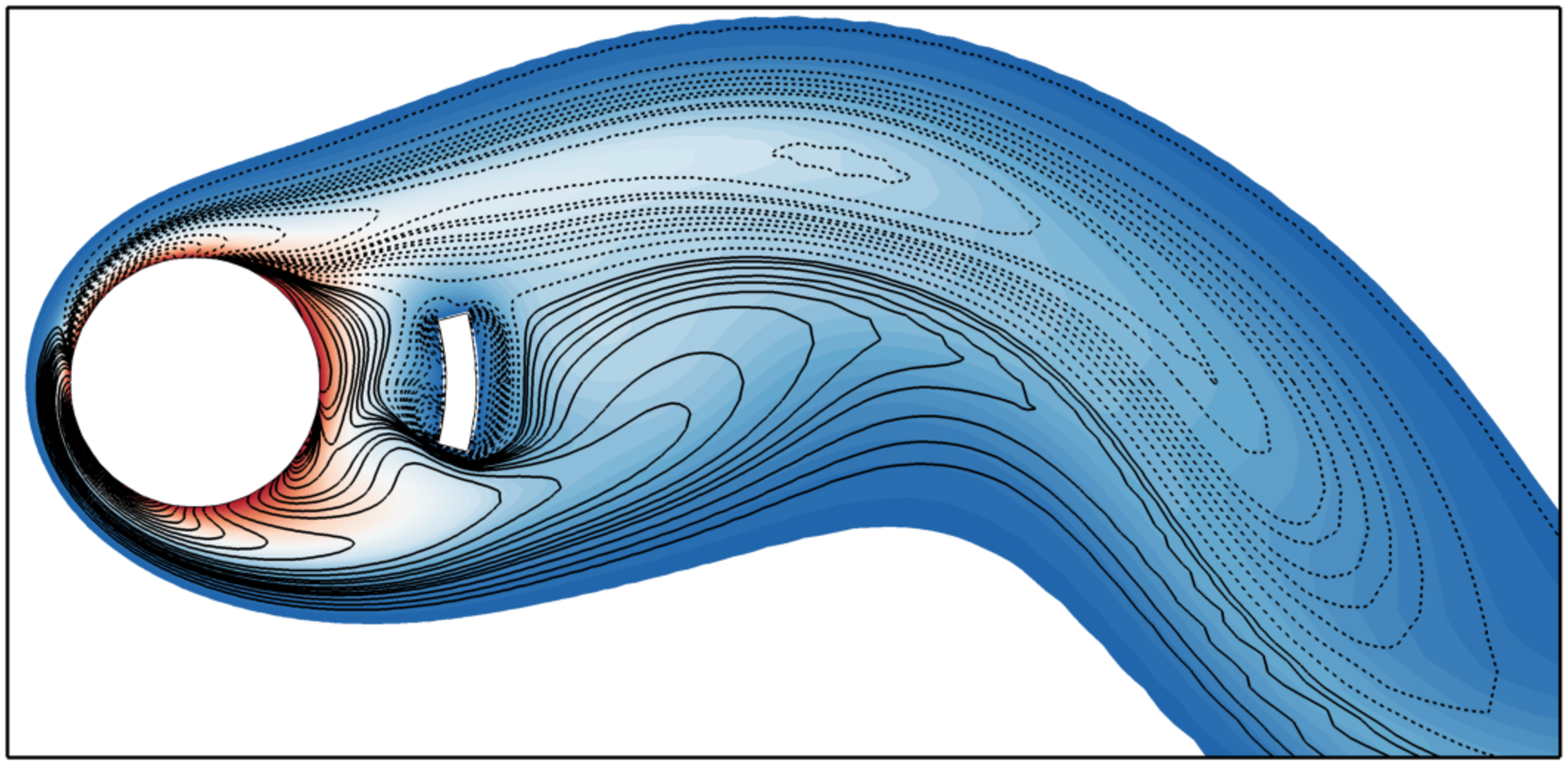}
\includegraphics[width=0.45\textwidth,trim={0.7cm 0.7cm 8cm 0.7cm},clip]{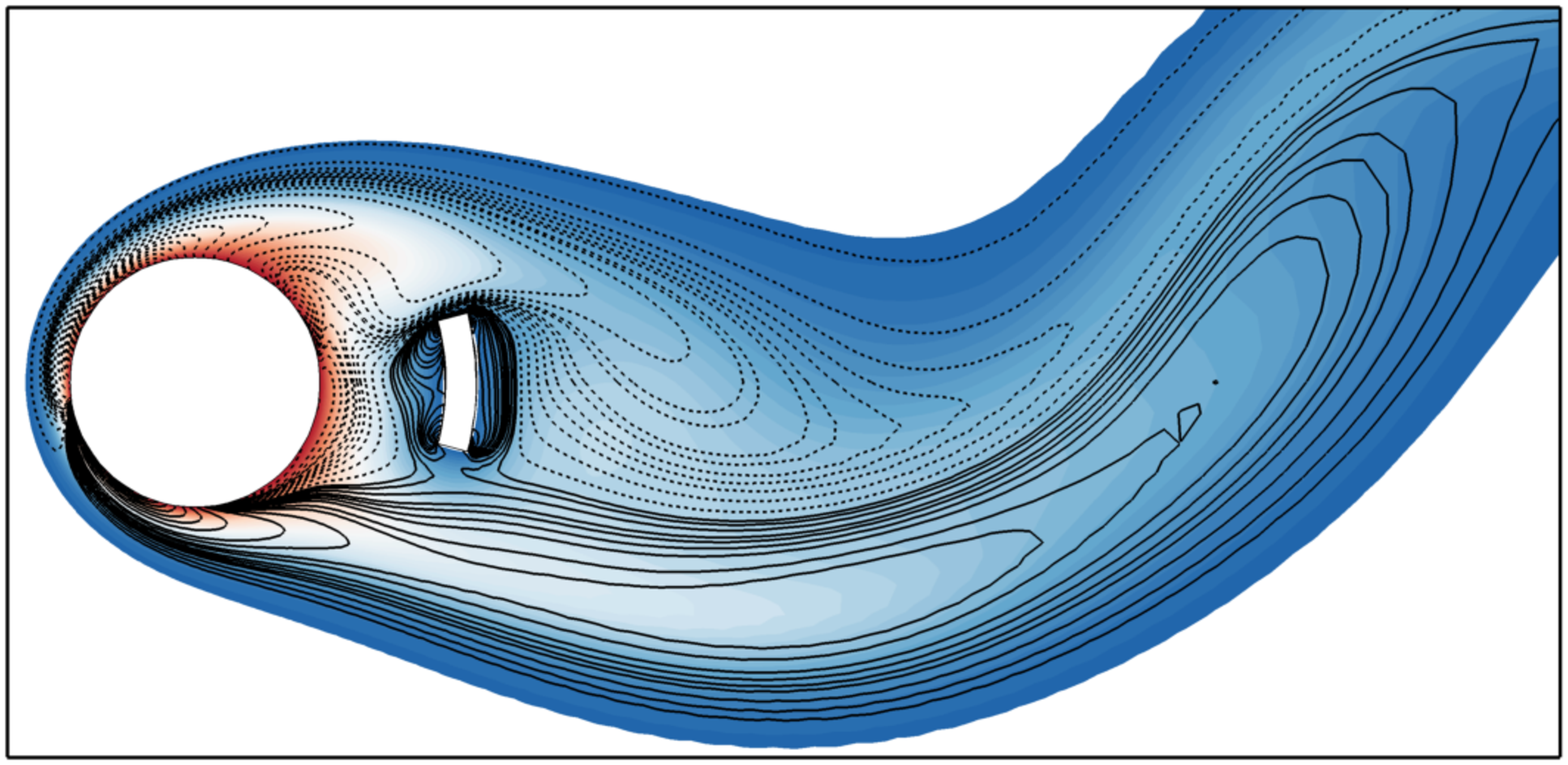}
\\
\scriptsize{$t=t_0+(1)T$}
\\
\includegraphics[width=0.45\textwidth,trim={0.7cm 0.7cm 8cm 0.7cm},clip]{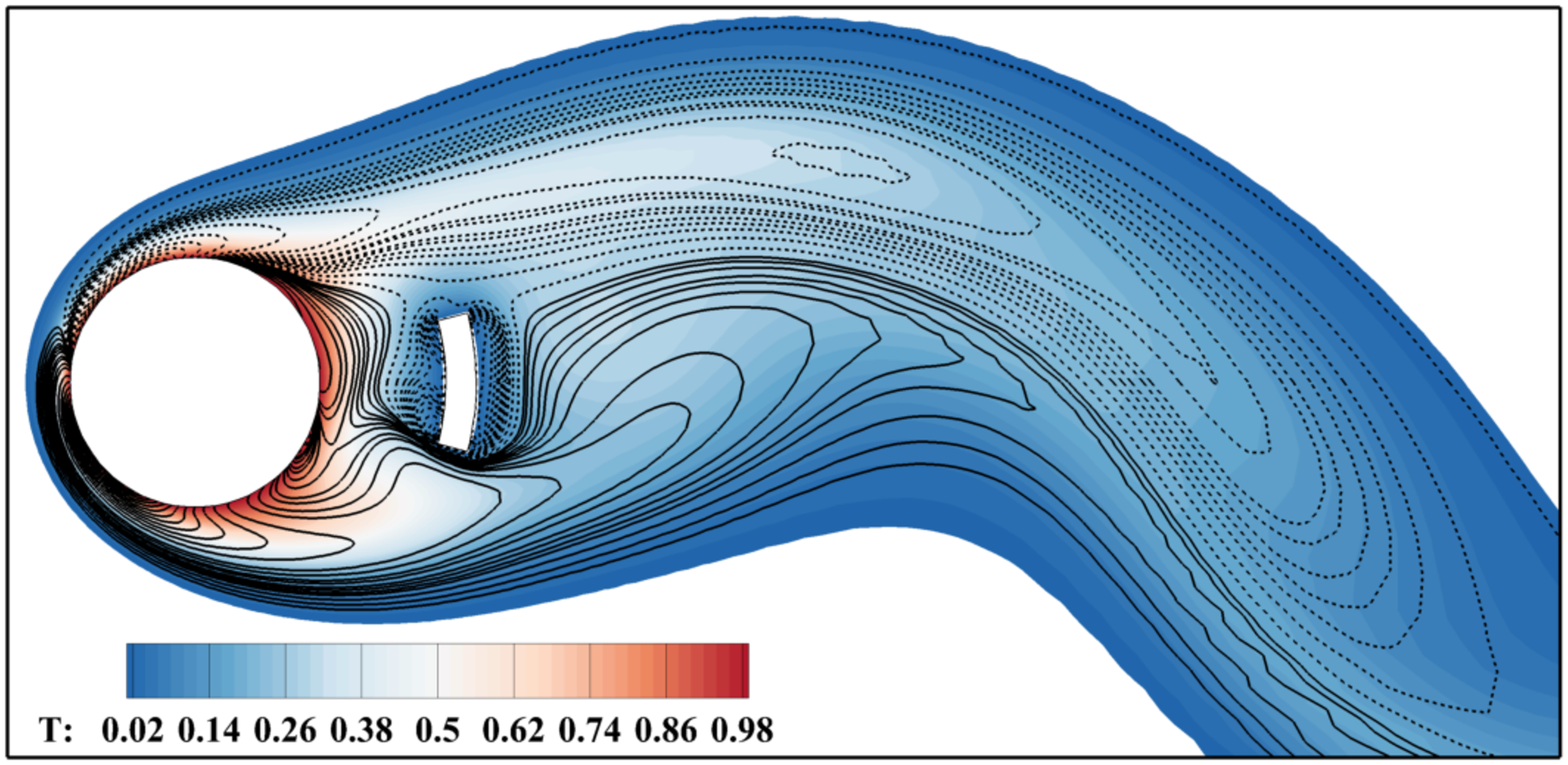}
\\
\caption{The distribution of the local Nusselt number $Nu$, over the surface of the cylinder and the control plate, and the isotherm contours during one oscillation period of the cylinder $T$, for $\alpha_m=0.5$ and $f/f_0=0.5$ with $d/R_0=1$ and $Re=150$.}
 \label{fig:d_1_a_0-5_f_0-5}
\end{figure*}

The local Nusselt number distribution and the isotherm contours are presented in \cref{fig:d_1_a_0-5_f_0-5} as the gap ratio of control plate is increased to $d/R_0=1$. At this gap ratio, the vortex shedding is locked-on for one oscillation period $T$, and the mode of vortex shedding is identified as $2S(T)$, as in the case of $d/R_0=0.5$. The heat convection mechanism is also similar to that of $d/R_0=0.5$. However, the differences are observed for the distribution of the local Nusselt number. The maximum peak is found near $\theta\approx \ang{180}$, and the maximum value is nearly constant throughout the oscillation period. Local maximum peaks of local Nusselt number distribution are observed within the areas $\ang{0}<\theta<\ang{60}$ and $\ang{300}<\theta<\ang{360}$. The local maximum peak sharply rises to maximum at $t=t_0+(1/4)T$ and $t_0+(3/4)T$. The distribution curves at these two phases are mirror images of one another, as seen for $d/R_0=0.5$. However, the maximum value of the local Nusselt number for $d/R_0=1$ is significantly lower than that of $d/R_0=0.5$. The local Nusselt number distribution on the surface of the control plate suggests that the heat absorption is maximum at the top ($CD$) and bottom ($AB$) surfaces of the control plate for $d/R_0=1$. However, the rate of heat absorption is less than that of $d/R_0=0.5$. It occurs as lower gap ratio has a higher heat absorption rate than higher gap ratio because there are denser isotherm contours close to the surface of the cylinder than further downstream.\\

\begin{figure*}[!t]
\centering
\includegraphics[width=0.45\textwidth,trim={0cm 0cm 0cm 0cm},clip]{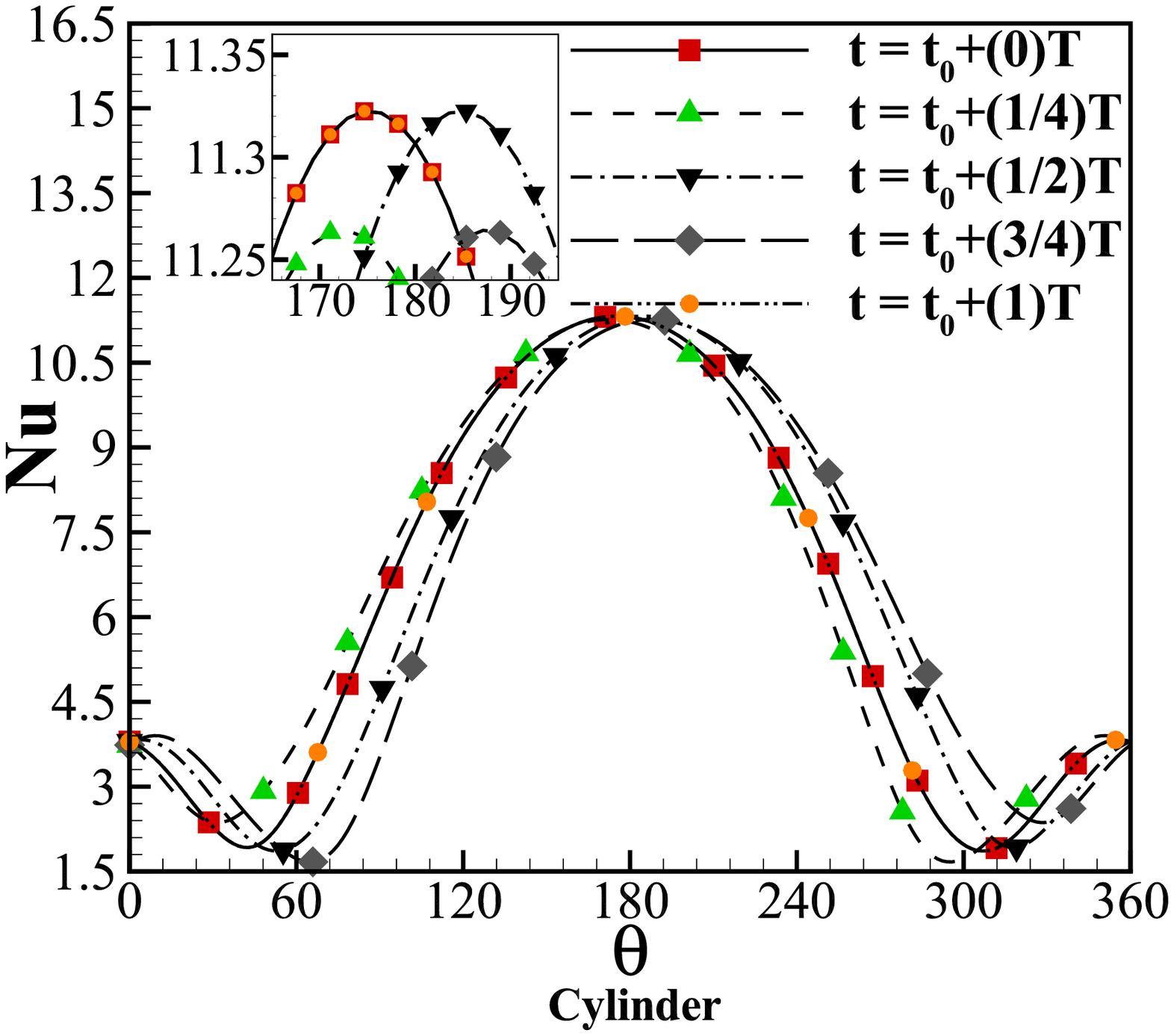}
\includegraphics[width=0.45\textwidth,trim={0cm 0cm 0cm 0cm},clip]{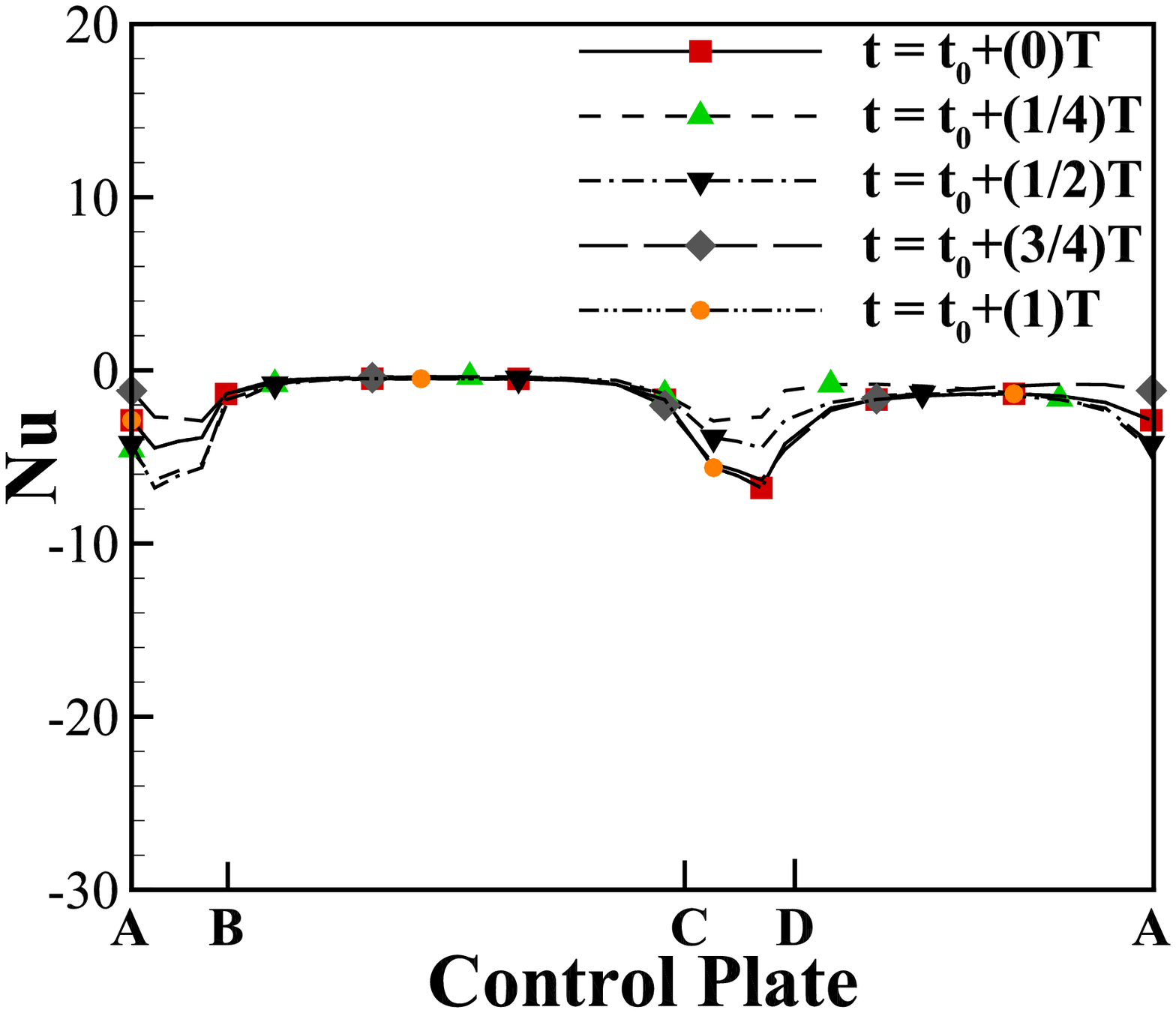}
\\
\scriptsize{$t=t_0+(0)T$}\hspace{4cm}\scriptsize{$t=t_0+(1/2)T$}
\\
\includegraphics[width=0.45\textwidth,trim={0.7cm 0.7cm 8cm 0.7cm},clip]{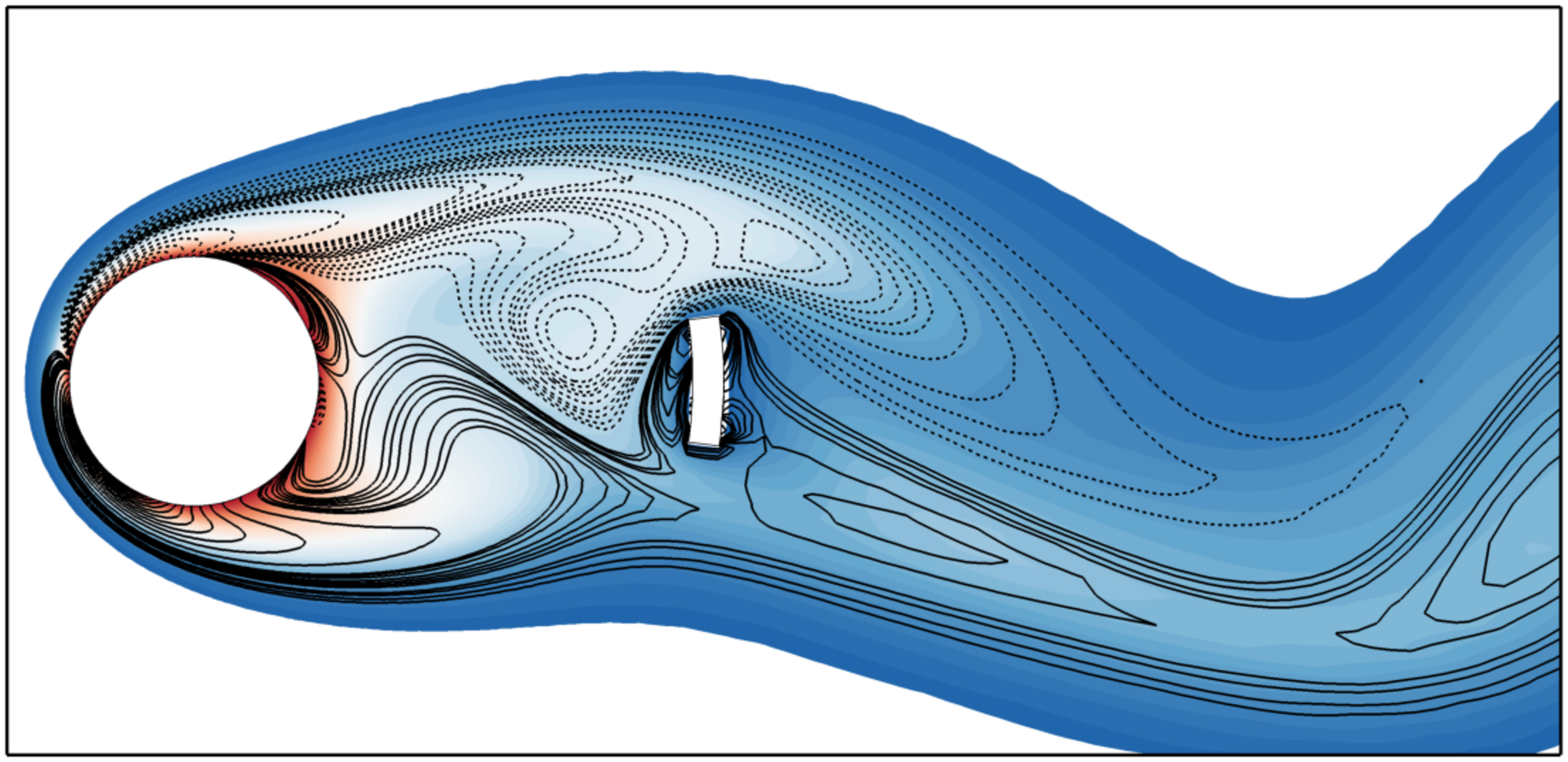}
\includegraphics[width=0.45\textwidth,trim={0.7cm 0.7cm 8cm 0.7cm},clip]{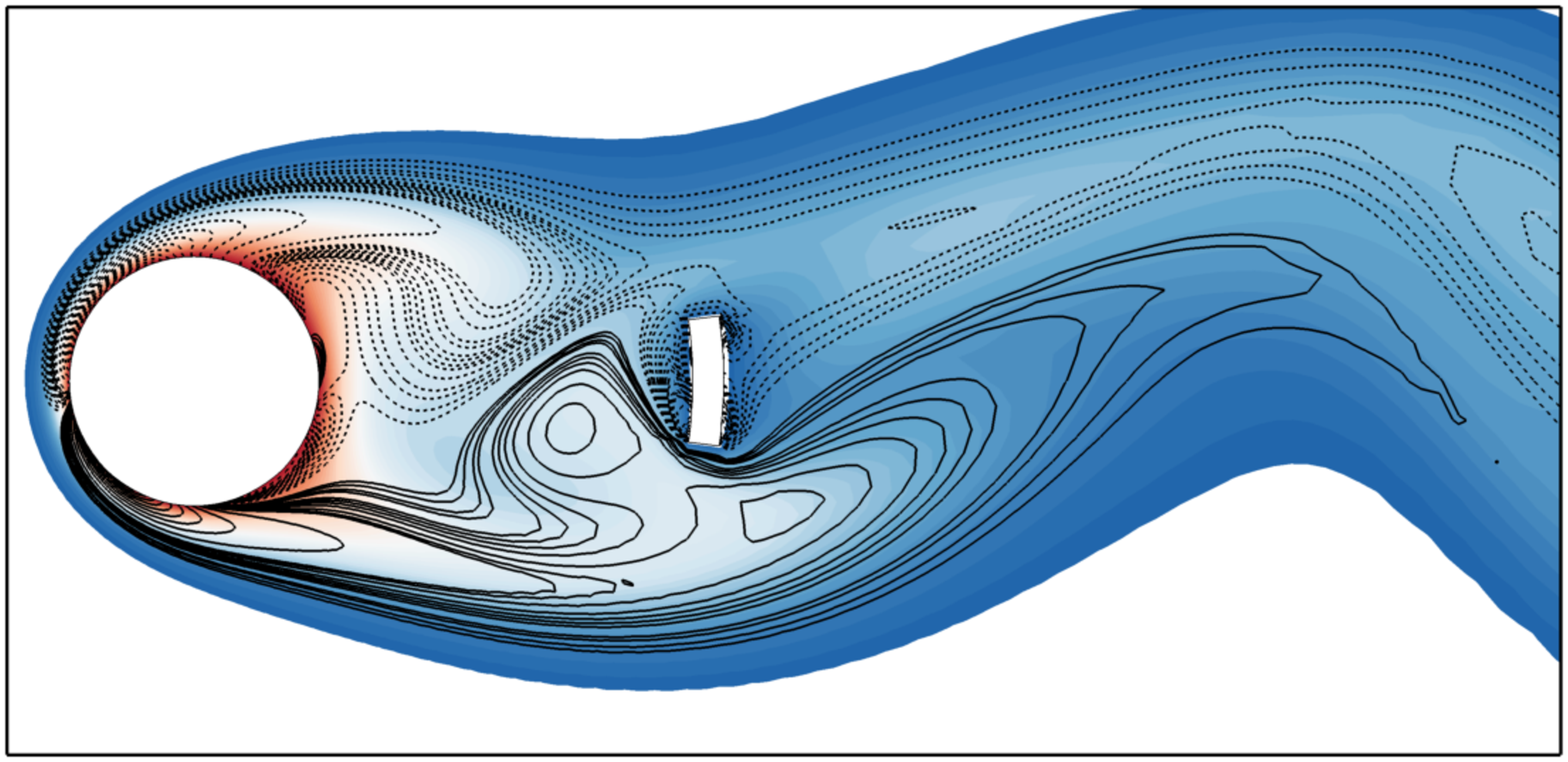}
\\
\scriptsize{$t=t_0+(1)T$}
\\
\includegraphics[width=0.45\textwidth,trim={0.7cm 0.7cm 8cm 0.7cm},clip]{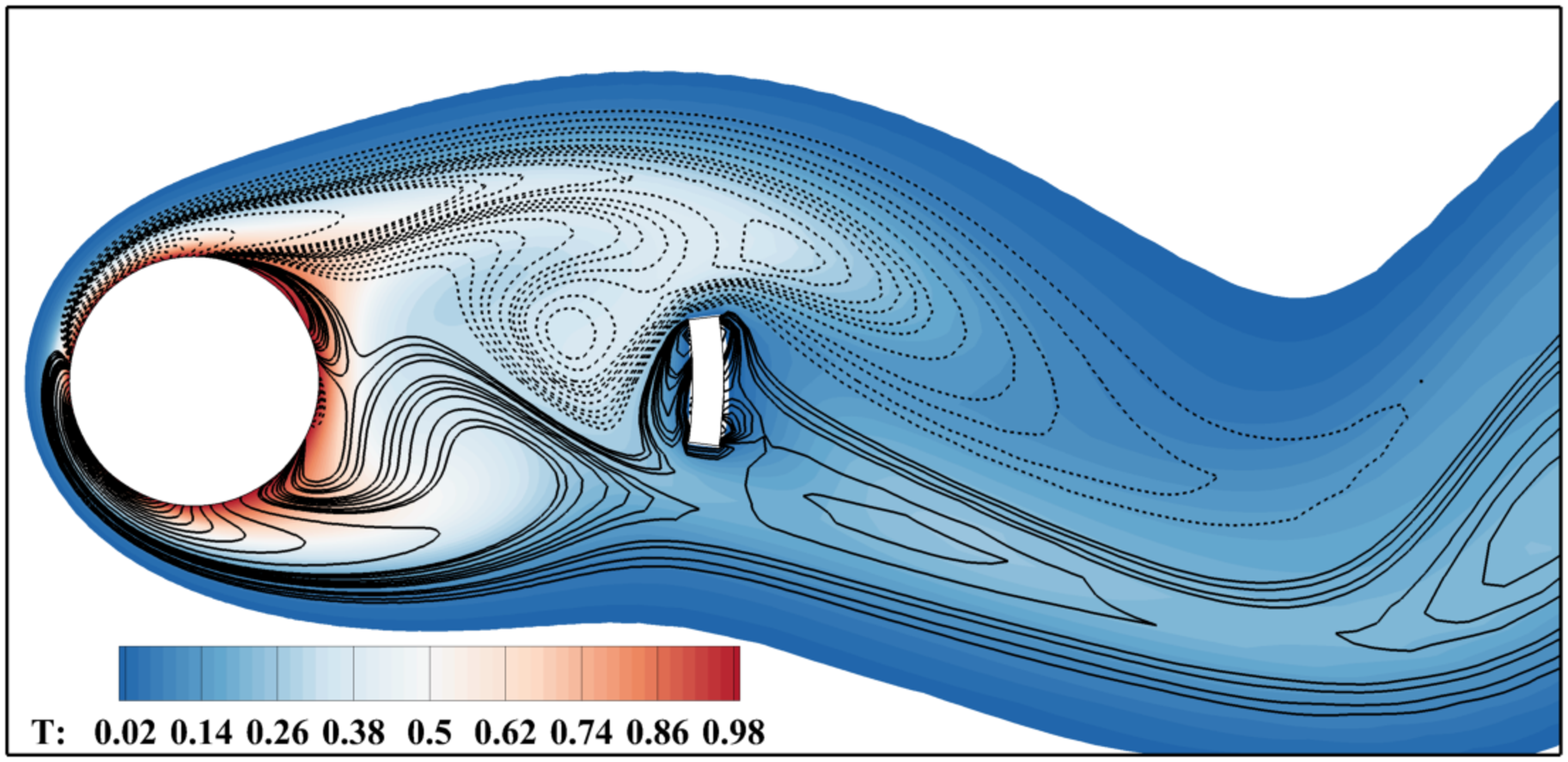}
\\
\caption{The distribution of the local Nusselt number $Nu$, over the surface of the cylinder and the control plate, and the isotherm contours during one oscillation period of the cylinder $T$, for $\alpha_m=0.5$ and $f/f_0=0.5$, with $d/R_0=3$ and $Re=150$.}
 \label{fig:d_3_a_0-5_f_0-5}
\end{figure*}

Further increasing the gap ratio to $d/R_0=3$ for the same rotary oscillation parameters, the local Nusselt number distribution and the isotherm contours are displayed in the \cref{fig:d_3_a_0-5_f_0-5} and similar vortex shedding mode of $2S(T)$ is observed. The vortices shed are smaller than that of $d/R_0=0.5,\ 1$. Two small vortices are seen to appear in the wake between the cylinder and the control plate in clockwise and counter-clockwise directions. It occurs when the shear layers coming from the surface of the cylinder are separated by the shear layers of the opposite sign attached to the control plate near wake. The reduction in the size of the vortices from $d/R_0=0.5,\ 1$ causes a reduction in total heat transfer from the surface of the cylinder. This is backed by the significant drop in the local maximum peak of $Nu$ in $\ang{0}<\theta<\ang{60}$ and $\ang{300}<\theta<\ang{360}$. The highest peak of $Nu$ oscillates near $\theta\approx \ang{180}$. Maximum value is only observed at the phases $t=t_0+(0)T$, $t_0+(1/2)T$ and $t_0+(1)T$. Local maximum peak of local Nusselt number is found in $\ang{0}\leq\theta\leq\ang{20}$ and $\ang{340}\leq\theta\leq\ang{360}$. Also, the density of the isotherm contours is highest near the corners of the control plate, which implies higher heat transfer relative to the remaining regions on the surface of the plate. For the gap ratio $d/R_0=3$, the local Nusselt number distribution on the control plate surface shows that heat transfer from the front ($BC$) and rear ($DA$) surfaces of the control plate is nearly zero. The local minima of the local Nusselt number near the corners, top ($CD$) and bottom ($AB$) surfaces of the control plate increased for $d/R_0=3$ compared to that of $d/R_0=1$. It implies that the heat absorption rates near the corners and on the top ($CD$) and bottom ($AB$) surfaces of the control plate are higher than the other surfaces but significantly lower than \cref{fig:d_1_a_0-5_f_0-5}.\\

\begin{figure*}[!htbp]
\centering
\includegraphics[width=0.3\textwidth,trim={0cm 0cm 0cm 0cm},clip]{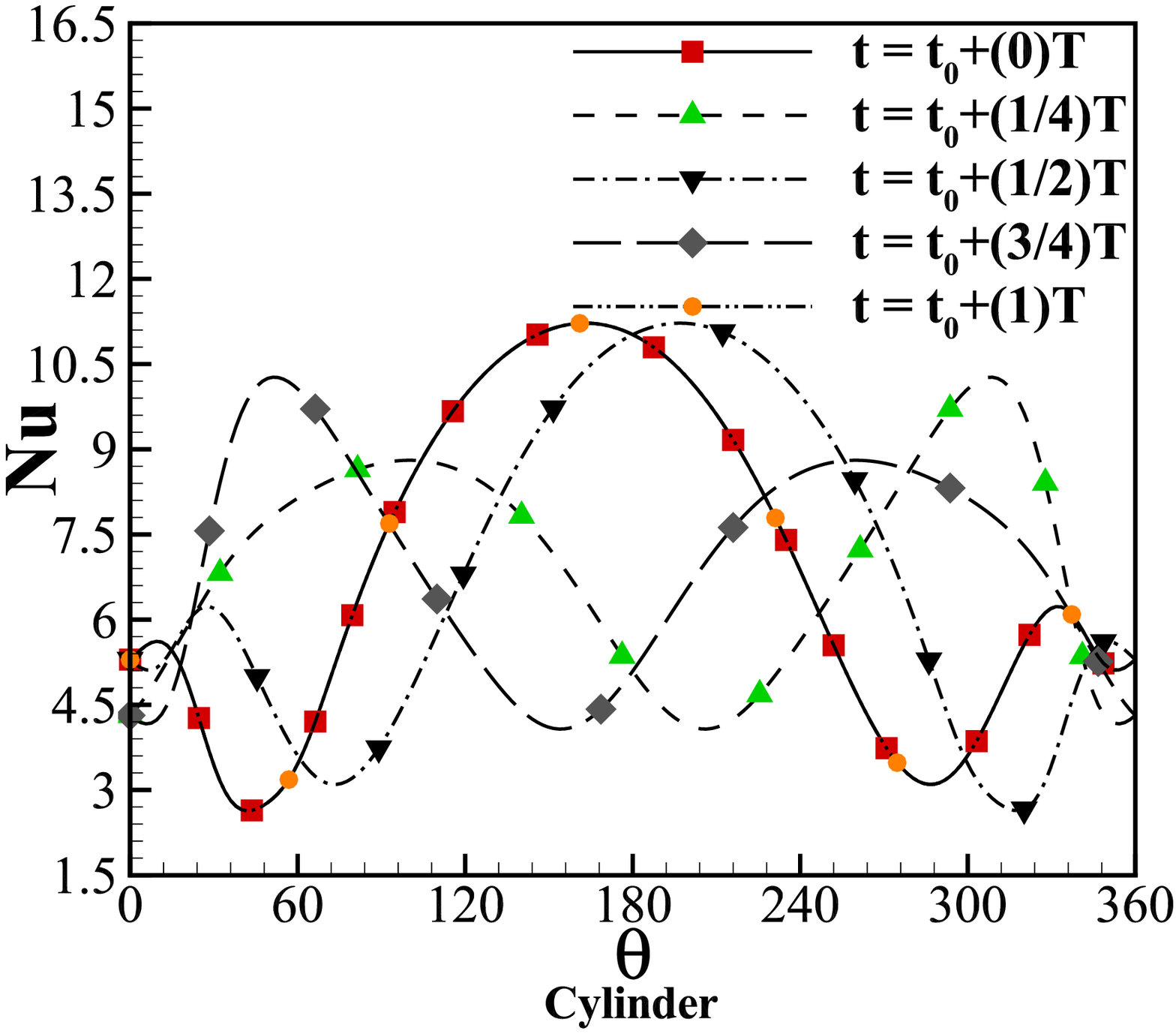}
\includegraphics[width=0.3\textwidth,trim={0cm 0cm 0cm 0cm},clip]{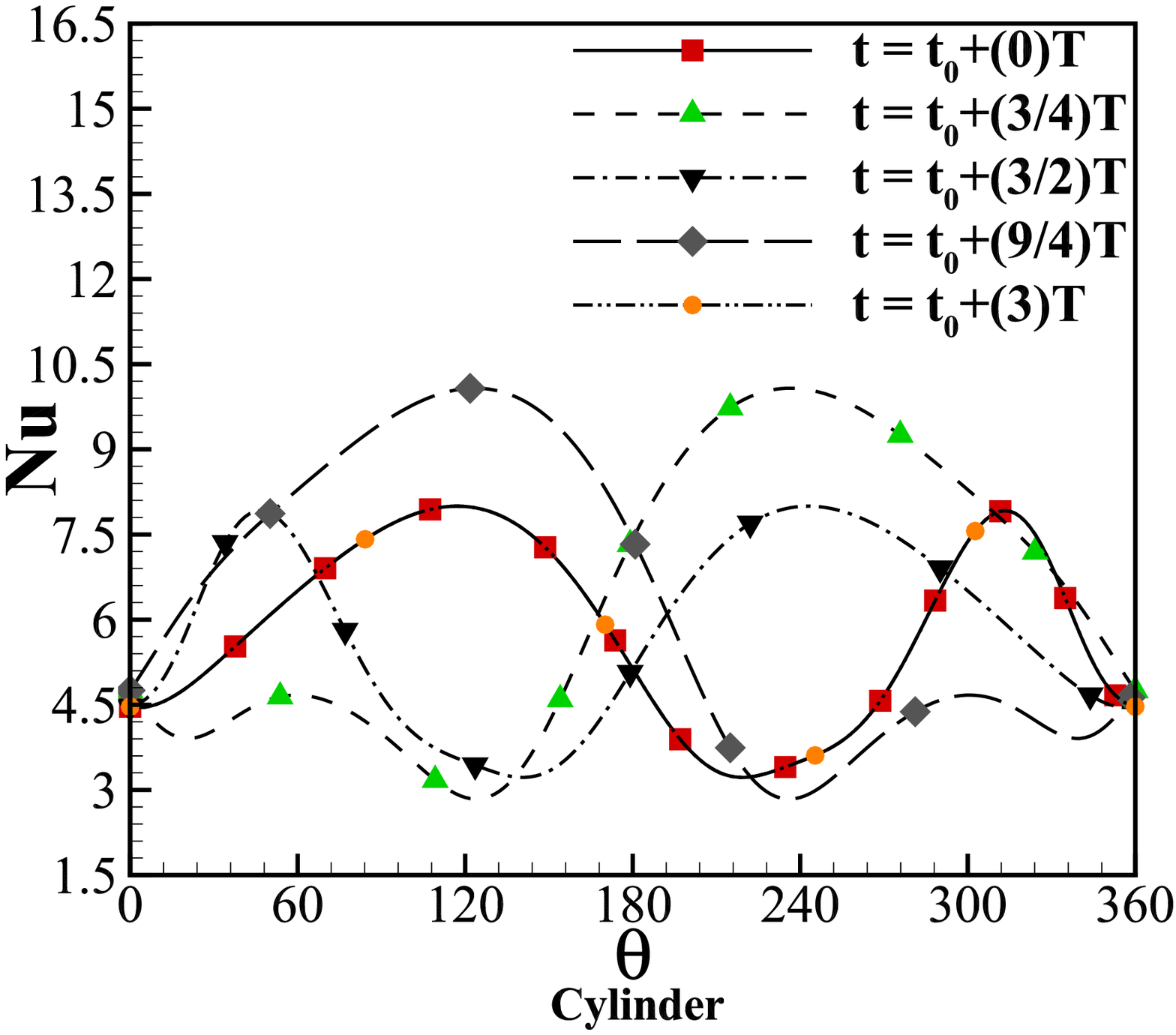}
\includegraphics[width=0.3\textwidth,trim={0cm 0cm 0cm 0cm},clip]{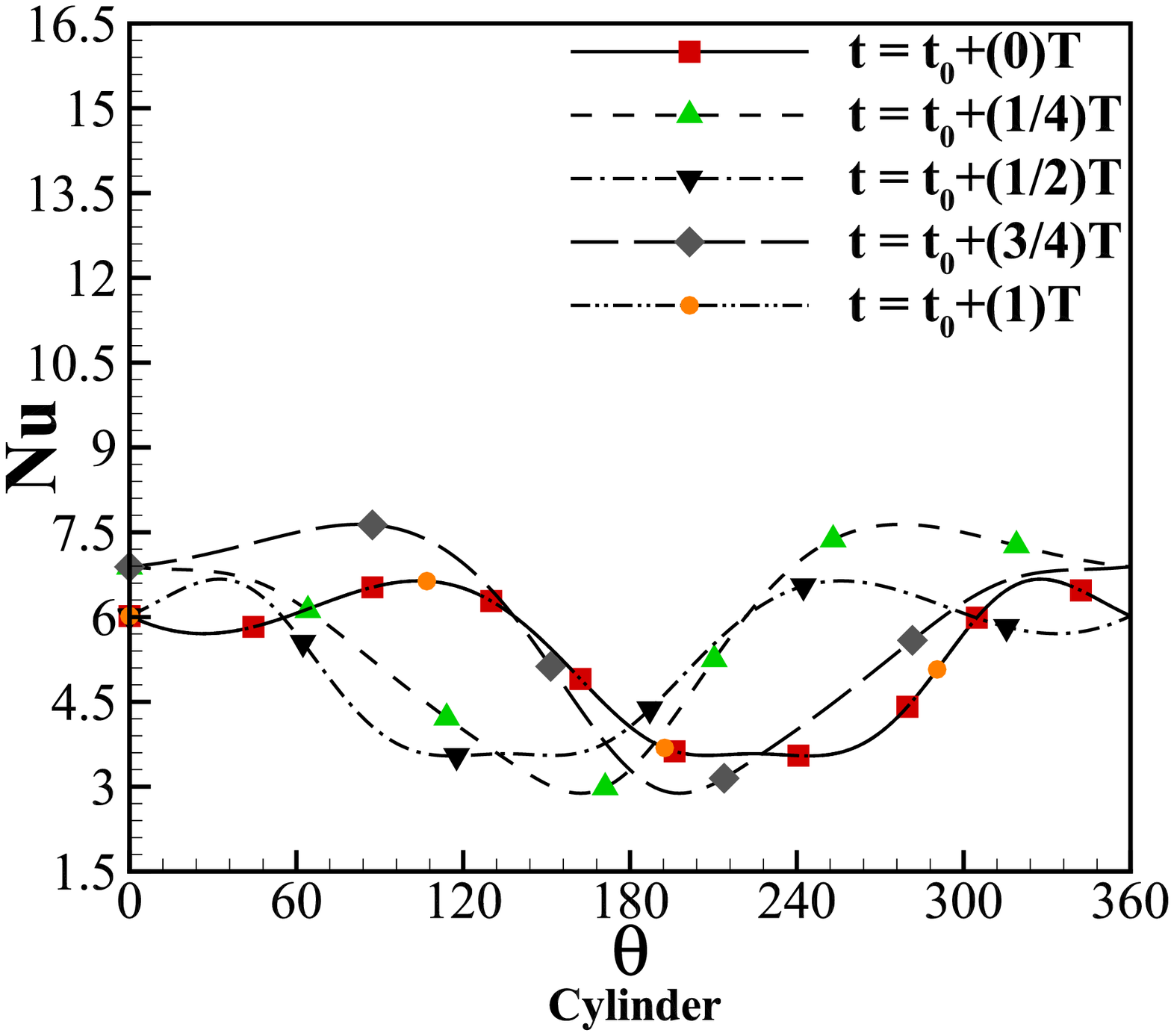}
\\
\hspace{0.5em}\scriptsize{$t=t_0+(1/4)T$}\hspace{9.5em}\scriptsize{$t=t_0+(3/4)T$}\hspace{9.5em}\scriptsize{$t=t_0+(1/4)T$}\hspace{0.5em}
\\
\includegraphics[width=0.3\textwidth,trim={0.7cm 0.7cm 8cm 0.7cm},clip]{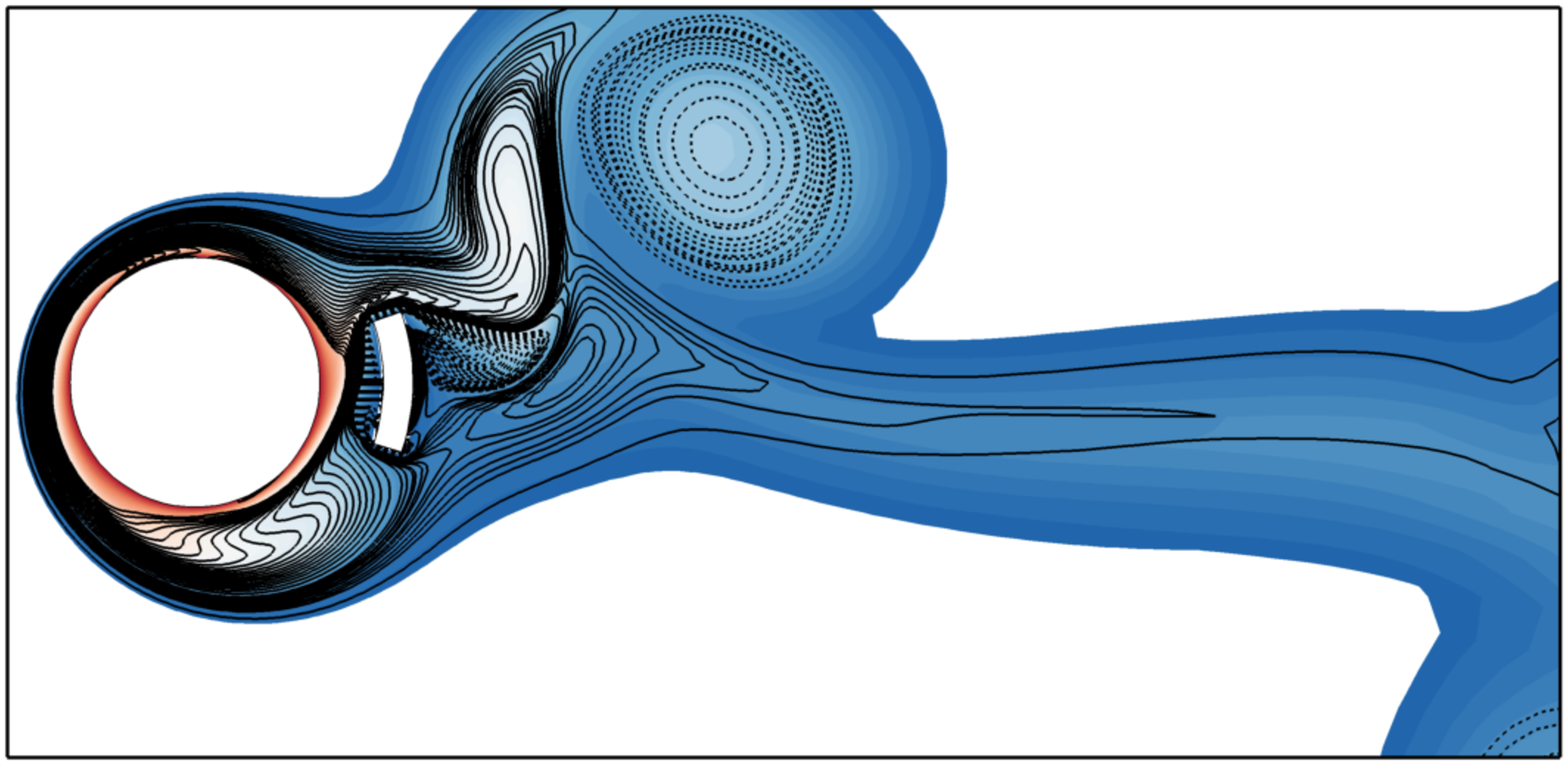}
\includegraphics[width=0.3\textwidth,trim={0.7cm 0.7cm 8cm 0.7cm},clip]{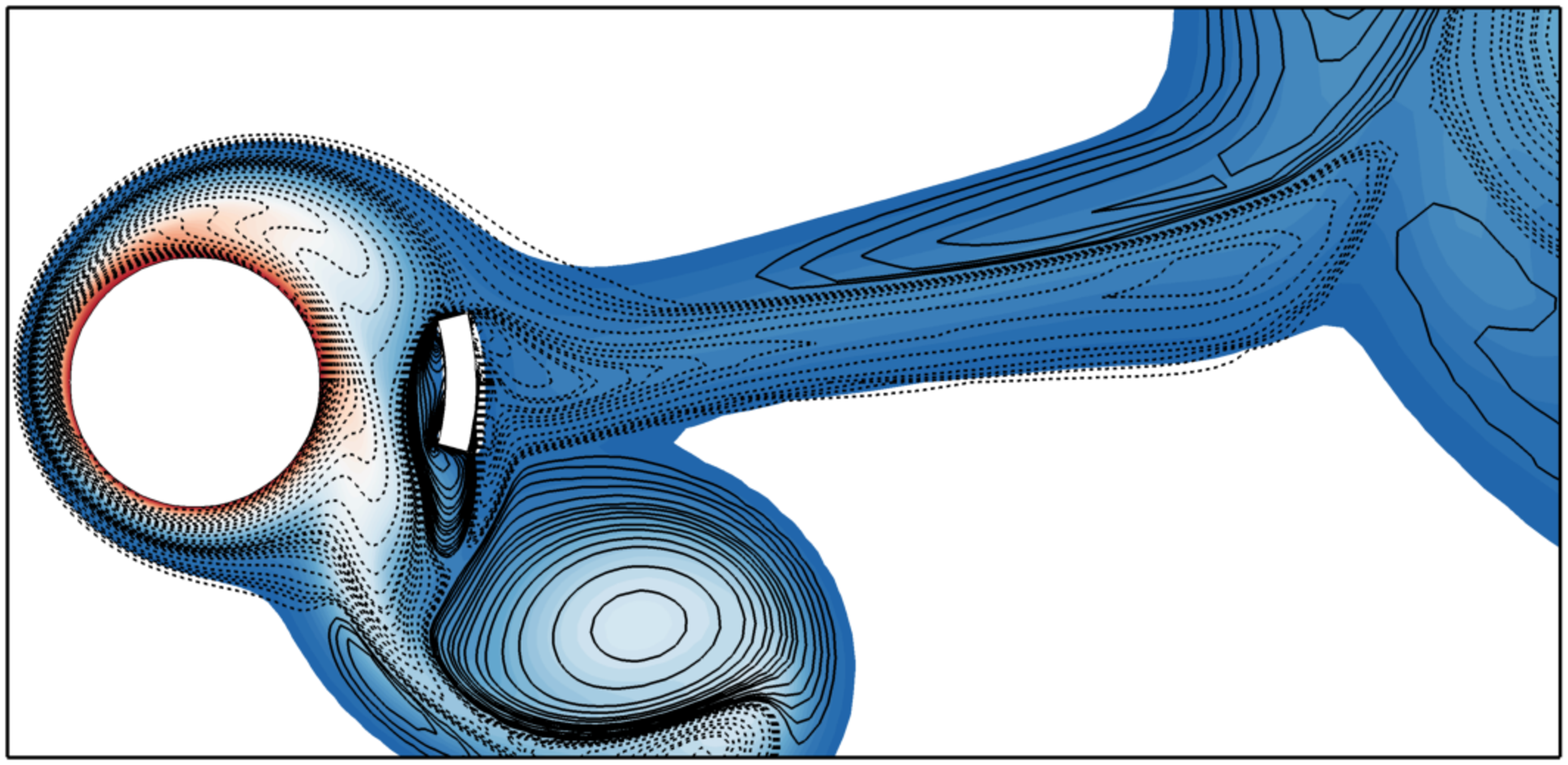}
\includegraphics[width=0.3\textwidth,trim={0.7cm 0.7cm 8cm 0.7cm},clip]{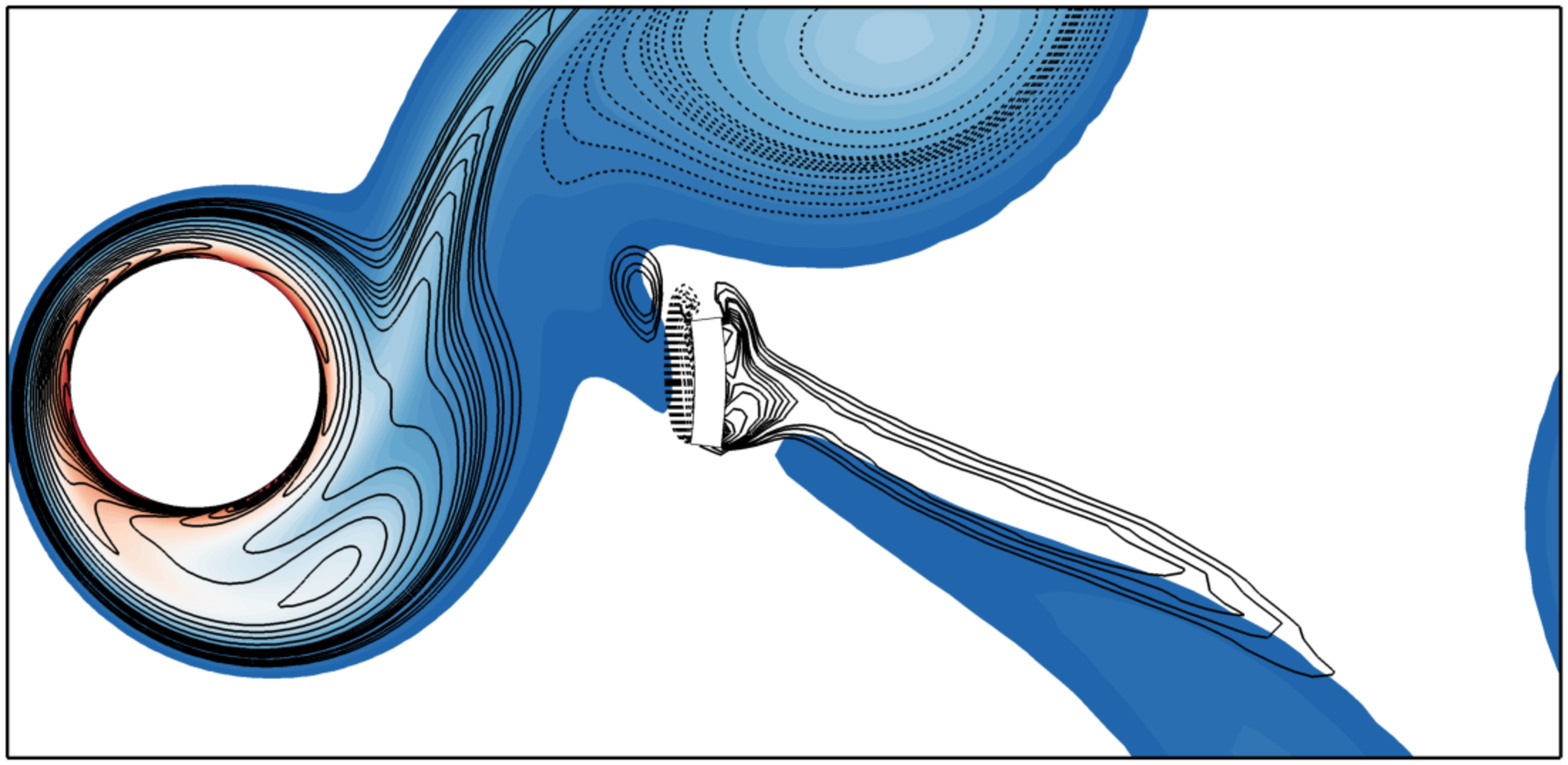}
\\
\hspace{0.5em}\scriptsize{$t=t_0+(1/2)T$}\hspace{9.5em}\scriptsize{$t=t_0+(3/2)T$}\hspace{9.5em}\scriptsize{$t=t_0+(1/2)T$}\hspace{0.5em}
\\
\includegraphics[width=0.3\textwidth,trim={0.7cm 0.7cm 8cm 0.7cm},clip]{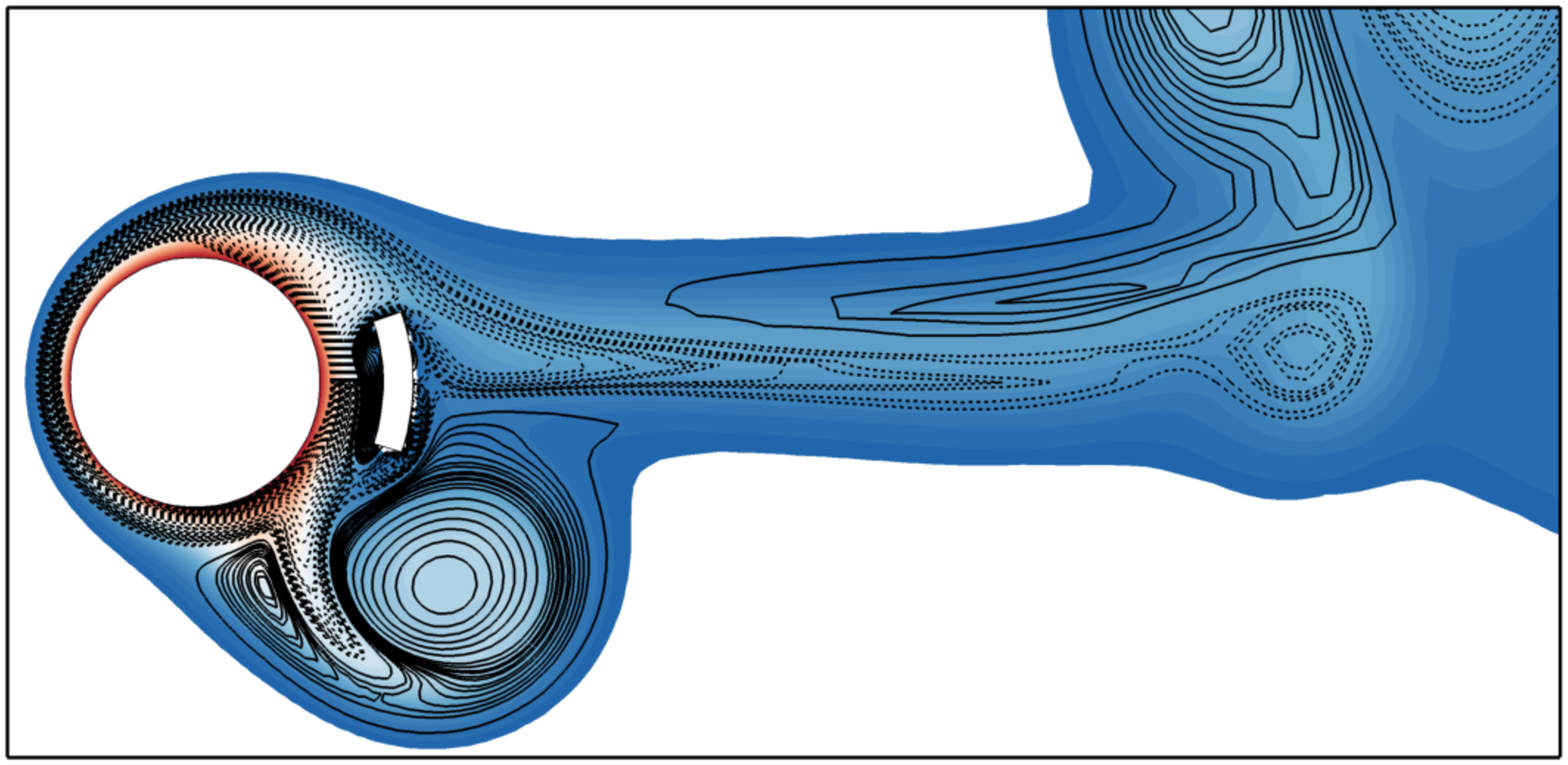}
\includegraphics[width=0.3\textwidth,trim={0.7cm 0.7cm 8cm 0.7cm},clip]{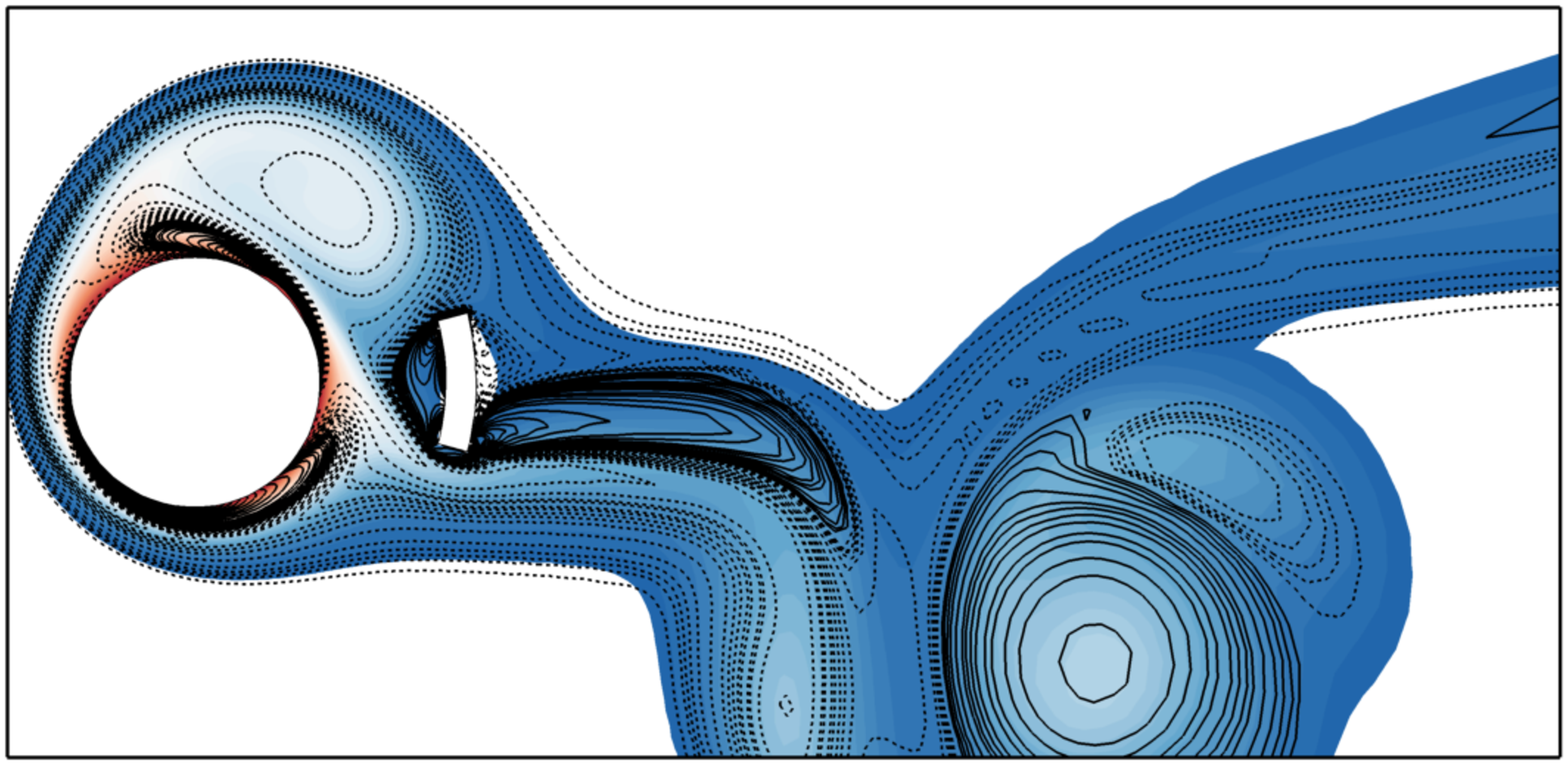}
\includegraphics[width=0.3\textwidth,trim={0.7cm 0.7cm 8cm 0.7cm},clip]{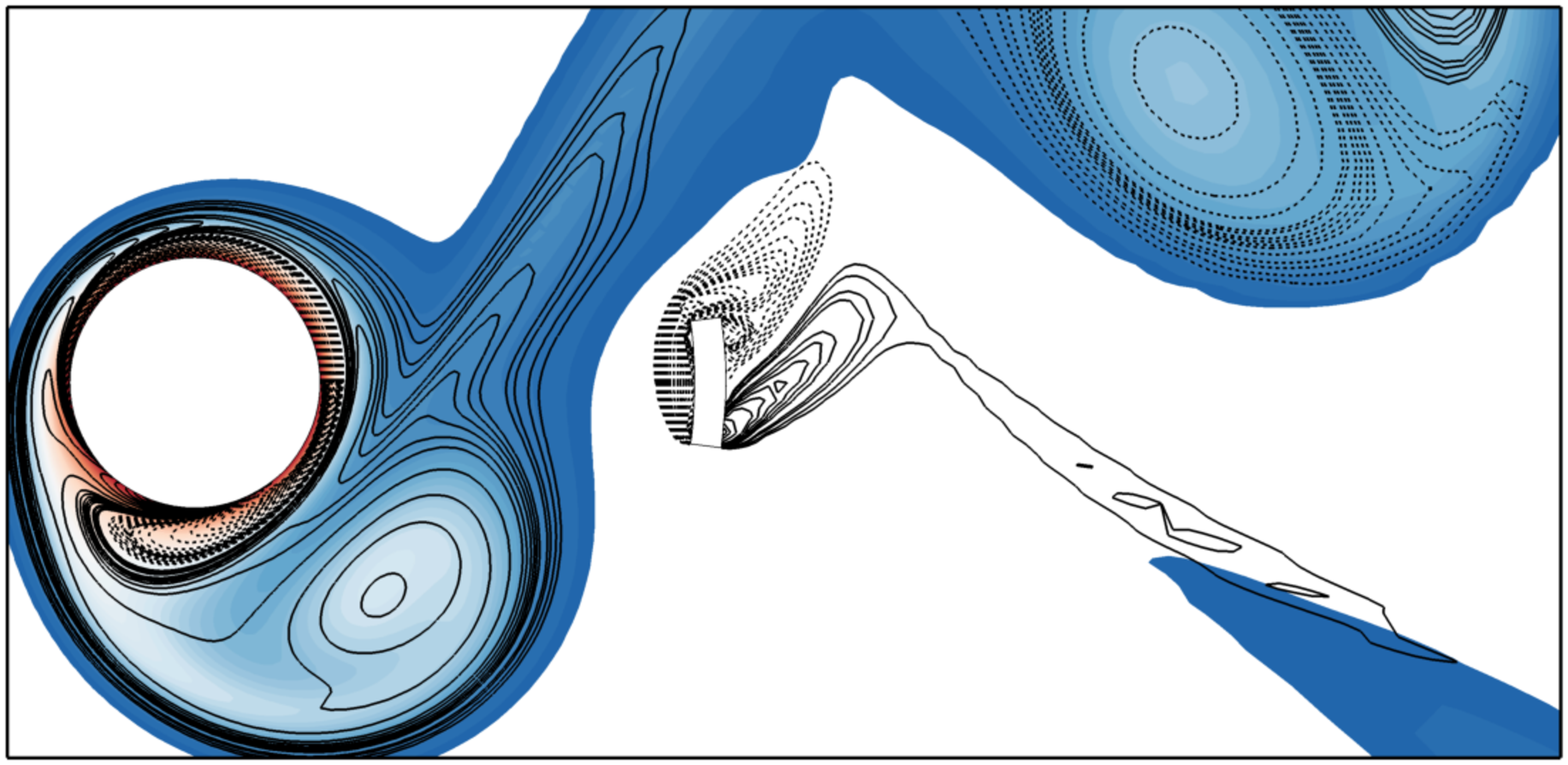}
\\
\hspace{0.5em}\scriptsize{$t=t_0+(3/4)T$}\hspace{9.5em}\scriptsize{$t=t_0+(9/4)T$}\hspace{9.5em}\scriptsize{$t=t_0+(3/4)T$}\hspace{0.5em}
\\
\includegraphics[width=0.3\textwidth,trim={0.7cm 0.7cm 8cm 0.7cm},clip]{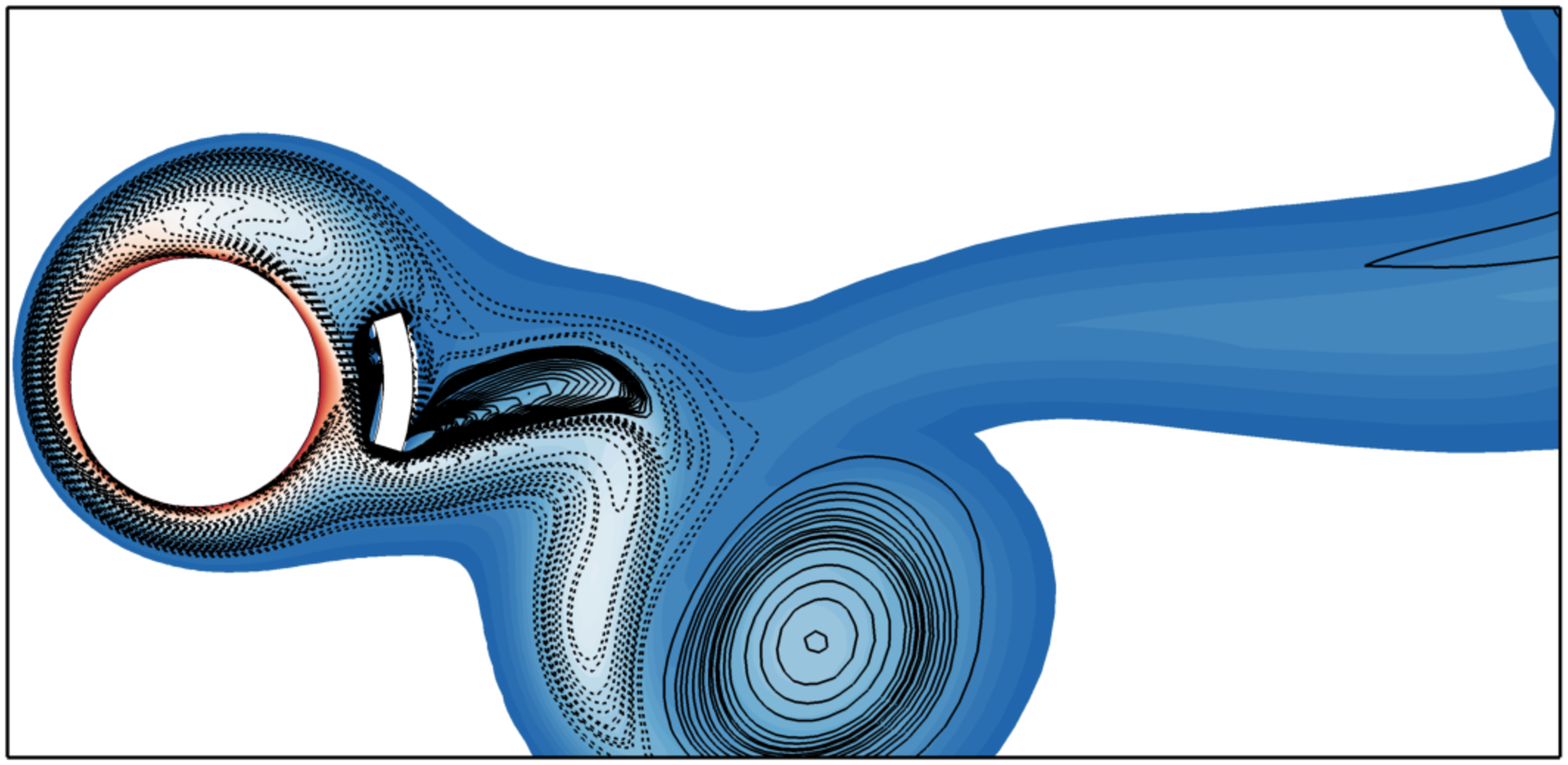}
\includegraphics[width=0.3\textwidth,trim={0.7cm 0.7cm 8cm 0.7cm},clip]{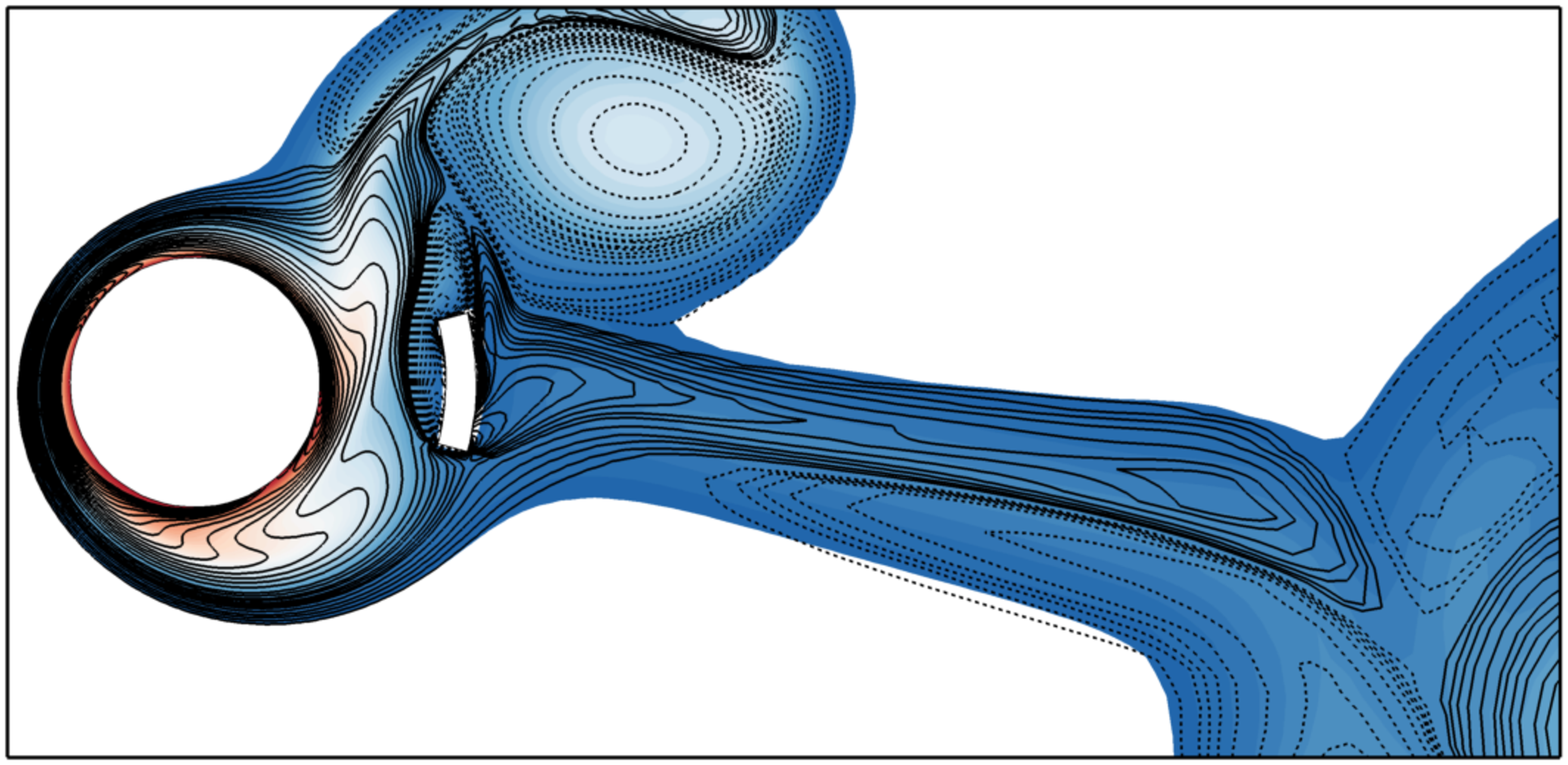}
\includegraphics[width=0.3\textwidth,trim={0.7cm 0.7cm 8cm 0.7cm},clip]{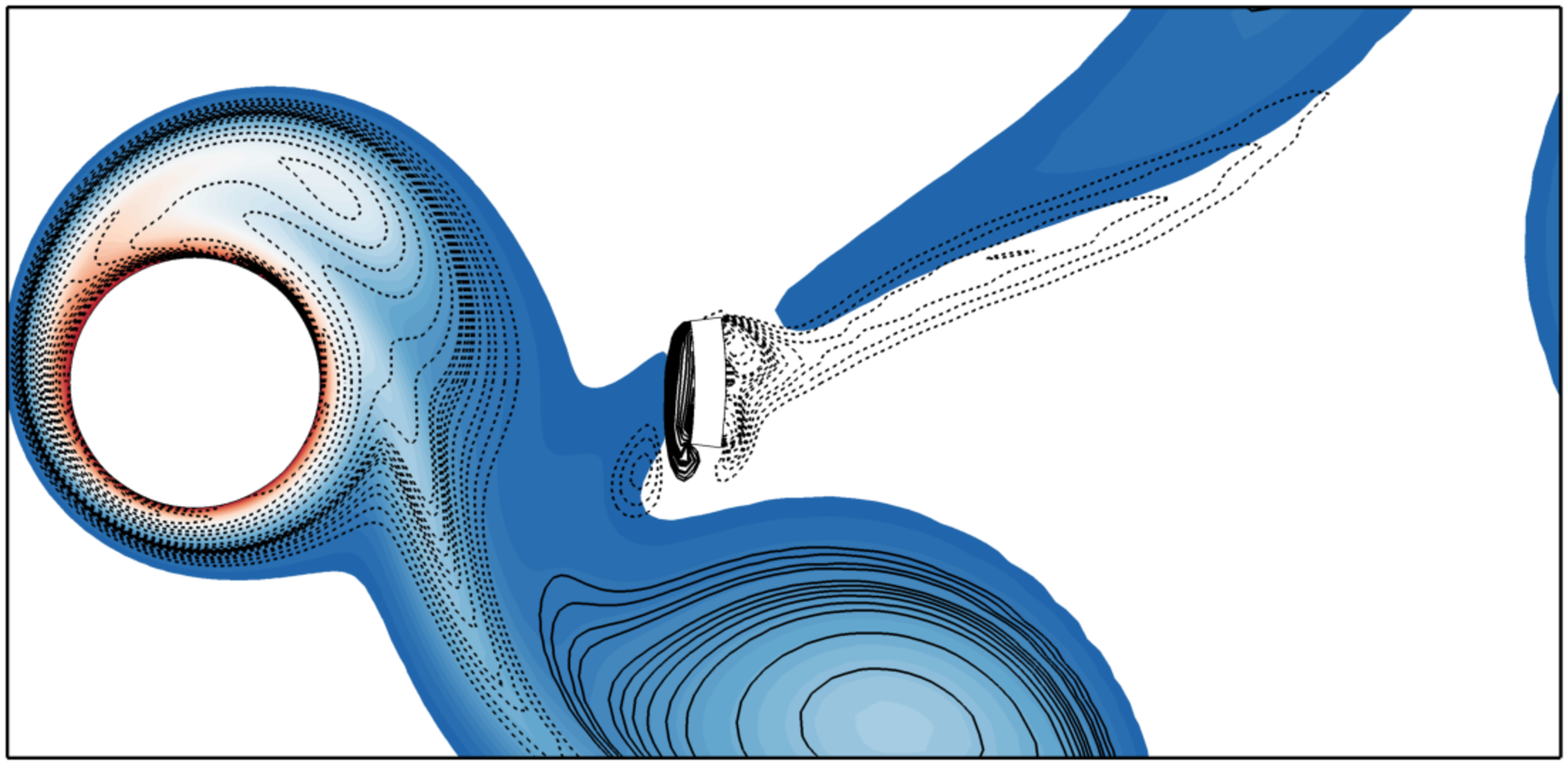}
\\
\hspace{0.5em}\scriptsize{$t=t_0+(1)T$}\hspace{9.5em}\scriptsize{$t=t_0+(3)T$}\hspace{9.5em}\scriptsize{$t=t_0+(1)T$}\hspace{0.5em}
\\
\includegraphics[width=0.3\textwidth,trim={0.7cm 0.7cm 8cm 0.7cm},clip]{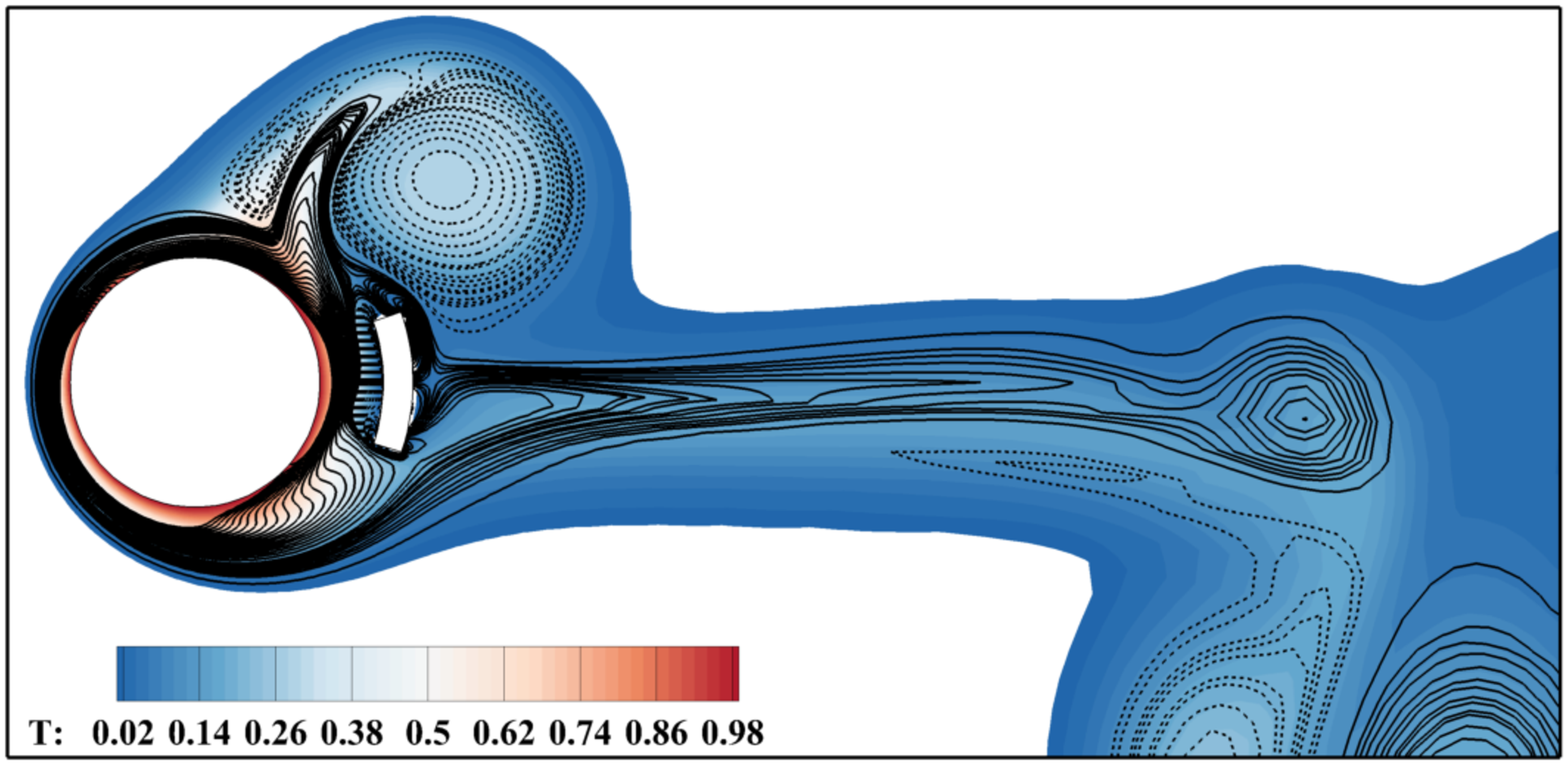}
\includegraphics[width=0.3\textwidth,trim={0.95cm 0.9cm 10cm 0.9cm},clip]{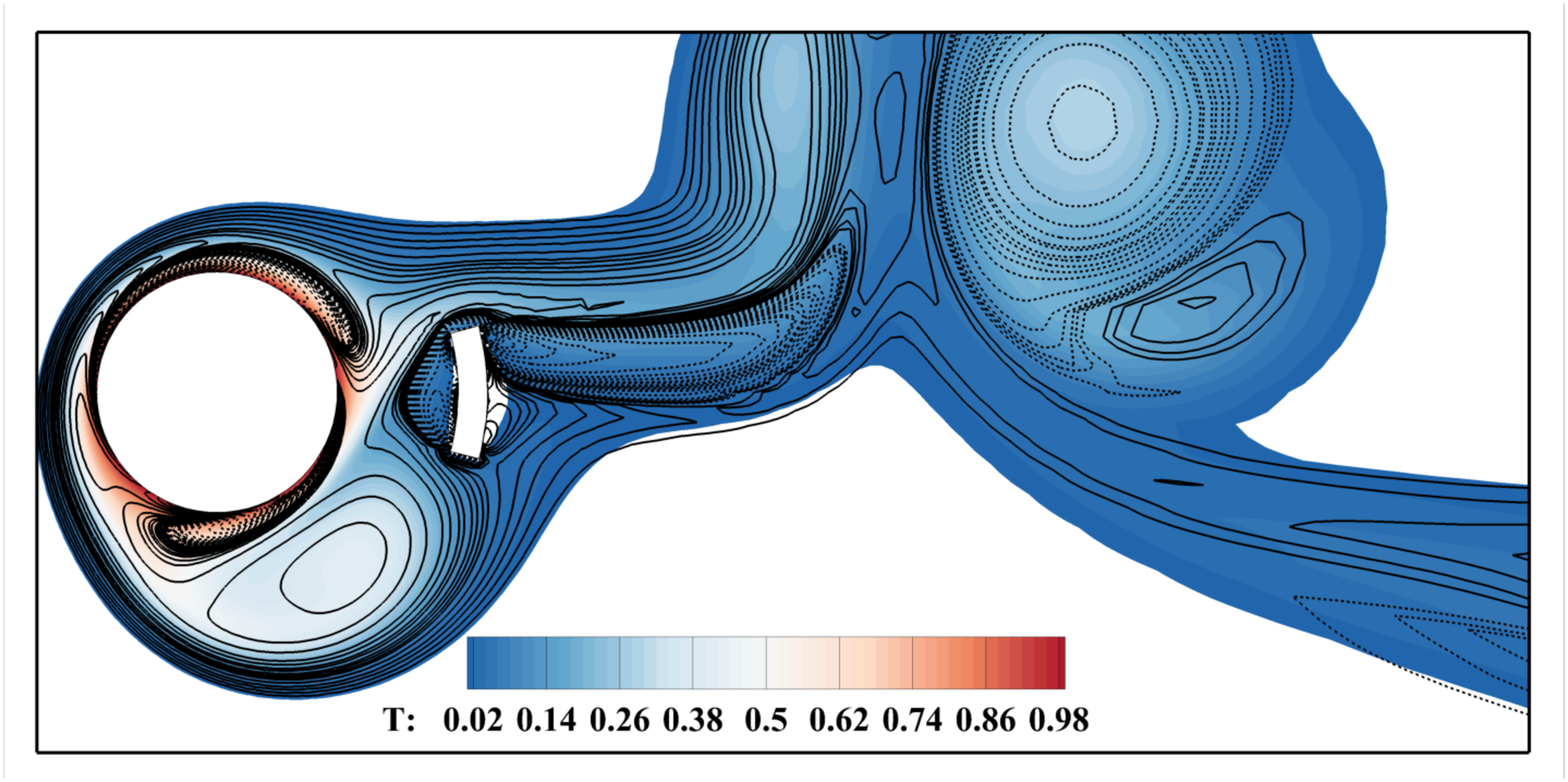}
\includegraphics[width=0.3\textwidth,trim={0.95cm 0.9cm 10cm 0.9cm},clip]{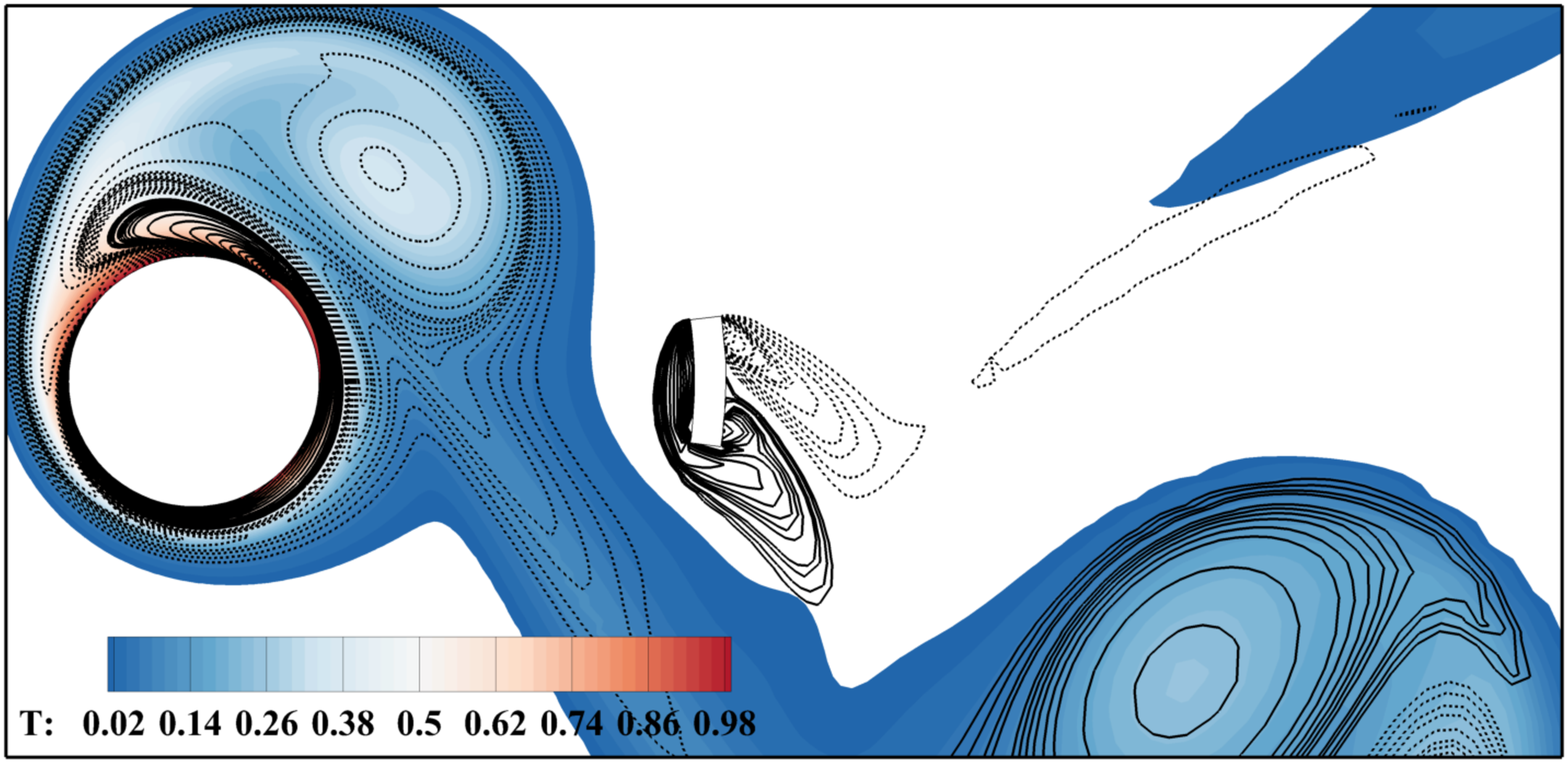}
\\
\hspace{2cm}(a) \hspace{4cm}(b) \hspace{4cm}(c)\hspace{2cm}
 \caption{The distribution of the local Nusselt number $Nu$, over the surface of the cylinder and the isotherm contours during (a) one oscillation period of the cylinder $T$, for $d/R_0=0.5$, (b) three oscillation period of the cylinder $3T$, for $d/R_0=1$ and (c) one oscillation period of the cylinder $T$, for $d/R_0=3$ with $\alpha_m=4$, $f/f_0=0.5$ and $Re=150$.}
 \label{fig:a_4_f_0-5}
\end{figure*}

\begin{figure*}[!t]
\centering
\includegraphics[width=0.45\textwidth,trim={0cm 0cm 0cm 0cm},clip]{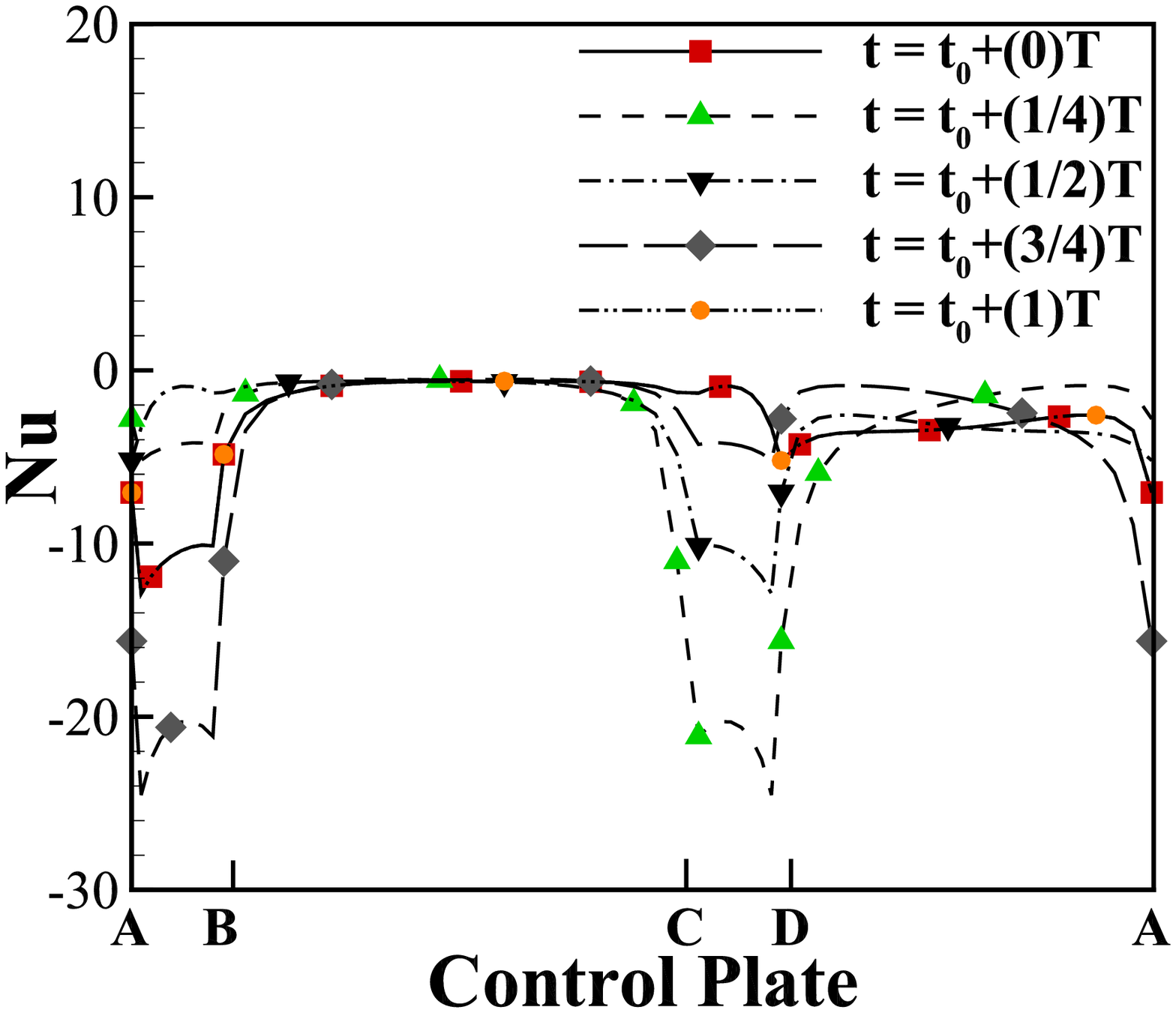}\\
(a)\\
\includegraphics[width=0.45\textwidth,trim={0cm 0cm 0cm 0cm},clip]{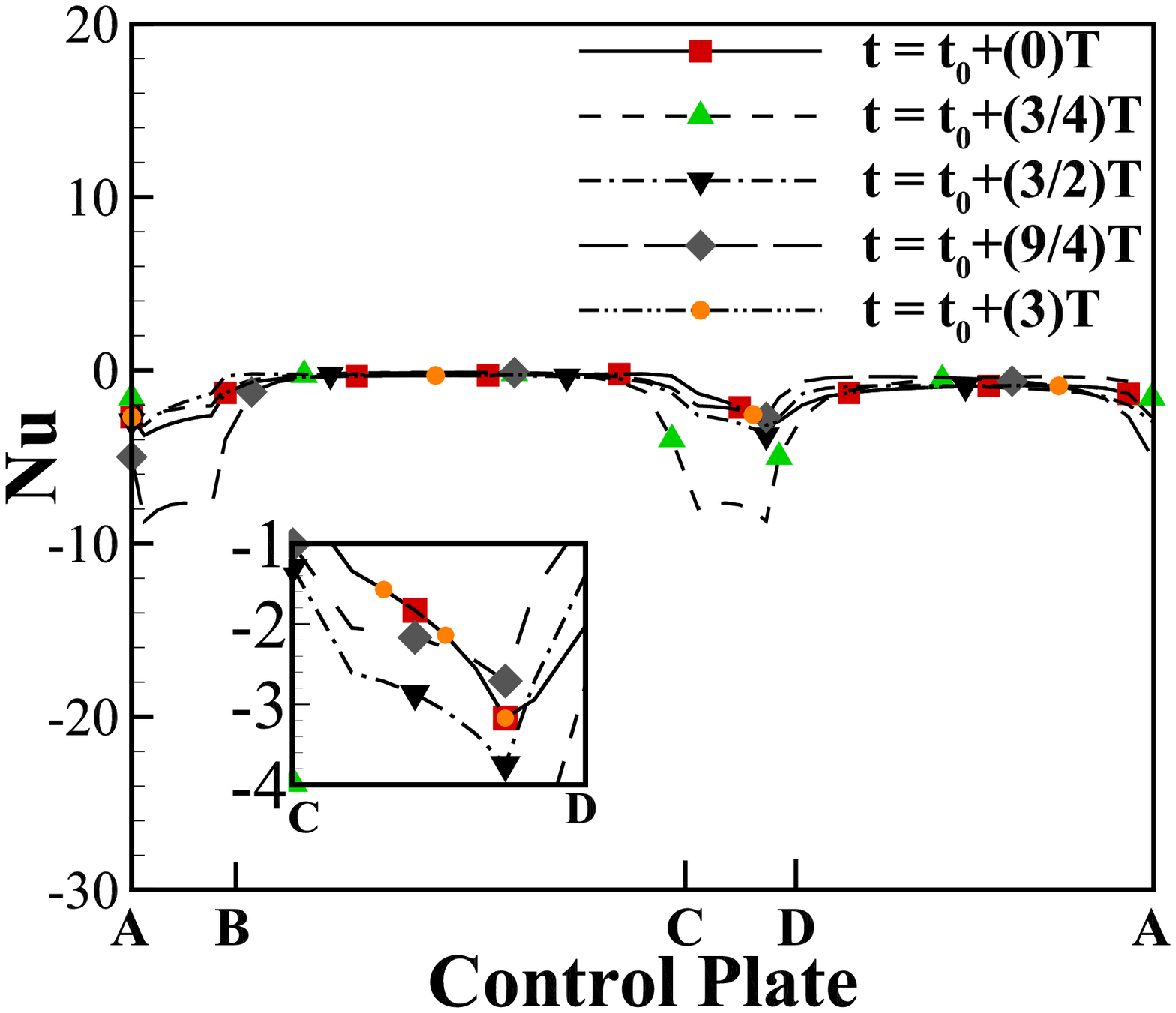}
\includegraphics[width=0.45\textwidth,trim={0cm 0cm 0cm 0cm},clip]{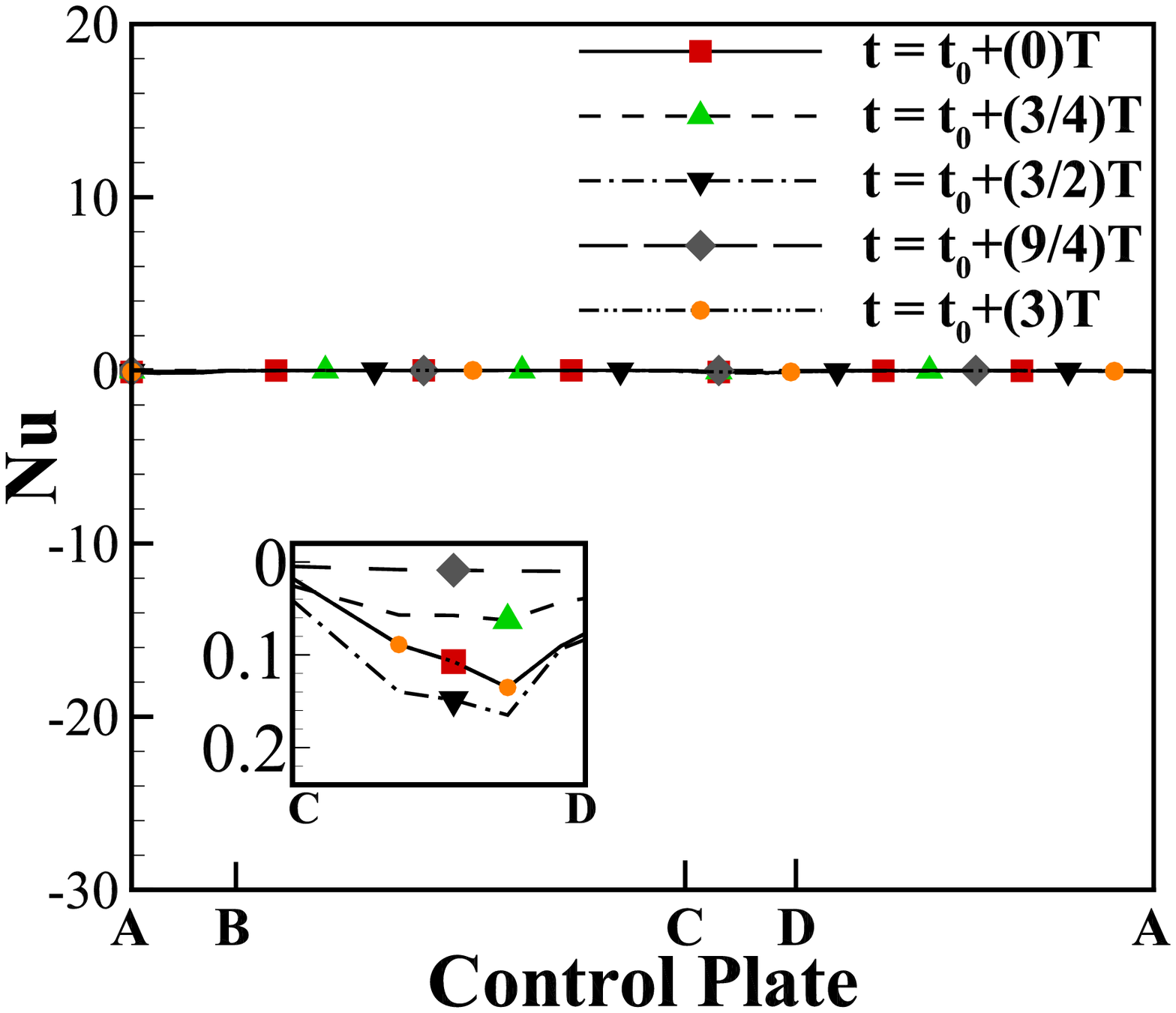}
\\
\hspace{2cm}(b) \hspace{6cm}(c)\hspace{2cm}
 \caption{The distribution of the local Nusselt number $Nu$, over the surface of the control plate during (a) one oscillation period of the cylinder $T$, for $d/R_0=0.5$, (b) three oscillation period of the cylinder $3T$, for $d/R_0=1$ and (c) one oscillation period of the cylinder $T$, for $d/R_0=3$ with $\alpha_m=4$, $f/f_0=0.5$ and $Re=150$.}
 \label{fig:plate_a_4_f_0-5}
\end{figure*}

The vortex shedding phenomenon becomes more complex as the amplitude of the rotary oscillation, $\alpha_m$, is increased from $0.5$ to $4$ for a fixed frequency ratio of $f/f_0=0.5$, and the corresponding isotherm contours and local Nusselt number distributions are shown in \cref{fig:a_4_f_0-5} with three different control plate placements, i.e., $d/R_0=0.5$, $1$ and $3$. The mode of vortex shedding for $d/R_0=0.5$ in \cref{fig:a_4_f_0-5}(a), is identified as $2P+2S$ and it is locked-on over one oscillation period $T$ of the cylinder. Two pairs of large vortices with opposing signs shed periodically from the top and bottom of the cylinder. Maximum peak of local Nusselt number on the surface of the cylinder is found in the region $\ang{120}<\theta<\ang{240}$ but the maximum value is lower than that of $\alpha_m=0.5$. The local Nusselt number distribution has its highest peak between, $\ang{0}< \theta<\ang{100}$ and $\ang{260}< \theta<\ang{360}$. In terms of the local Nusselt number distribution, the gap between the local maximum peak and the global peak is relatively low. The mode of vortex shedding remains the same as $2P+2S$ as the gap ratio of the control plate is increased to $d/R_0=1$, but it is locked-on over three periods of cylinder oscillation, i.e., $3T$, as shown in \cref{fig:a_4_f_0-5}(b). The thermal boundary layer is thickened at $\theta\approx \ang{180}$ by the large oscillation amplitude of the cylinder and increased gap ratio of the control plate. This leads to less heat being transferred by forced convection in this region. It is supported by the local Nusselt number distribution. Maximum peak of local Nusselt number is found in the regions $\ang{100}<\theta<\ang{150}$ and $\ang{210}<\theta<\ang{260}$ at phases $t=t_0+(9/4)T$, $t_0+(3/4)T$ respectively. With the growing gap ratio of the control plate, there are now larger gaps between the highest peaks of local Nusselt numbers. Local maximum peaks of local Nusselt number are found in regions, $\ang{0}<\theta<\ang{80}$ and $\ang{280}<\theta<\ang{360}$. As the gap ratio is increased to $d/R_0=3$, the number of vortices shed downstream decreases, as seen in \cref{fig:a_4_f_0-5}(c). The vortex shedding mode is identified as $2P$ and it is locked-on over one oscillation period of cylinder, $T$. Two pairs of vortices of opposite signs are periodically shed from either side of the cylinder. The vortex shedding process causes the density of the vorticity contours near $\theta\approx\ang{180}$ to be significantly lower than the previous two cases. It implies that the thermal boundary layer in this area has more thickness relative to the cases with gap ratios, $d/R_0=0.5,\ 1$. As a result, the heat convection in this region drops significantly. The distibution of the local Nusselt number on the surface of the cylinder also suggests that the maximum peak is shifted from $\theta\approx\ang{180}$ to $\theta\approx\ang{0}$ or $\ang{360}$. It is found in the areas $\ang{0}<\theta<\ang{120}$ and $\ang{240}<\theta<\ang{360}$. The maximum value of the local Nusselt number is seen to be at its lowest for the $d/R_0=3$, when compared to the other two locations of the control plate. In addition, there are local maximum peaks of the Nusselt number discovered close to $\theta\approx\ang{0}$ or $\ang{360}$. \cref{fig:plate_a_4_f_0-5} displays the distribution of local Nusselt number on the surface of the control plate for $d/R_0=0.5$, $1$, $3$ and $(\alpha_m,\ f/f_0)=(4,\ 0.5)$. This higher maximum angular velocity ($\alpha_m$) of the cylinder enhances the movement of the flow around the cylinder, which effectively enhances the heat absorption rate around the top ($CD$) and bottom ($AB$) surfaces of the control plate for $d/R_0=0.5$. Heat absorption decreases as the gap ratio of the control plate increases to $d/R_0 = 1$, and it approaches zero at $d/R_0 = 3$. When the heat absorption completely vanishes, the cold control plate effects the heat transfer only by influencing the vortex shedding process.\\

\begin{figure*}[!htbp]
\centering
\includegraphics[width=0.3\textwidth,trim={0cm 0cm 0cm 0cm},clip]{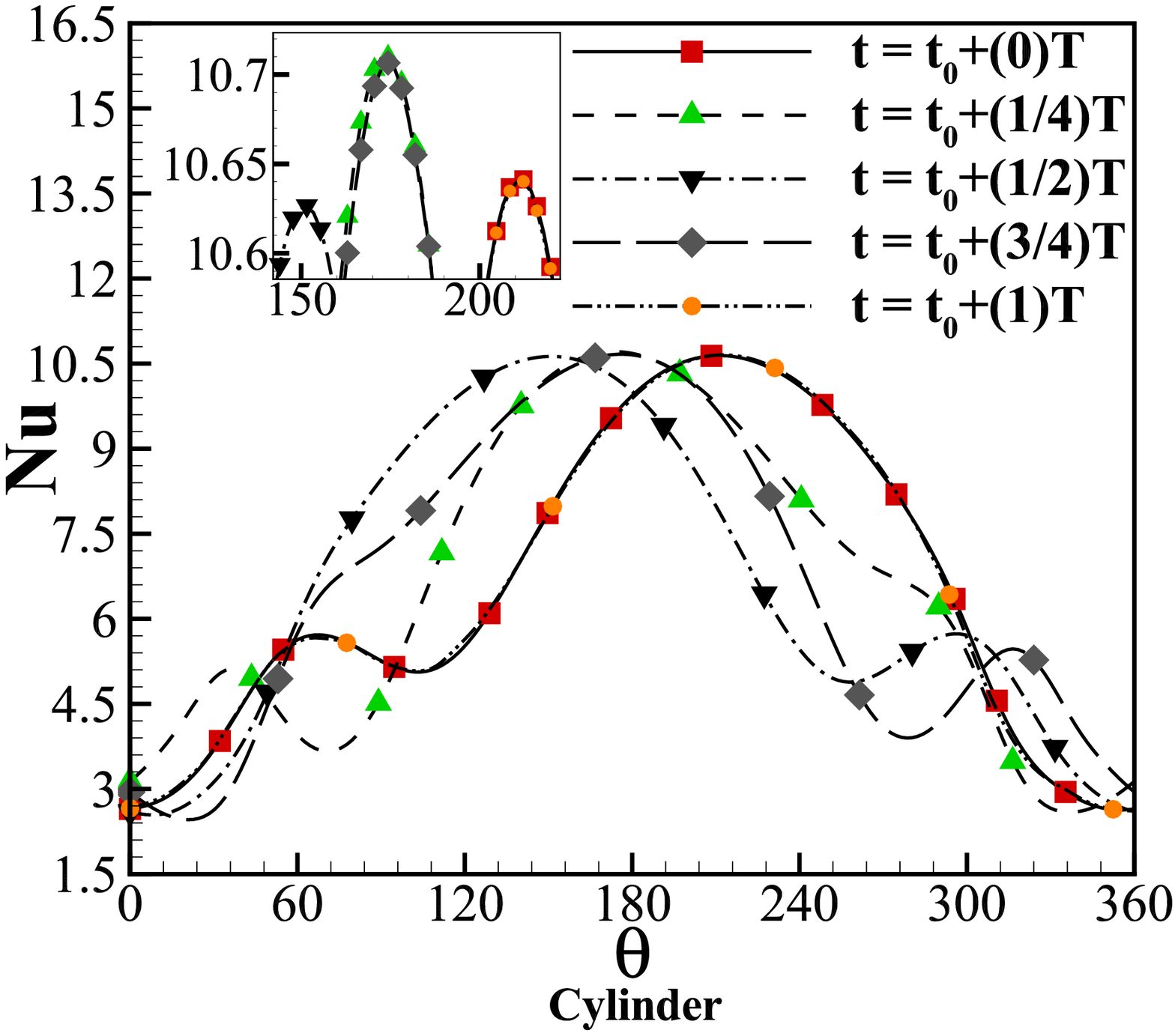}
\includegraphics[width=0.3\textwidth,trim={0cm 0cm 0cm 0cm},clip]{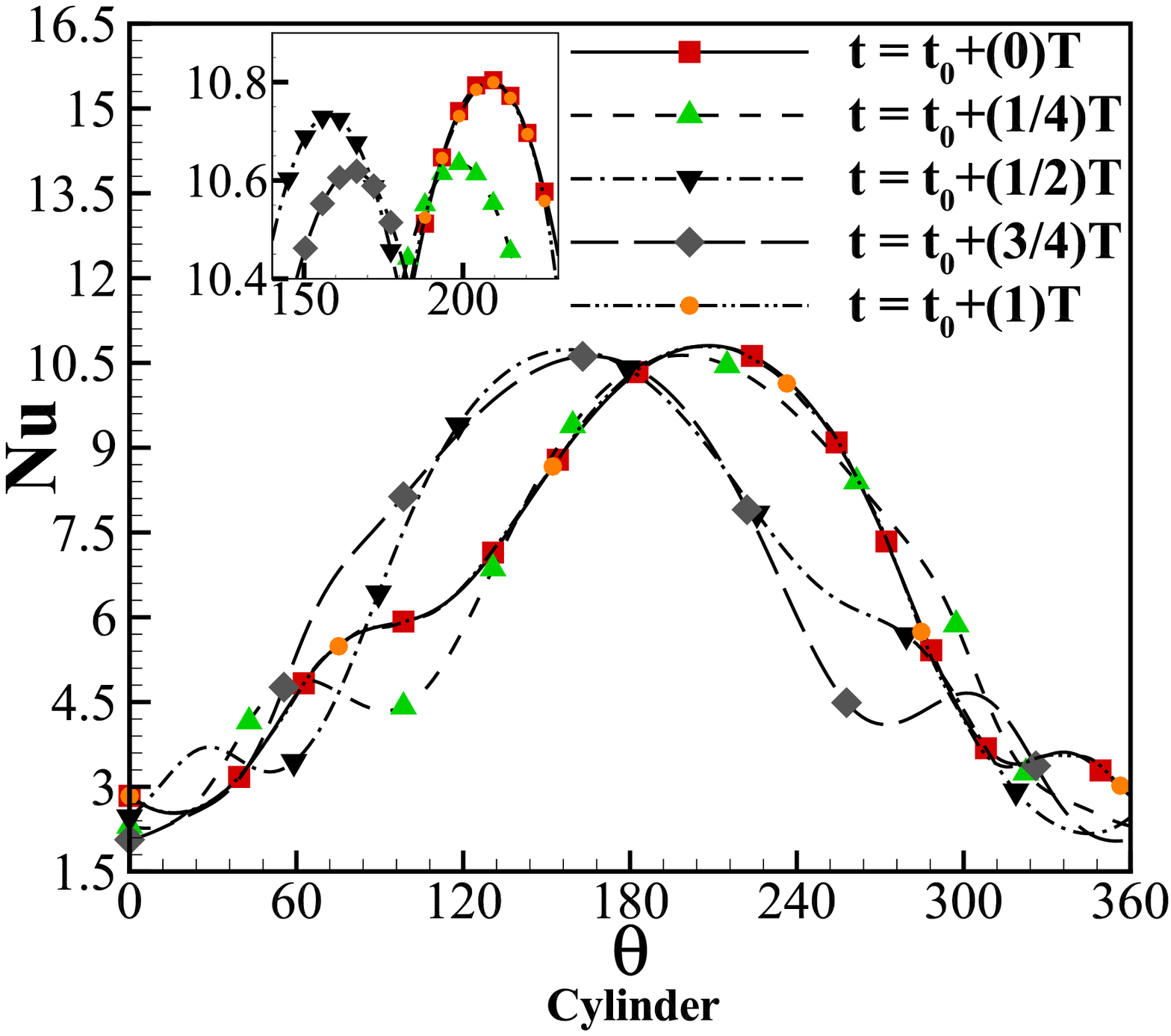}
\includegraphics[width=0.3\textwidth,trim={0cm 0cm 0cm 0cm},clip]{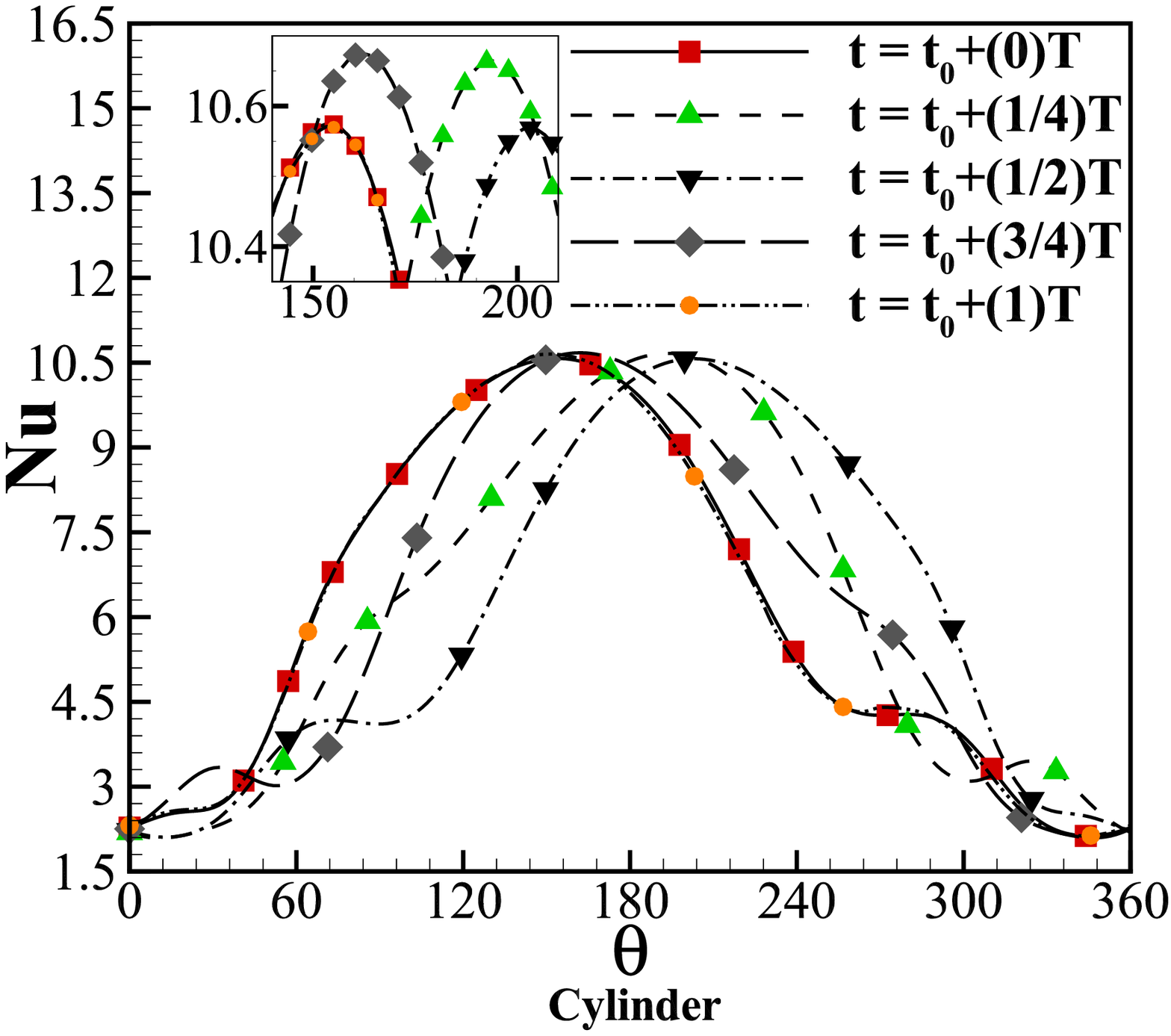}
\\
\hspace{0.5em}\scriptsize{$t=t_0+(1/4)T$}\hspace{9.5em}\scriptsize{$t=t_0+(1/4)T$}\hspace{9.5em}\scriptsize{$t=t_0+(1/4)T$}\hspace{0.5em}
\\
\includegraphics[width=0.3\textwidth,trim={0.7cm 0.7cm 8cm 0.7cm},clip]{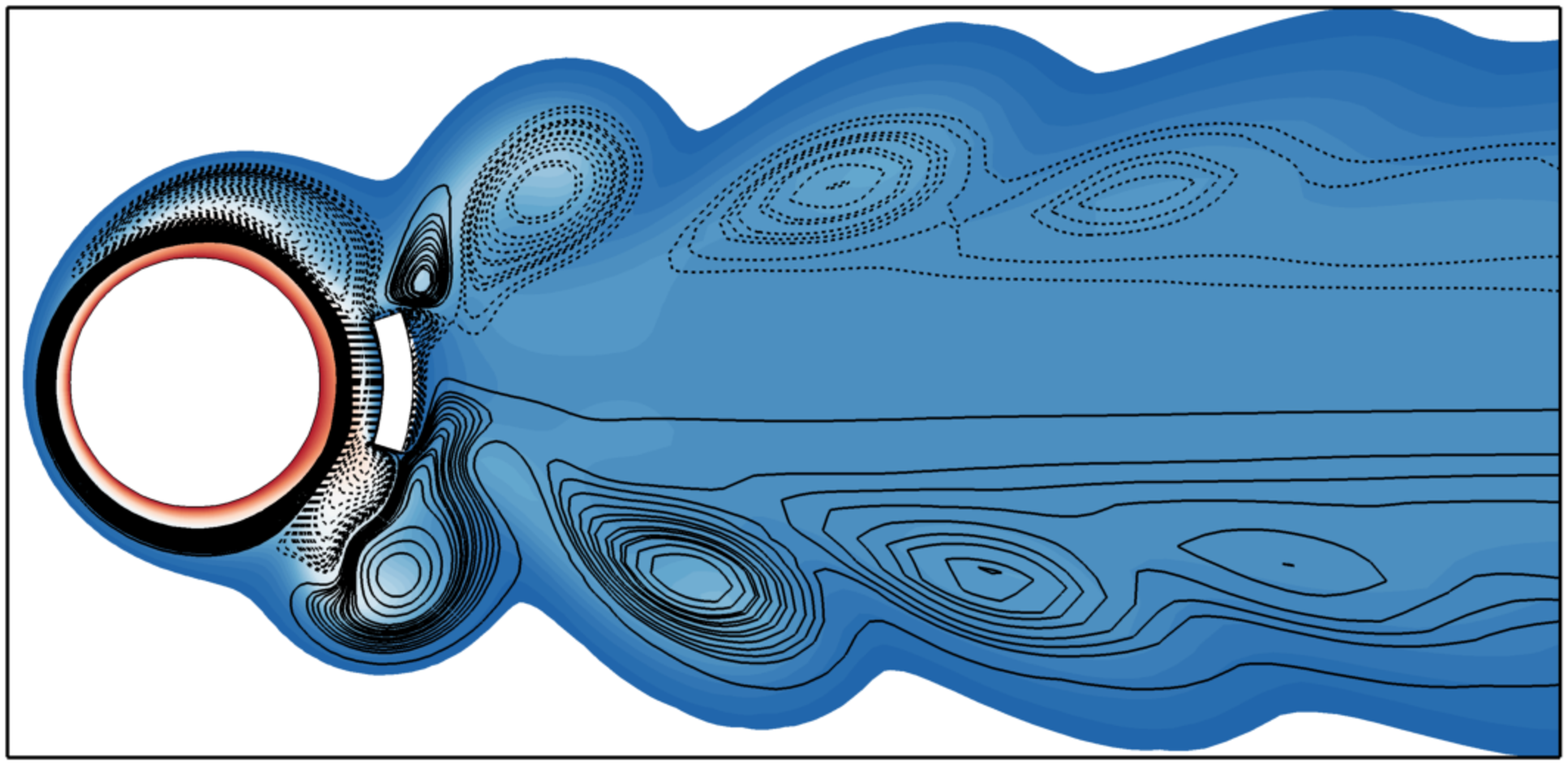}
\includegraphics[width=0.3\textwidth,trim={0.7cm 0.7cm 8cm 0.7cm},clip]{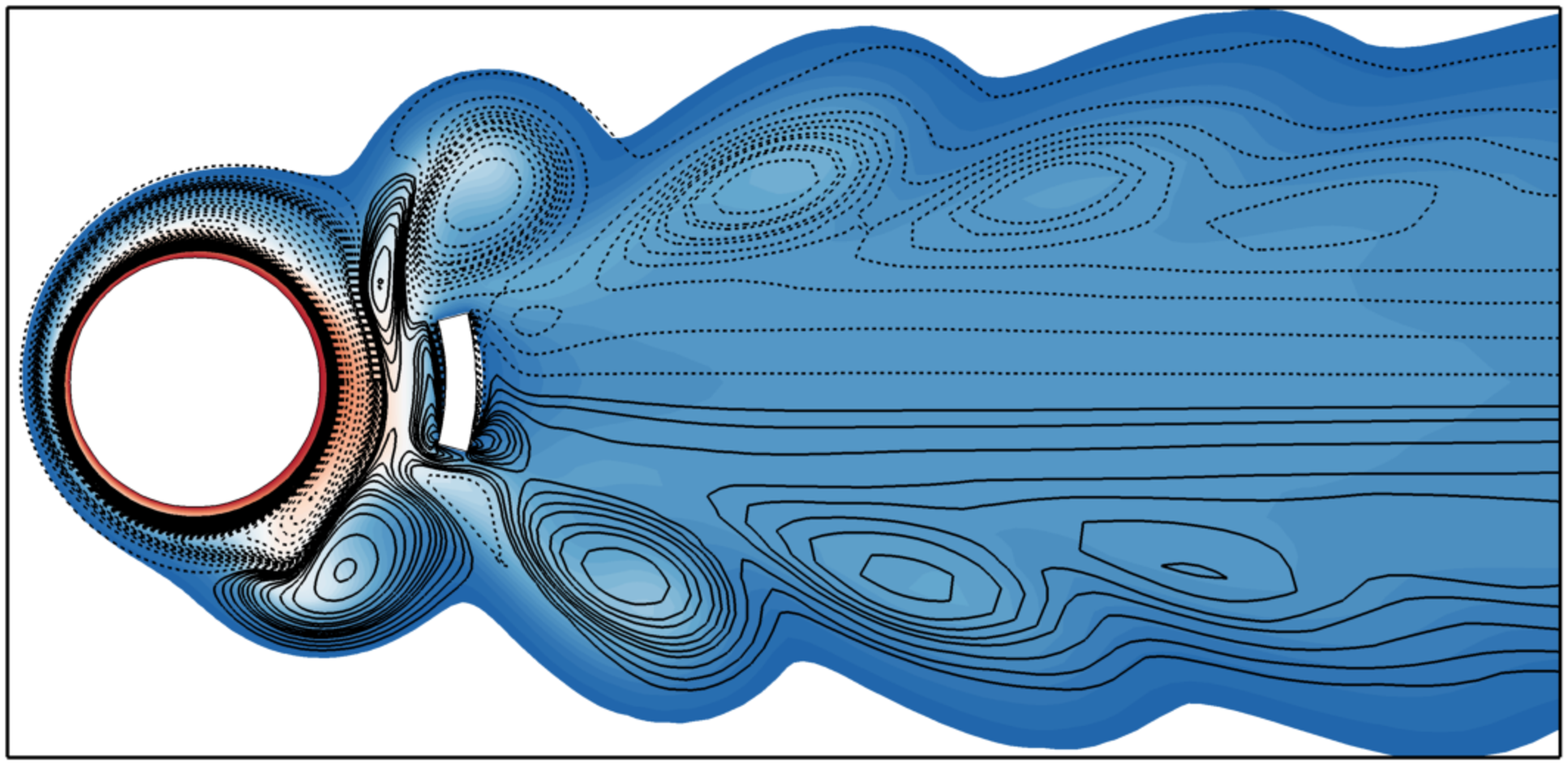}
\includegraphics[width=0.3\textwidth,trim={0.7cm 0.7cm 8cm 0.7cm},clip]{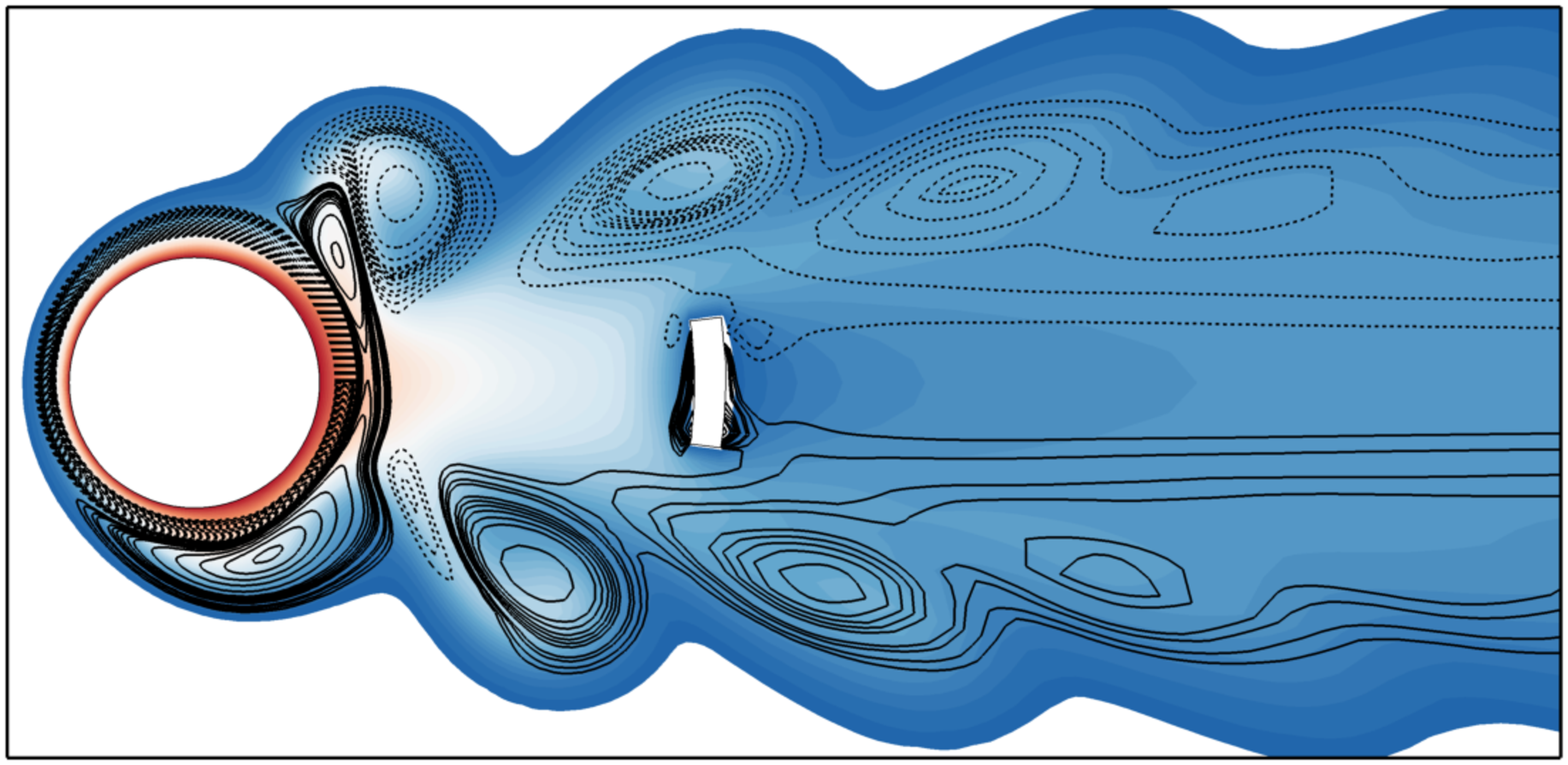}
\\
\hspace{0.5em}\scriptsize{$t=t_0+(1/2)T$}\hspace{9.5em}\scriptsize{$t=t_0+(1/2)T$}\hspace{9.5em}\scriptsize{$t=t_0+(1/2)T$}\hspace{0.5em}
\\
\includegraphics[width=0.3\textwidth,trim={0.7cm 0.7cm 8cm 0.7cm},clip]{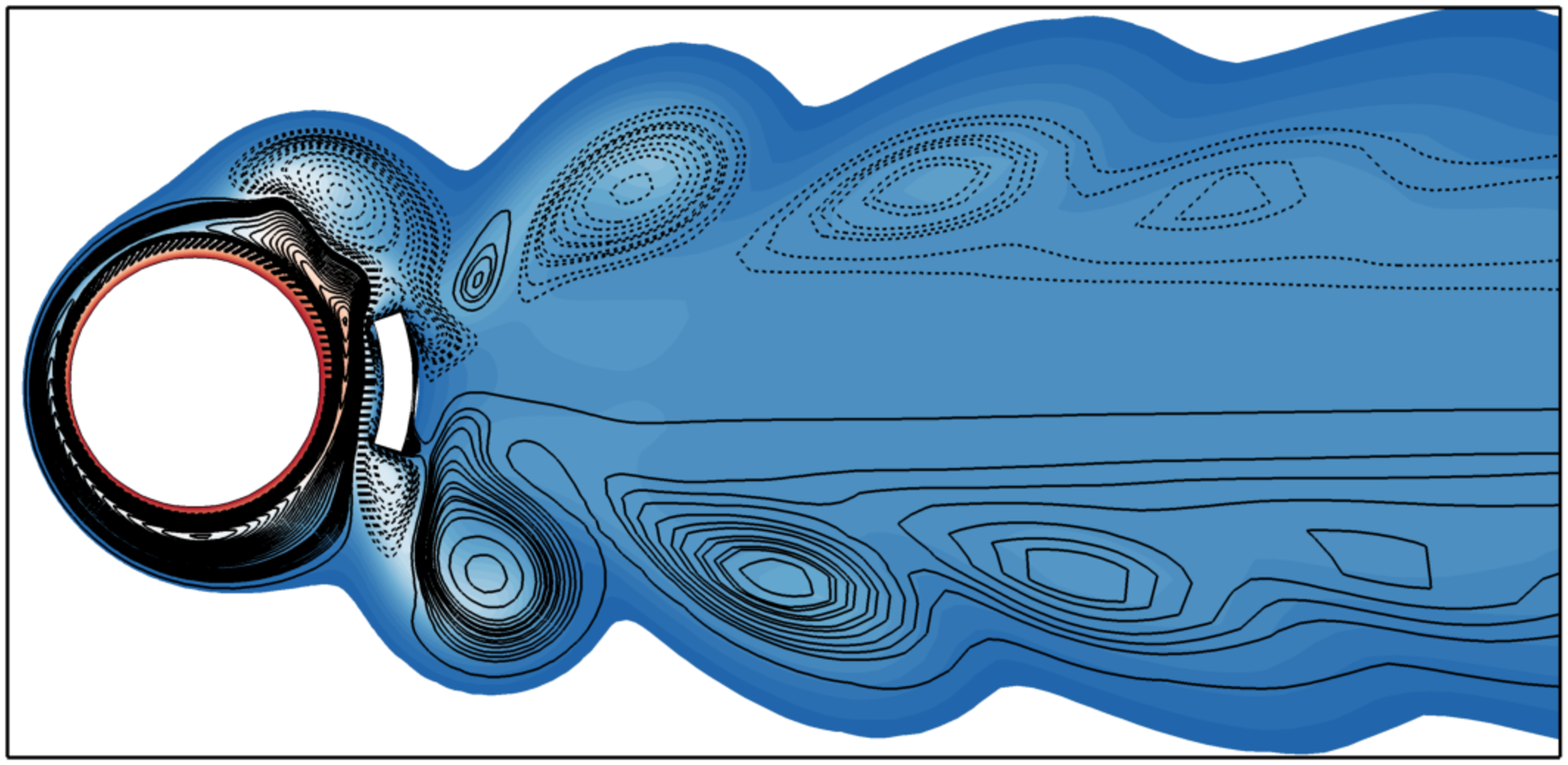}
\includegraphics[width=0.3\textwidth,trim={0.7cm 0.7cm 8cm 0.7cm},clip]{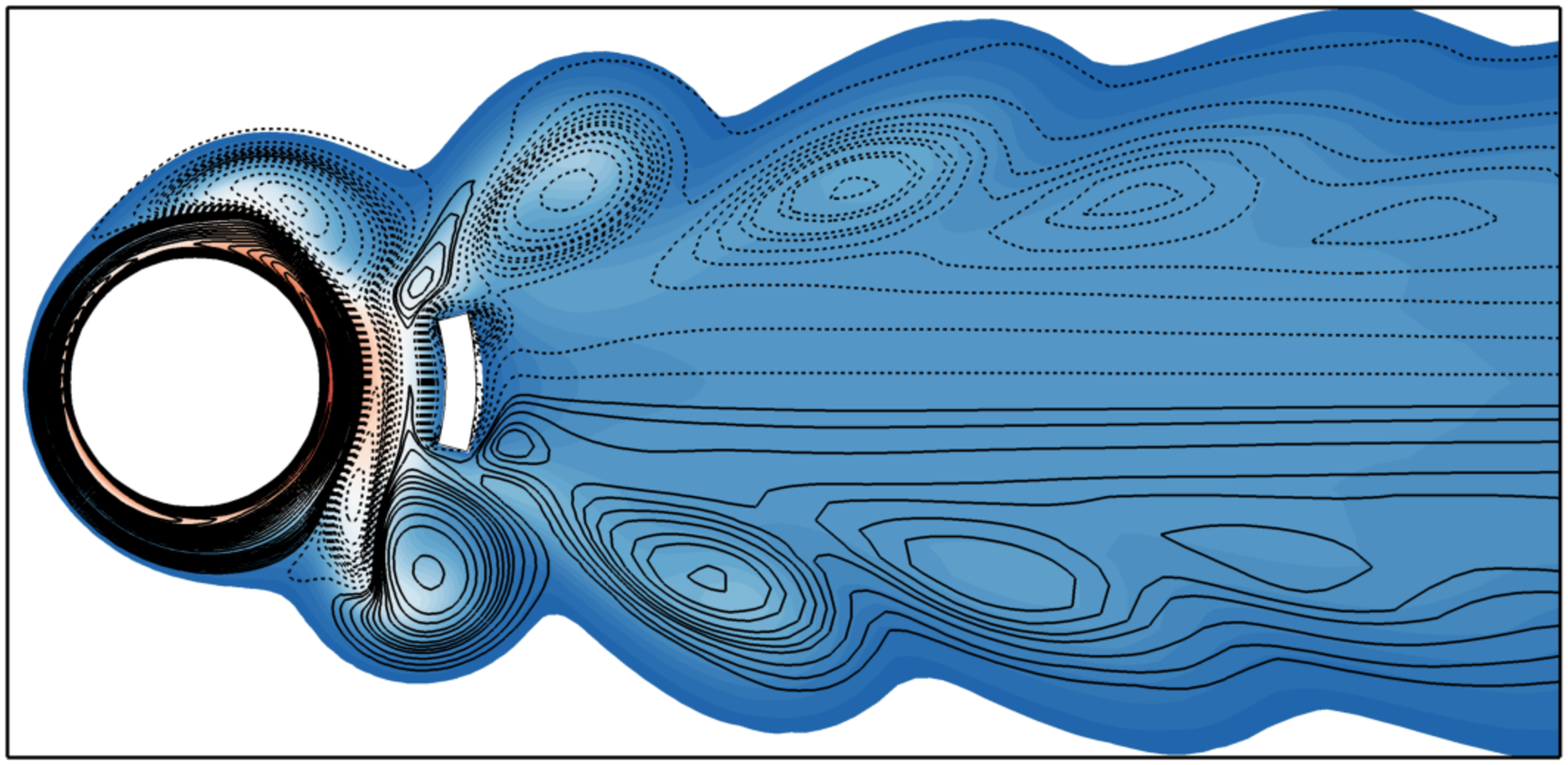}
\includegraphics[width=0.3\textwidth,trim={0.7cm 0.7cm 8cm 0.7cm},clip]{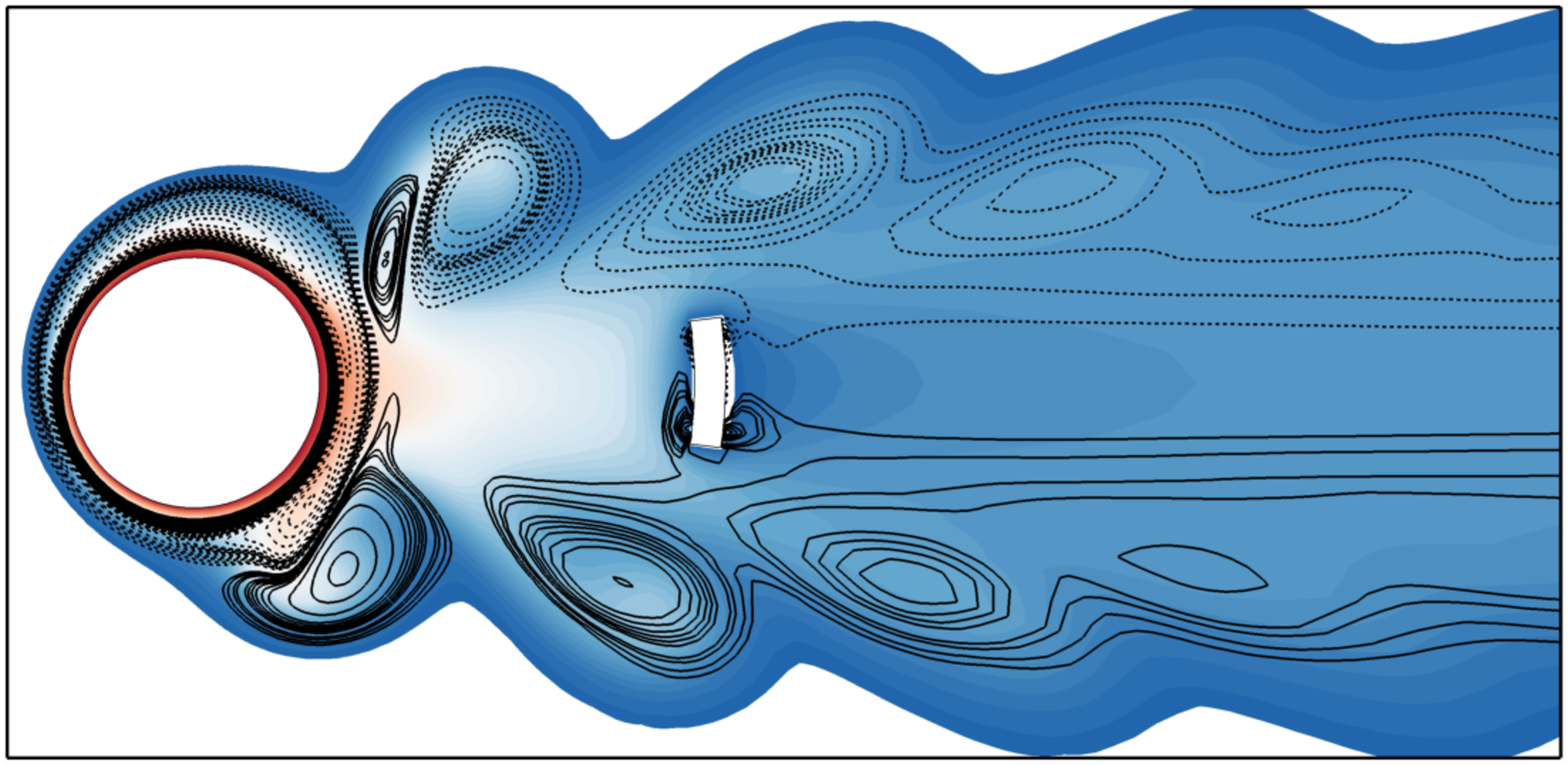}
\\
\hspace{0.5em}\scriptsize{$t=t_0+(3/4)T$}\hspace{9.5em}\scriptsize{$t=t_0+(3/4)T$}\hspace{9.5em}\scriptsize{$t=t_0+(3/4)T$}\hspace{0.5em}
\\
\includegraphics[width=0.3\textwidth,trim={0.7cm 0.7cm 8cm 0.7cm},clip]{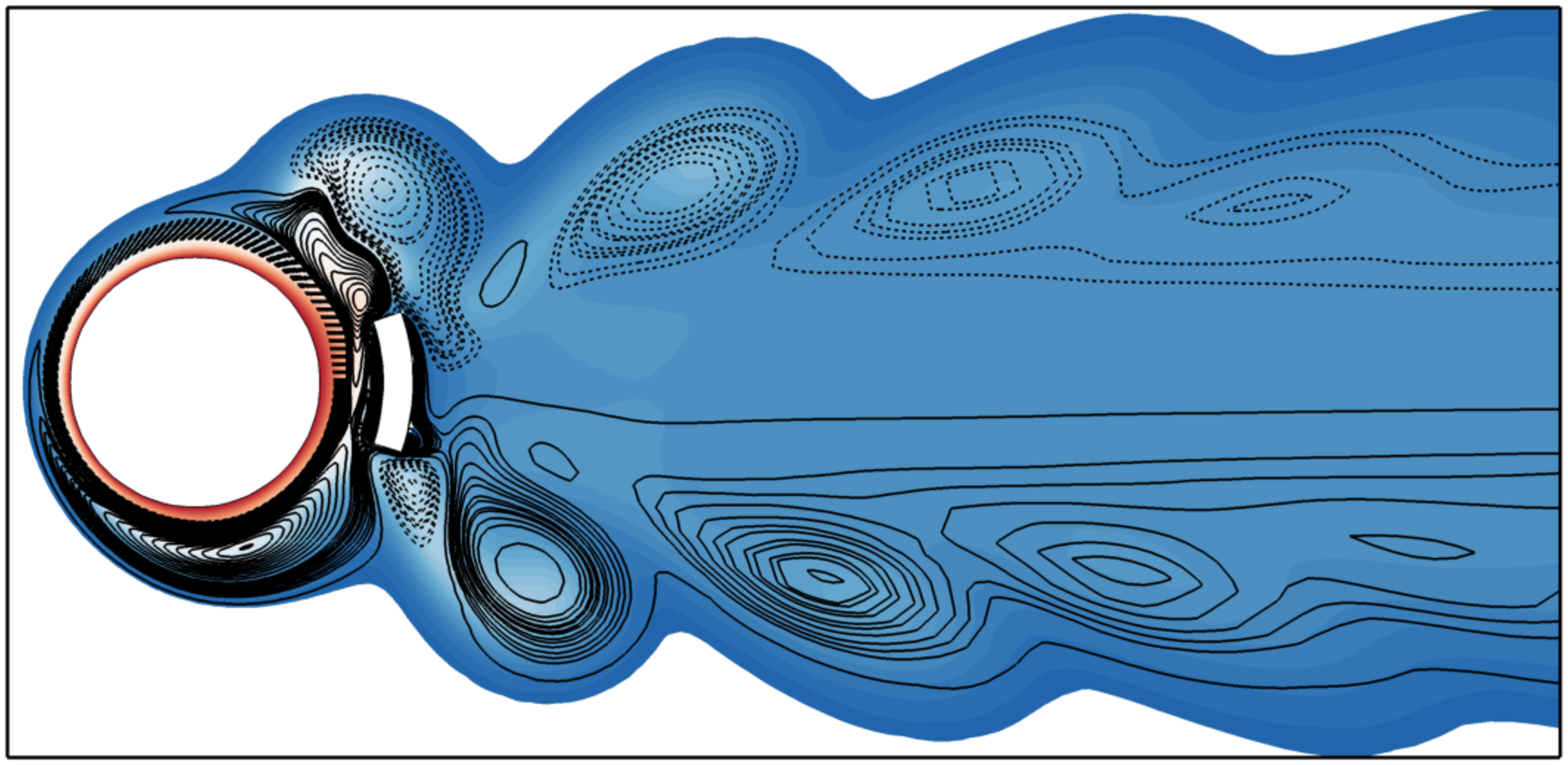}
\includegraphics[width=0.3\textwidth,trim={0.7cm 0.7cm 8cm 0.7cm},clip]{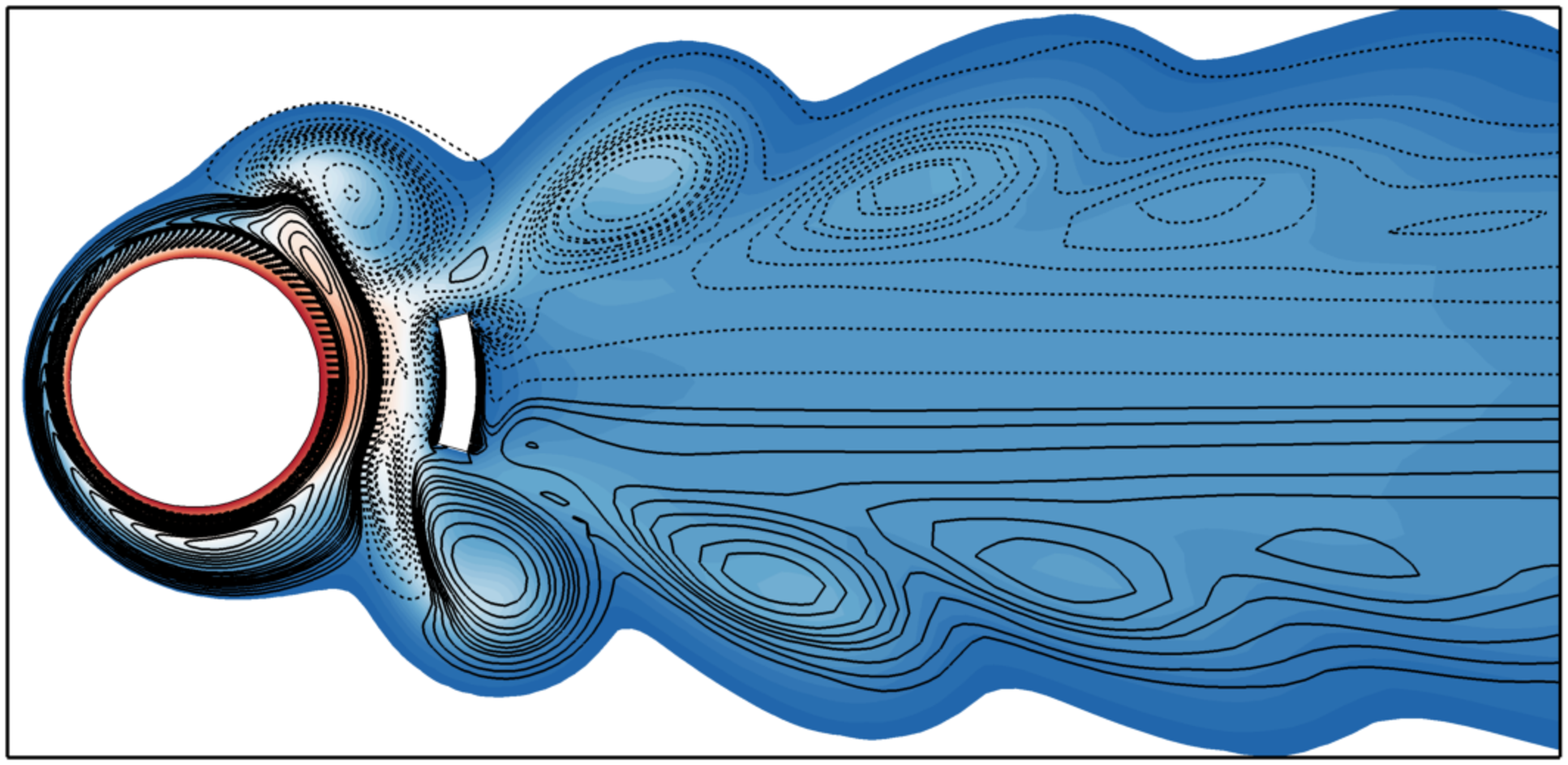}
\includegraphics[width=0.3\textwidth,trim={0.7cm 0.7cm 8cm 0.7cm},clip]{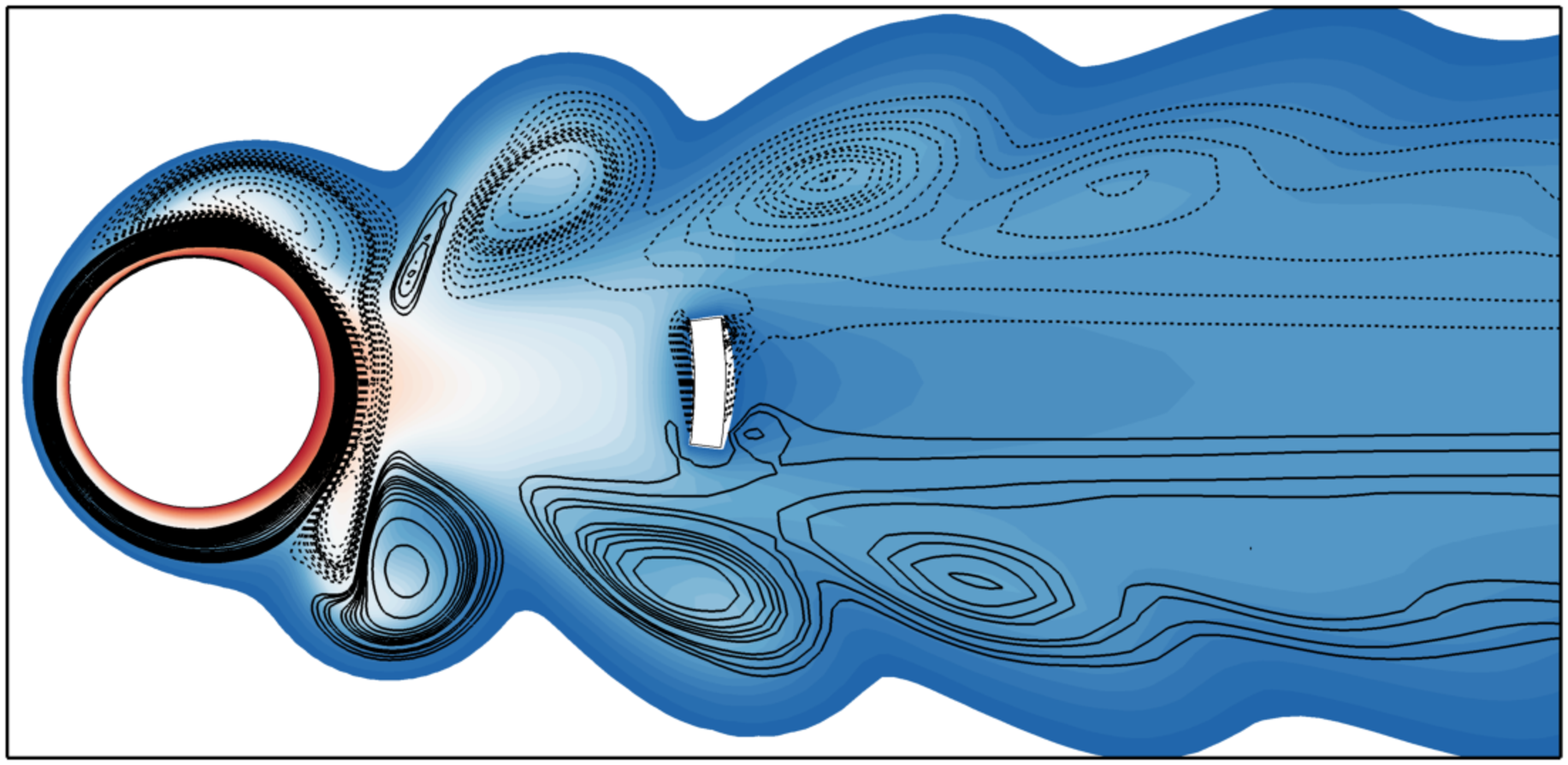}
\\
\hspace{0.5em}\scriptsize{$t=t_0+(1)T$}\hspace{9.5em}\scriptsize{$t=t_0+(1)T$}\hspace{9.5em}\scriptsize{$t=t_0+(1)T$}\hspace{0.5em}
\\
\includegraphics[width=0.3\textwidth,trim={0.7cm 0.65cm 8cm 0.7cm},clip]{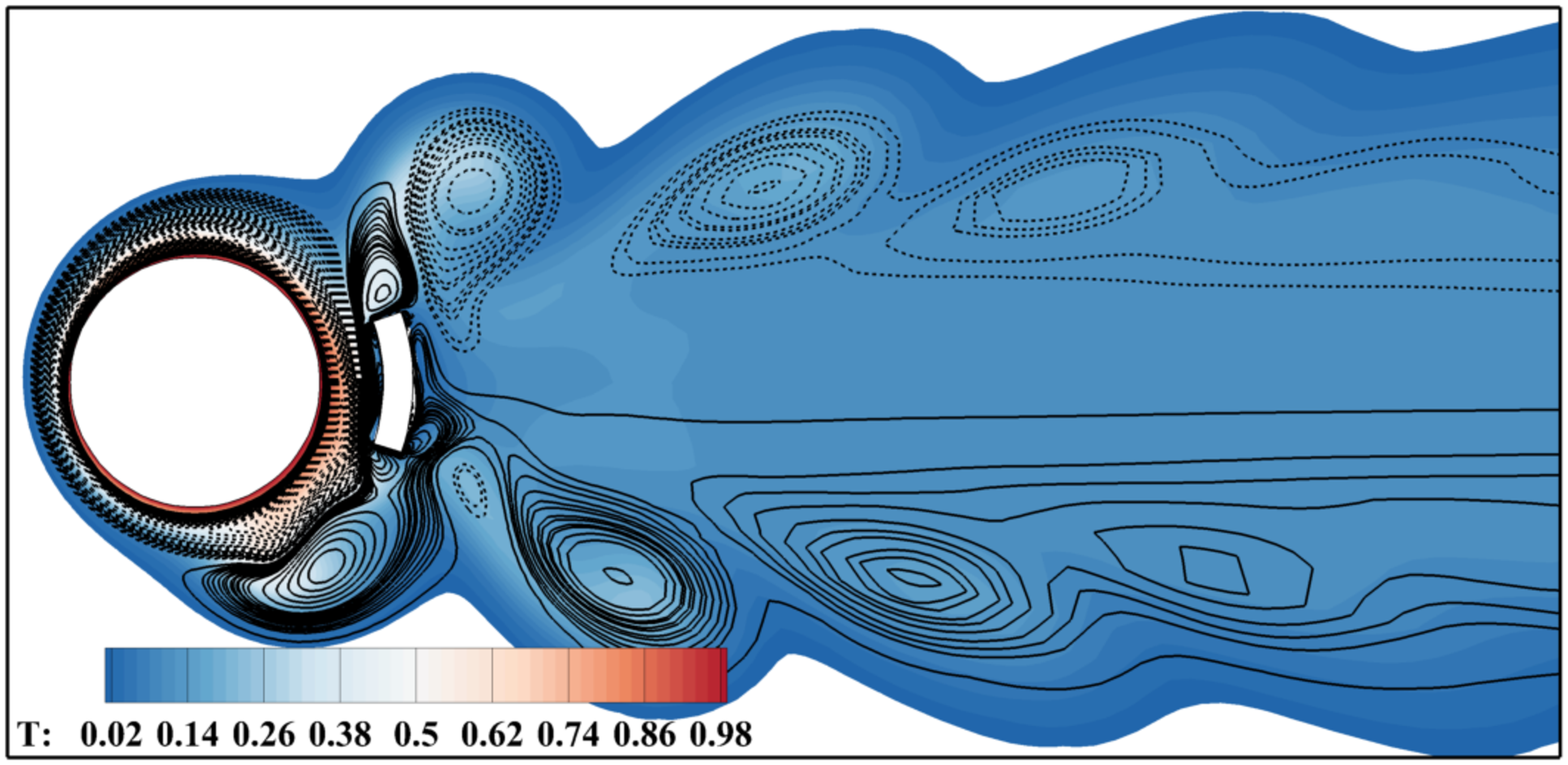}
\includegraphics[width=0.3\textwidth,trim={0.7cm 0.65cm 8cm 0.7cm},clip]{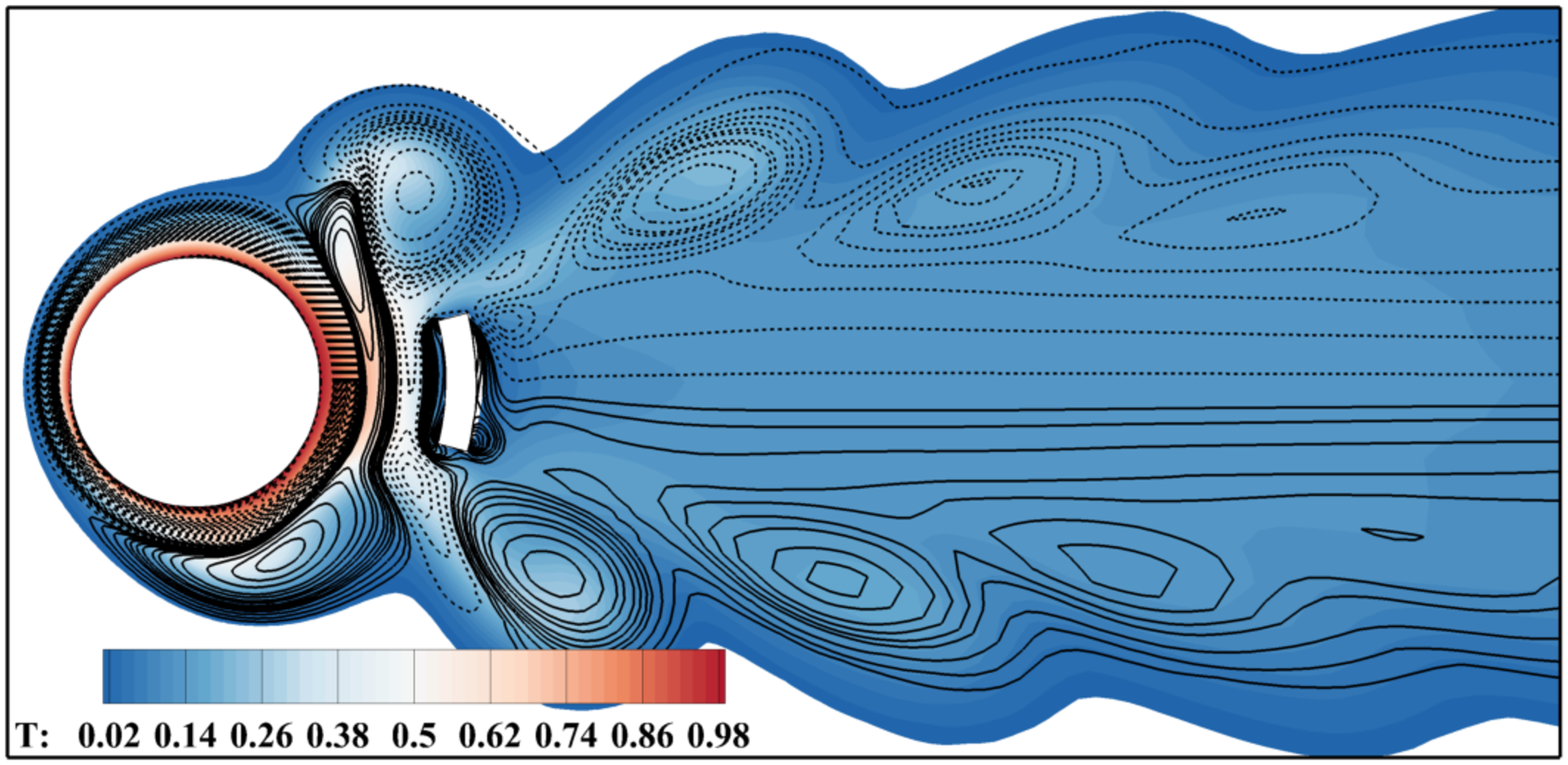}
\includegraphics[width=0.3\textwidth,trim={0.7cm 0.65cm 8cm 0.7cm},clip]{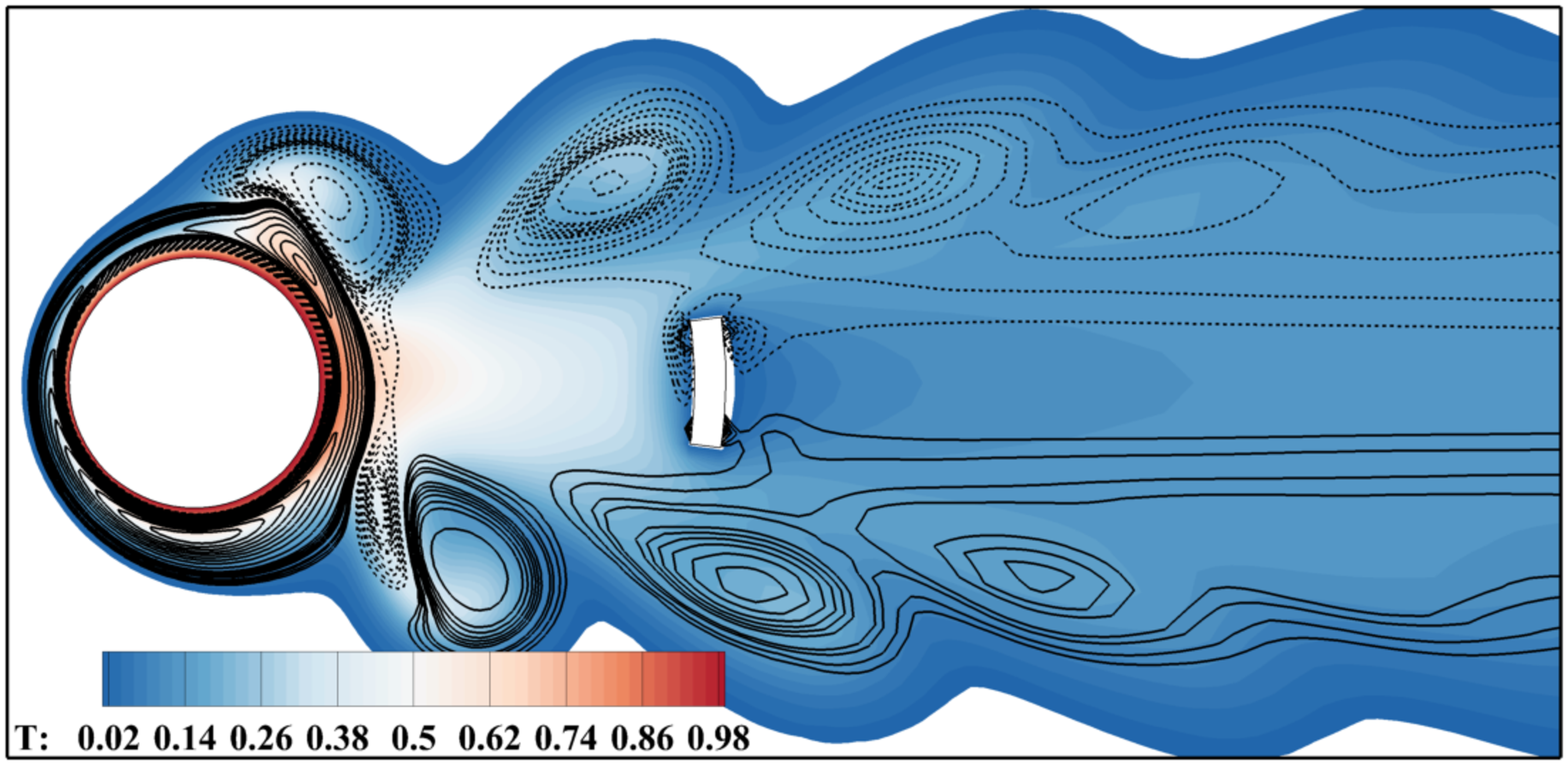}
\\
\hspace{2cm}(a) \hspace{4cm}(b) \hspace{4cm}(c)\hspace{2cm}
 \caption{The distribution of the local Nusselt number $Nu$, over the surface of the cylinder and the isotherm contours during (a) one oscillation period of the cylinder $T$, for $d/R_0=0.5$, (b) one oscillation period of the cylinder $T$, for $d/R_0=1$ and (c) one oscillation period of the cylinder $T$, for $d/R_0=3$ with $\alpha_m=4$, $f/f_0=3$ and $Re=150$.}
 \label{fig:a_4_f_3}
\end{figure*}

\begin{figure*}[!t]
\centering
\includegraphics[width=0.45\textwidth,trim={0cm 0cm 0cm 0cm},clip]{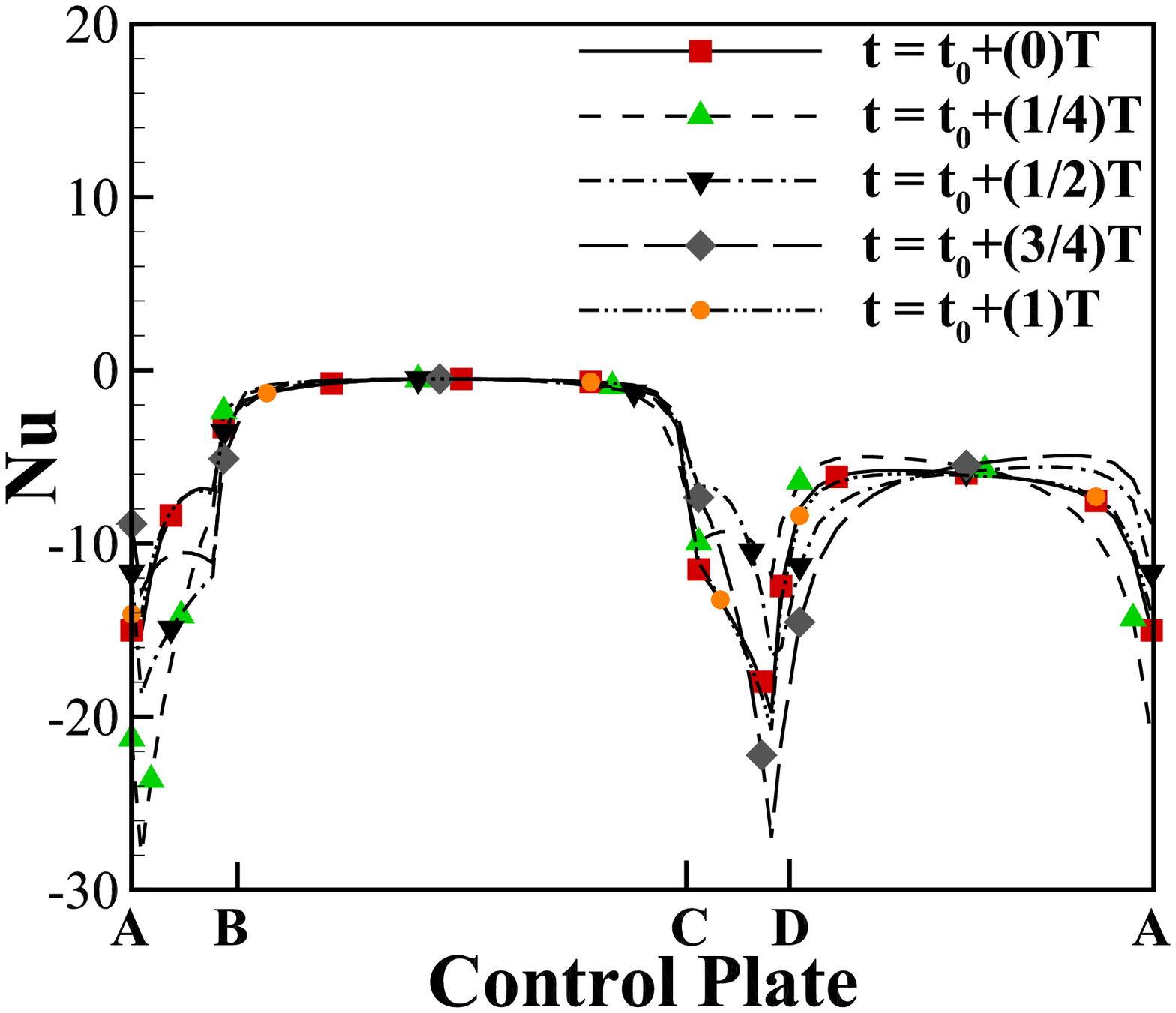}\\
(a)\\
\includegraphics[width=0.45\textwidth,trim={0cm 0cm 0cm 0cm},clip]{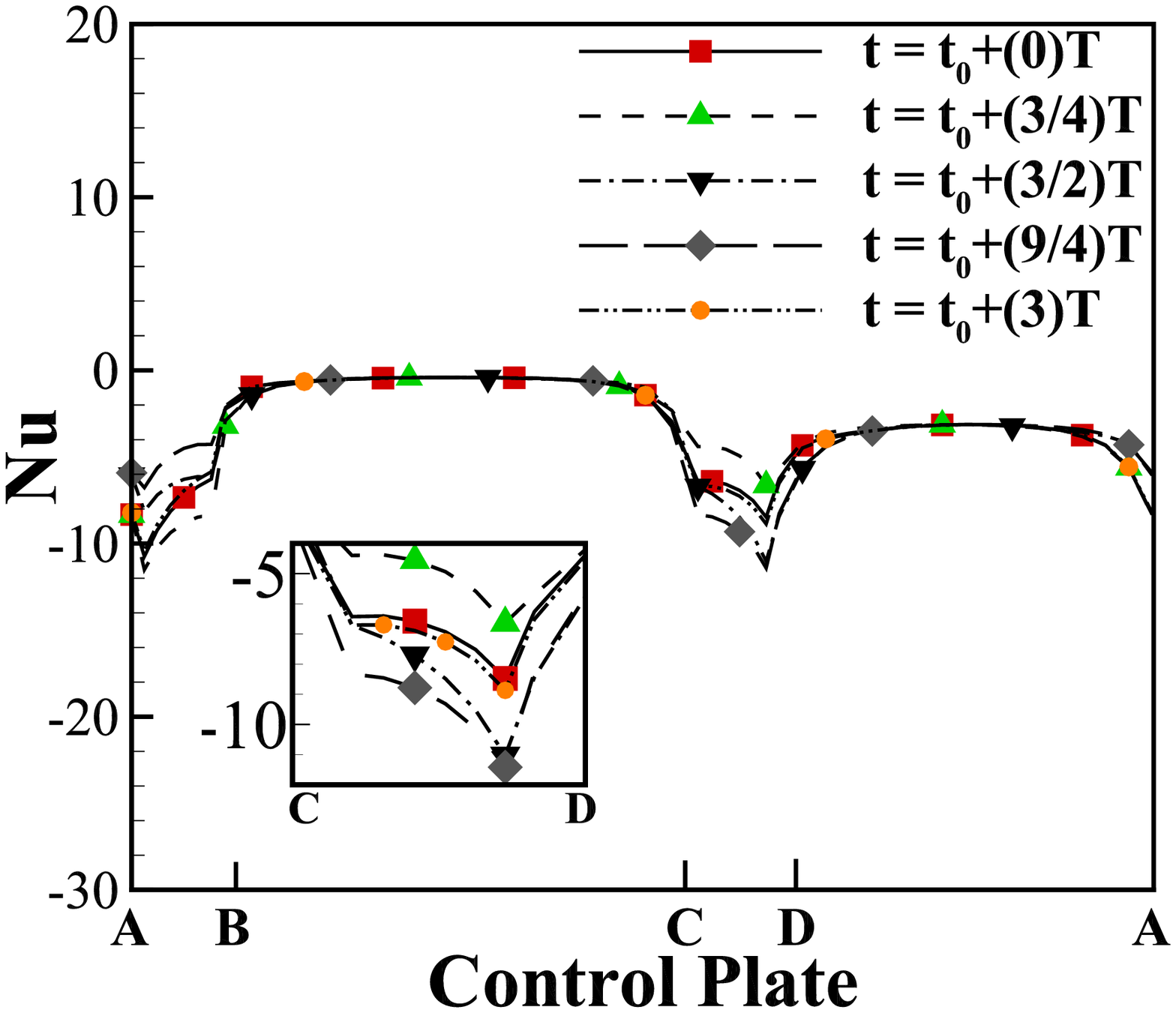}
\includegraphics[width=0.45\textwidth,trim={0cm 0cm 0cm 0cm},clip]{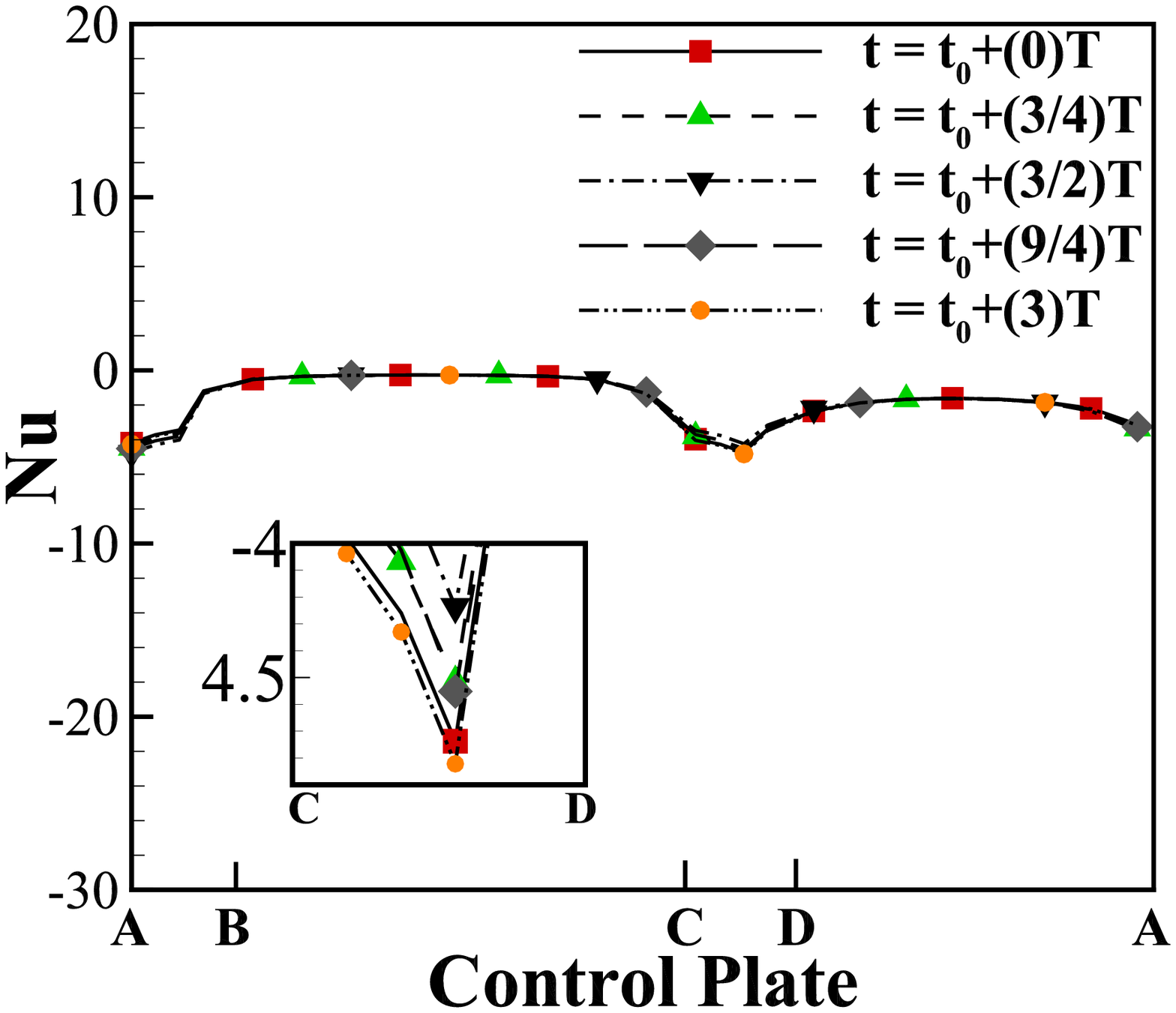}
\\
\hspace{2cm}(b) \hspace{6cm}(c)\hspace{2cm}
 \caption{The distribution of the local Nusselt number $Nu$, over the surface of the control plate during (a) one oscillation period of the cylinder $T$, for $d/R_0=0.5$, (b) one oscillation period of the cylinder $T$, for $d/R_0=1$ and (c) one oscillation period of the cylinder $T$, for $d/R_0=3$ with $\alpha_m=4$, $f/f_0=3$ and $Re=150$.}
 \label{fig:plate_a_4_f_3}
\end{figure*}

In \cref{fig:a_4_f_3}, when the frequency ratio is increased to $f/f_0=3$ with $\alpha_m=4$, the vortex shedding modes are identified as $2S(T)$ for all three positions of control plate, i.e. $d/R_0=0.5$, $1$ and $3$. For $d/R_0=1$, the density of the vorticity contours rises, increasing the downstream heat convection. The highest peak of the local Nusselt number distribution varies within the region, $\ang{120}<\theta<\ang{240}$. Dense isotherm contours can be seen here more than anywhere else on the surface, providing further evidence that this region is the most active in terms of heat convection. Additional local maximum peaks are observed between $\ang{0}$ and $\ang{90}$ and between $\ang{270}$ and $\ang{360}$, but they gradually diminish as $d/R_0$ increases. Highest amount of heat absorption is found near the corners $A$ and $D$, in \cref{fig:plate_a_4_f_3}(a) for $d/R_0=0.5$, $\alpha_m=4$ and $f/f_0=3$. The heat absorption rate decreases as the gap ratio of the control plate increases as seen in \cref{fig:plate_a_4_f_3}(b) and \cref{fig:plate_a_4_f_3}(c). The rear ($DA$) surface of the control plate absorbs very little heat from the fluid, but the front ($BC$) surface of the plate absorbs a little bit more heat.\\

\begin{figure*}[!t]
\centering
\includegraphics[width=0.45\textwidth,trim={0cm 0cm 0cm 0cm},clip]{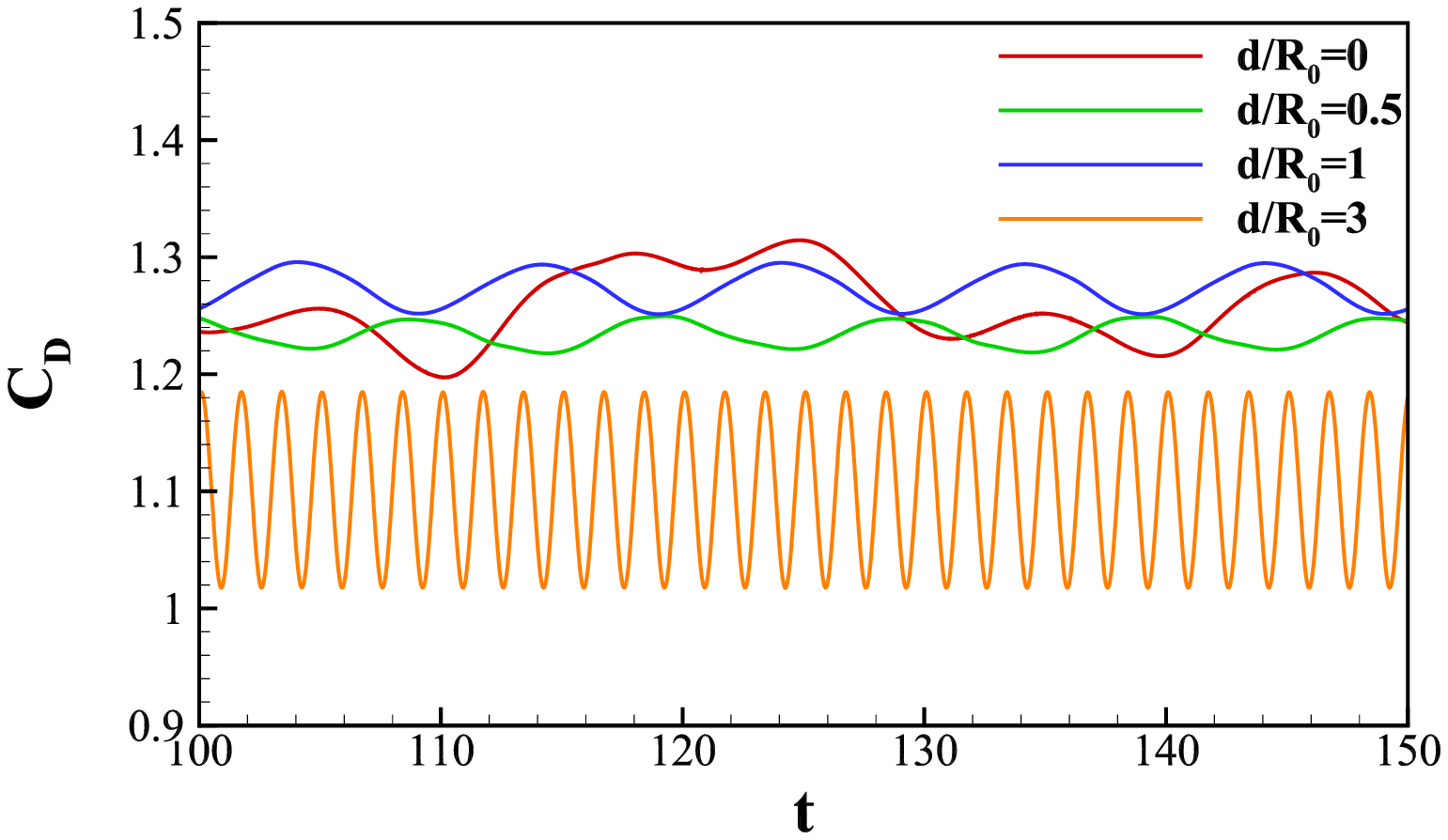}
\includegraphics[width=0.45\textwidth,trim={0cm 0cm 0cm 0cm},clip]{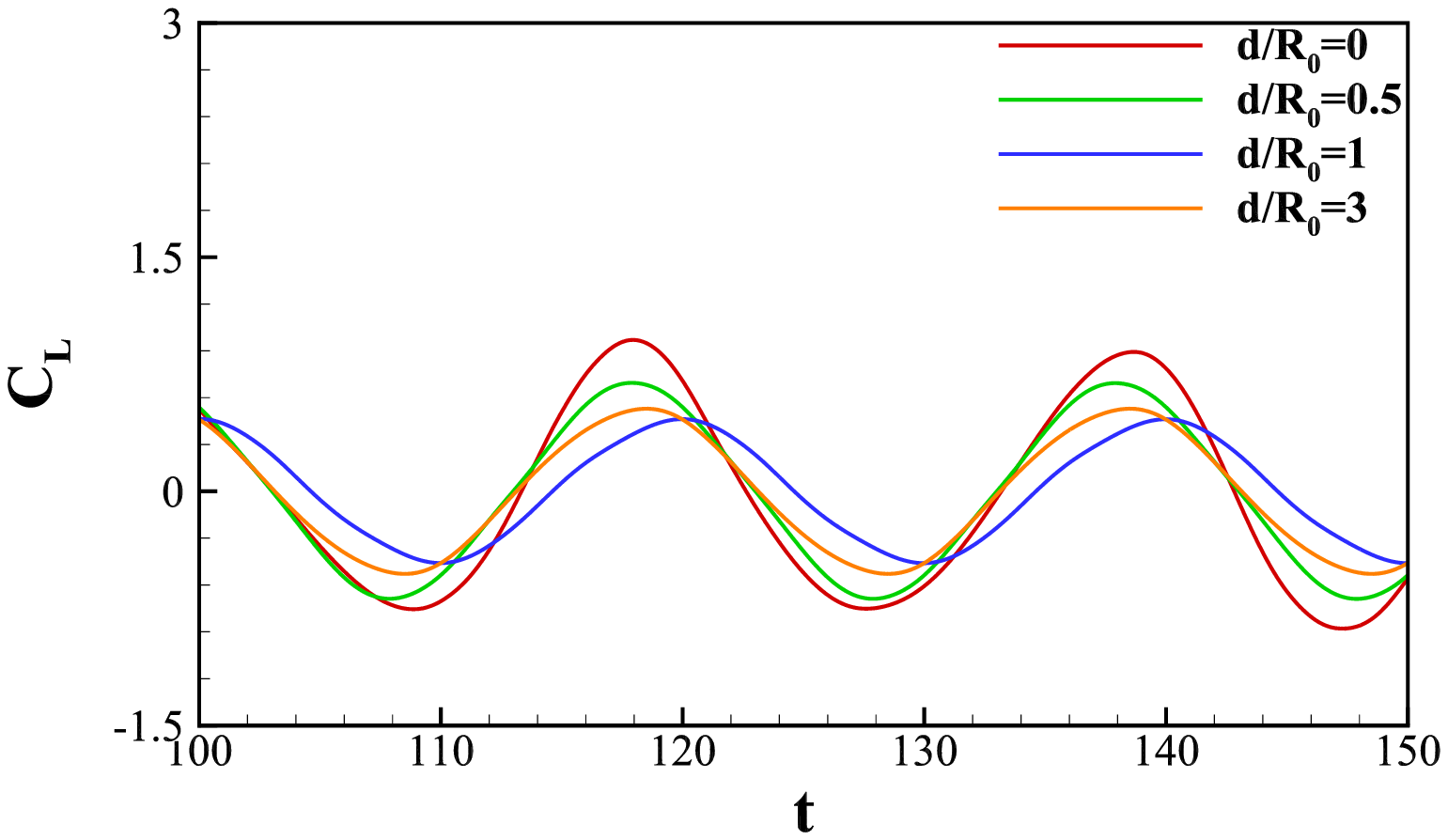}
\\
\hspace{2cm}(a) \hspace{5.5cm}(b)\hspace{2cm}
\\
\includegraphics[width=0.45\textwidth,trim={0cm 0cm 0cm 0cm},clip]{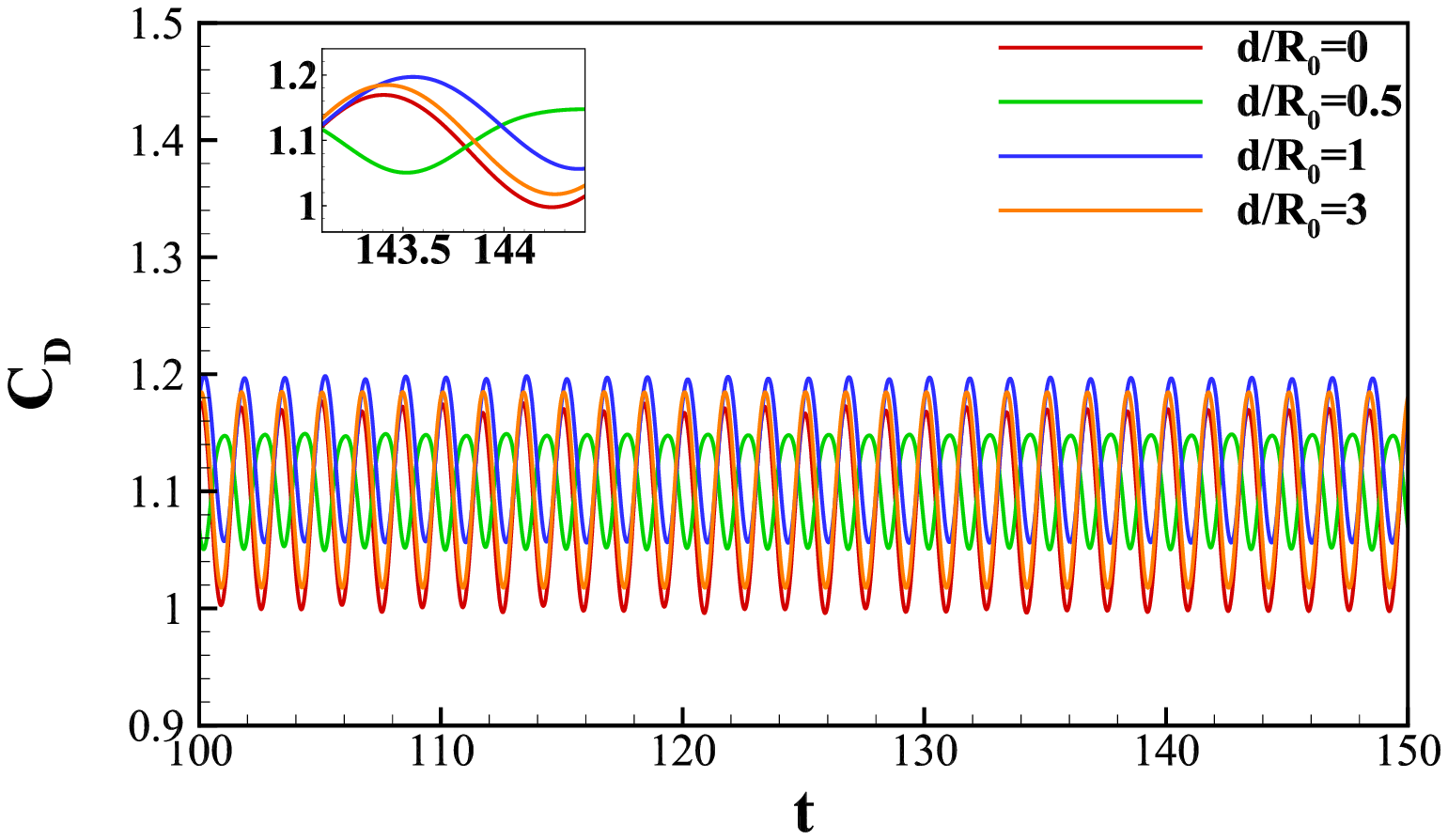}
\includegraphics[width=0.45\textwidth,trim={0cm 0cm 0cm 0cm},clip]{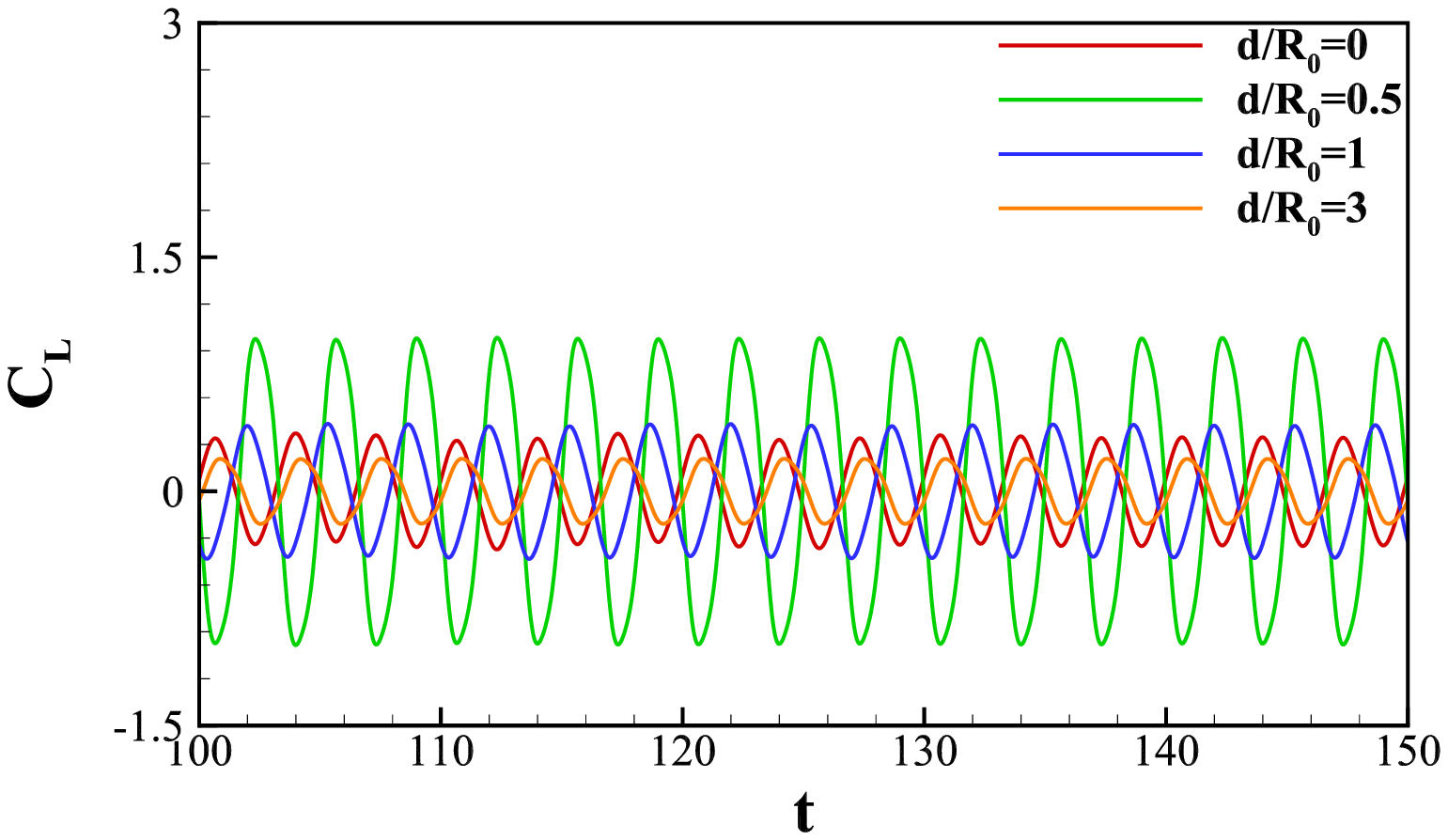}
\\
\hspace{2cm}(c) \hspace{5.5cm}(d)\hspace{2cm}
 \caption{The variation of (a) drag coefficient $C_D$ with $(\alpha_m,\ f/f_0)=(0.5,\ 0.5)$; (b) lift coefficient $C_L$ with $(\alpha_m,\ f/f_0)=(0.5,\ 0.5)$; (c) drag coefficient $C_D$ with $(\alpha_m,\ f/f_0)=(4,\ 3)$ and (d) lift coefficient $C_L$ with $(\alpha_m,\ f/f_0)=(4,\ 3)$ and $Re=150$.}
 \label{fig:drag_lift}
\end{figure*}

\cref{fig:drag_lift} displays the drag ($C_D$) and lift ($C_L$) coefficients for varying $d/R_0$ with rotary oscillation parameters, $(\alpha_m,\ f/f_0)=(0.5,\ 0.5)$ ((a) and (b)); $(\alpha_m,\ f/f_0)=(4,\ 3)$ ((c) and (d)). $d/R_0 = 0$ is assumed to define the case without the control plate. The maximum value of the drag coefficient is found to be reduced by $4.97\%$ after the introduction of the control plate at $d/R_0 = 0.5$ in \cref{fig:drag_lift}(a). After positioning the control plate at a gap ratio of $d/R_0=3$, the peak value of the drag coefficient has decreased by $9.87\%$ when compared to the no plate case. The amplitude and highest value of the lift coefficient steadily declines with increasing $d/R_0$ within the range of $0\leq d/R_0\leq 1$ in \cref{fig:drag_lift}(b). However, the peak value slightly increases at $d/R_0=3$ than that of $d/R_0=1$. With the increase in rotary oscillation parameters to $\alpha_m=4$ and $f/f_0=3$, the maximum value of the drag coefficient decreases by $1.92\%$ at a gap ratio of $d/R_0=0.5$  from $d/R_0=0$, as seen in \cref{fig:drag_lift}(c). With the exception of $d/R_0=0.5$, no change is seen for varied control plate gap ratios at this rotary oscillation. When $d/R_0=0.5$, the magnitude of $C_D$ is found to be the least. Variation of lift coefficient, $C_L$ for this high rotary oscillation is displayed in \cref{fig:drag_lift}(d). It can be observed that the maximum value and magnitude are significantly increased when the control plate is positioned at the gap ratio of $d/R_0=0.5$. Due to the small gap between the cylinder and the control plate, the shear layers around the cylinder are pushed back when the vortices shed from the cylinder encounter resistance from the control plate.\\

\begin{figure*}[!t]
\centering
\includegraphics[width=0.45\textwidth,trim={0cm 0cm 0cm 0cm},clip]{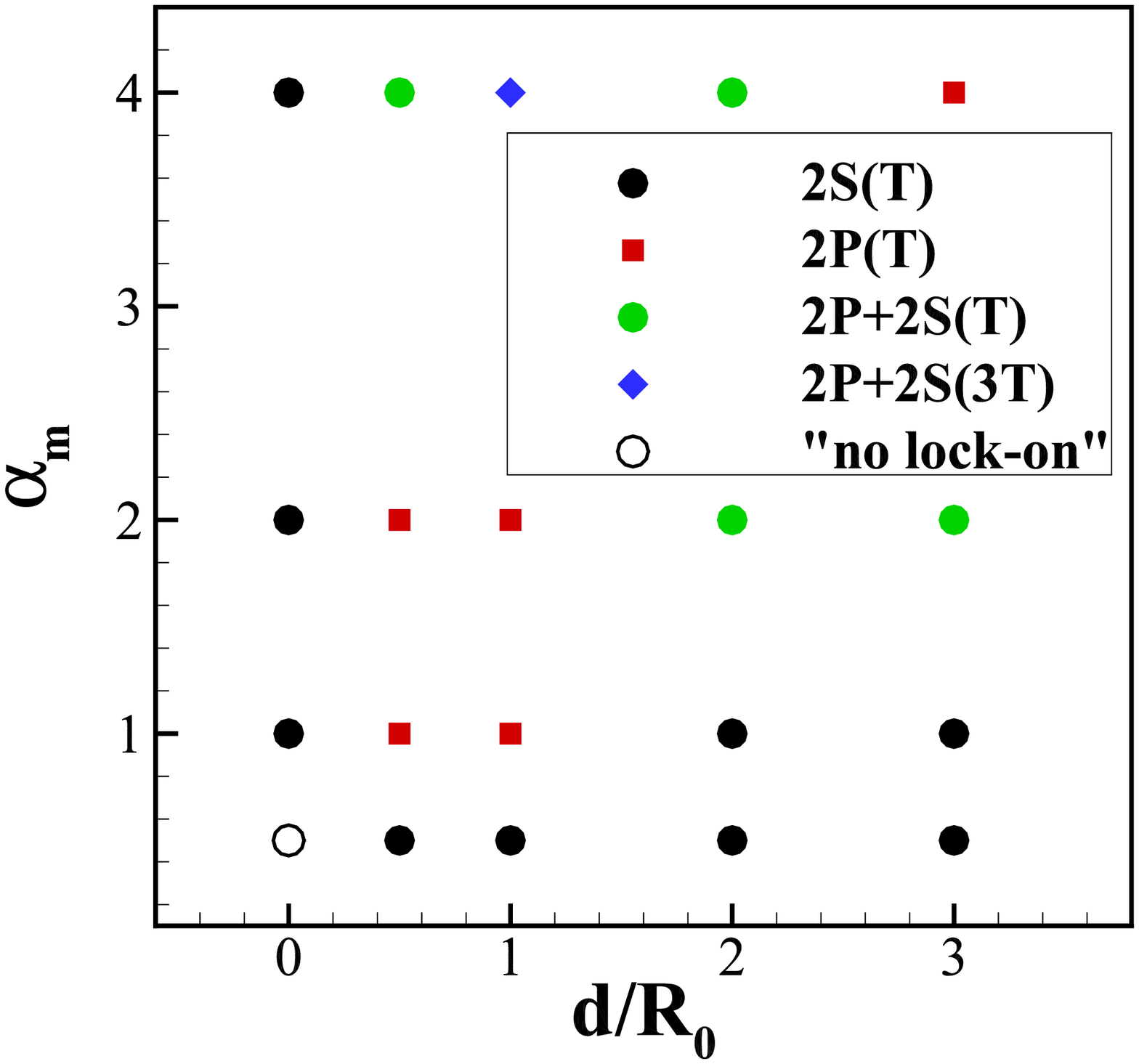}
\includegraphics[width=0.45\textwidth,trim={0cm 0cm 0cm 0cm},clip]{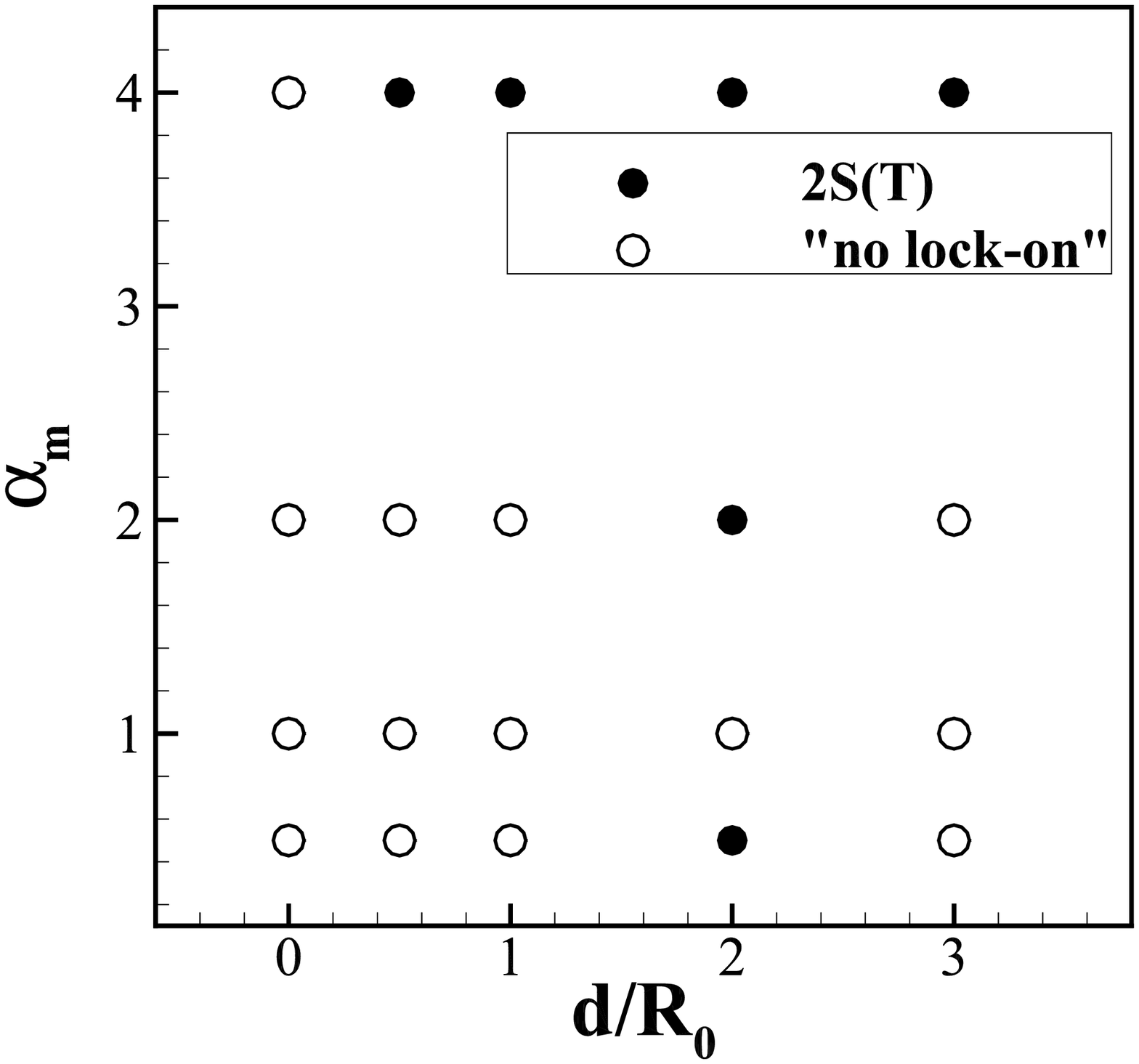}
\\
\hspace{2cm}(a) \hspace{5.5cm}(b)\hspace{2cm}
 \caption{An overview of the vortex-shedding modes for (a) $f=0.5$ and (b) $f=3$ with $\alpha_m\in [0.5,\ 4]$, $d/R_0\in [0,\ 3]$ and $Re=150$.}
 \label{fig:vortex_mode}
\end{figure*}

\cref{fig:vortex_mode} summarizes the vortex shedding modes for (a) $f/f_0=0.5$ and (b) $f/f_0=3$ with $\alpha_m\in [0.5,\ 4]$, $d/R_0\in[0,\ 3]$ and $Re=150$. For $(\alpha_m,\ f/f_0)=(0.5,\ 0.5)$ and $0.5\leq d/R_0\leq 3$, the modes for shedding vortices are are locked-on and identified as $2S(T)$ but no lock-on is observed for $d/R_0=0$. For $\alpha_m\geq 2$ and $d/R_0\geq 0.5$, more than two vortices are shed from each side of the cylinder at regular intervals. The various types of lock-on modes identified in \cref{fig:vortex_mode}(a) are $2S(T)$, $2P(T)$, $2P+2S(T)$ and $2P+2S(3T)$. When $f/f_0$ is increased to $3$, the lock-on vortex shedding mode is only observed for $(\alpha_m,\ d/R_0)=$ $\{(0.5,\ 2),$ $(2,\ 2),$ $(4,\ 0.5),$ $(4,\ 1),$ $(4,\ 2),$ $(4,\ 3)\}$. All of the lock-on modes for vortex shedding are classified as $2S(T)$.\\

\begin{figure*}[!htbp]
\centering
\includegraphics[width=0.45\textwidth,trim={0cm 0cm 0cm 0cm},clip]{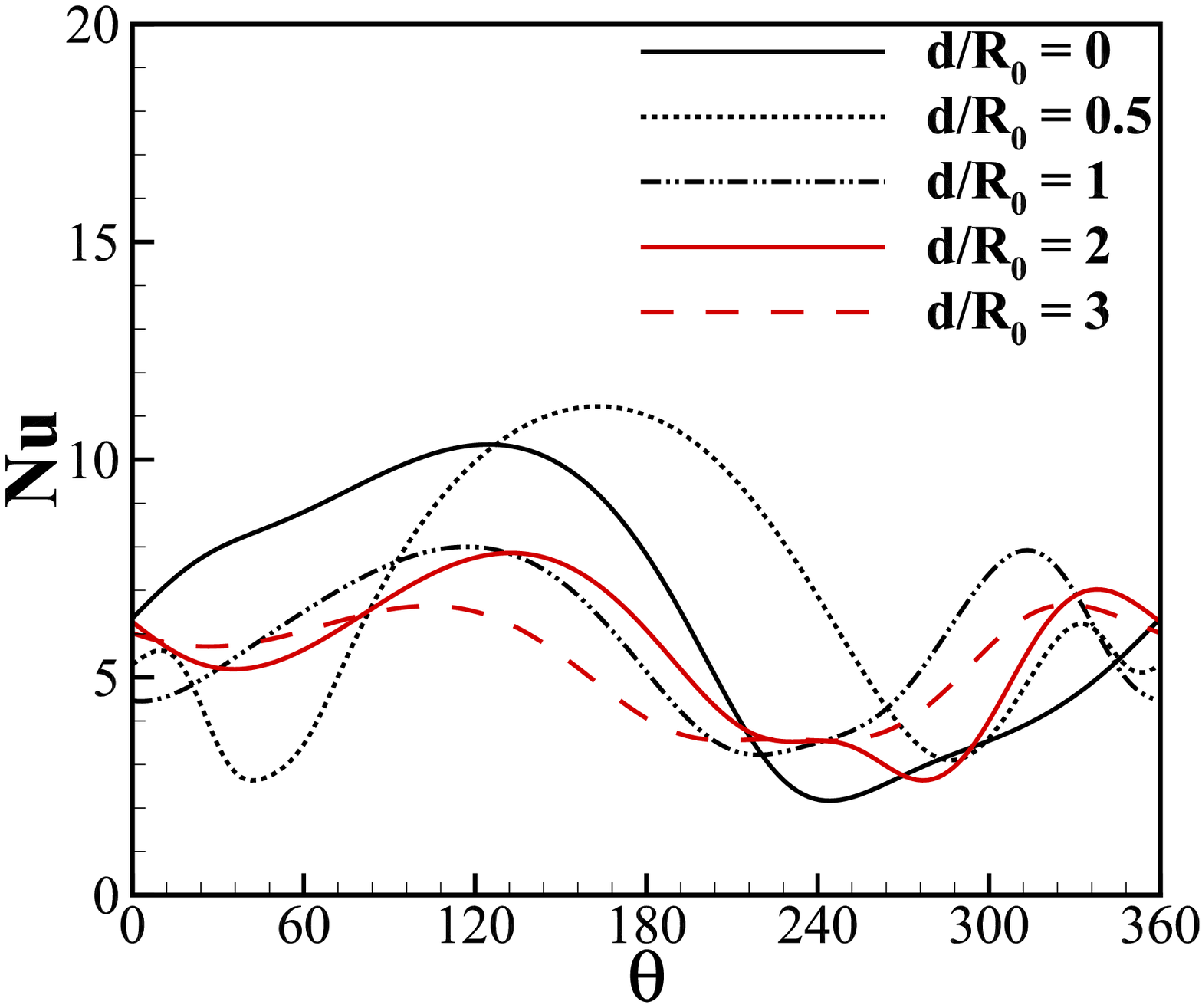}
\includegraphics[width=0.45\textwidth,trim={0.1cm 0.1cm 0.1cm 0.1cm},clip]{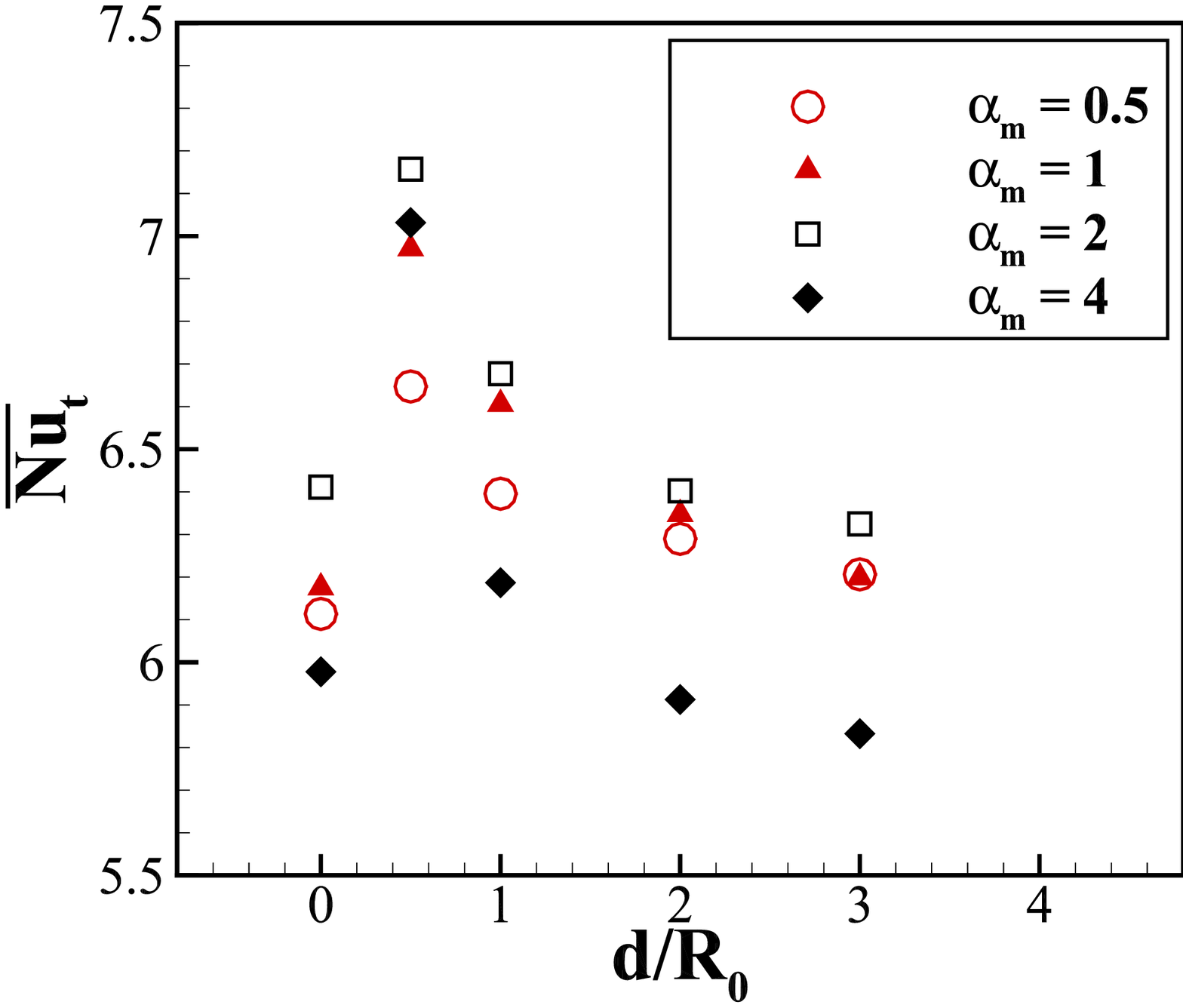}
\\
\hspace{2cm}(a) \hspace{5.5cm}(b)\hspace{2cm}
 \caption{The variation of (a) $Nu$ at $t=t_0+3T$ for various $d/R_0$ values with $\alpha_m=4$ and (b) the time average Nusselt number $\overline{Nu_t}$ on the surface of the cylinder in relation to $\alpha_m$ for $Re=150$, $Pr=0.7$, $f/f_0=0.5$.}
 \label{fig:total_Nu_f_0-5}
\end{figure*}

The instantaneous local Nusselt number distribution for $\alpha_m=4$ and time-averaged Nusselt number with varying gap ratios of control plate are displayed in \cref{fig:total_Nu_f_0-5}(a) and \cref{fig:total_Nu_f_0-5}(b) respectively for $Re=150$, $Pr=0.7$ and $f/f_0=0.5$. \cref{fig:total_Nu_f_0-5}(a) shows that the maximum peak of the local Nusselt number rises near $\theta\approx\ang{180}$ after the introduction of the control plate at $d/R_0=0.5$ compared to no plate arrangement. The maximum peak value of $Nu$ near $\theta\approx\ang{180}$ gradually decreases with increasing gap ratio of the control plate within the range of $1\leq d/R_0 \leq3$. Local maximum peak of $Nu$ near $\theta\approx\ang{360}$ is found to be largest for $d/R_0=1$. Time average Nusselt number $\overline{Nu}_t$ increases after introduction of the control plate at a gap ratio of $d/R_0=0.5$ by a significant amount for all maximum angular velocity, $\alpha_m$ as seen in \cref{fig:total_Nu_f_0-5}(b). At all gap ratios except $d/R_0=0.5$, $\overline{Nu}_t$ is found to be the lowest for $\alpha_m=4$. The reduction of heat transfer process at high maximum angular velocity of rotary oscillation is caused by the thickening of thermal boundary layer. Similar results are also reported for flow past a rotary oscillating cylinder without any control plate \cite{ghazanfarian2009numerical,mittal2018numerical} where the increased rotary oscillating motion of the cylinder thickens the thermal boundary layer and, as a result, reduces the rate of heat transfer from the cylinder surface. $\overline{Nu}_t$ gradually decreases for each $\alpha_m$ as the gap ratio, $d/R_0$, increases from $0.5$ to $3$.\\

\begin{figure*}[!htbp]
\centering
\includegraphics[width=0.45\textwidth,trim={0cm 0cm 0cm 0cm},clip]{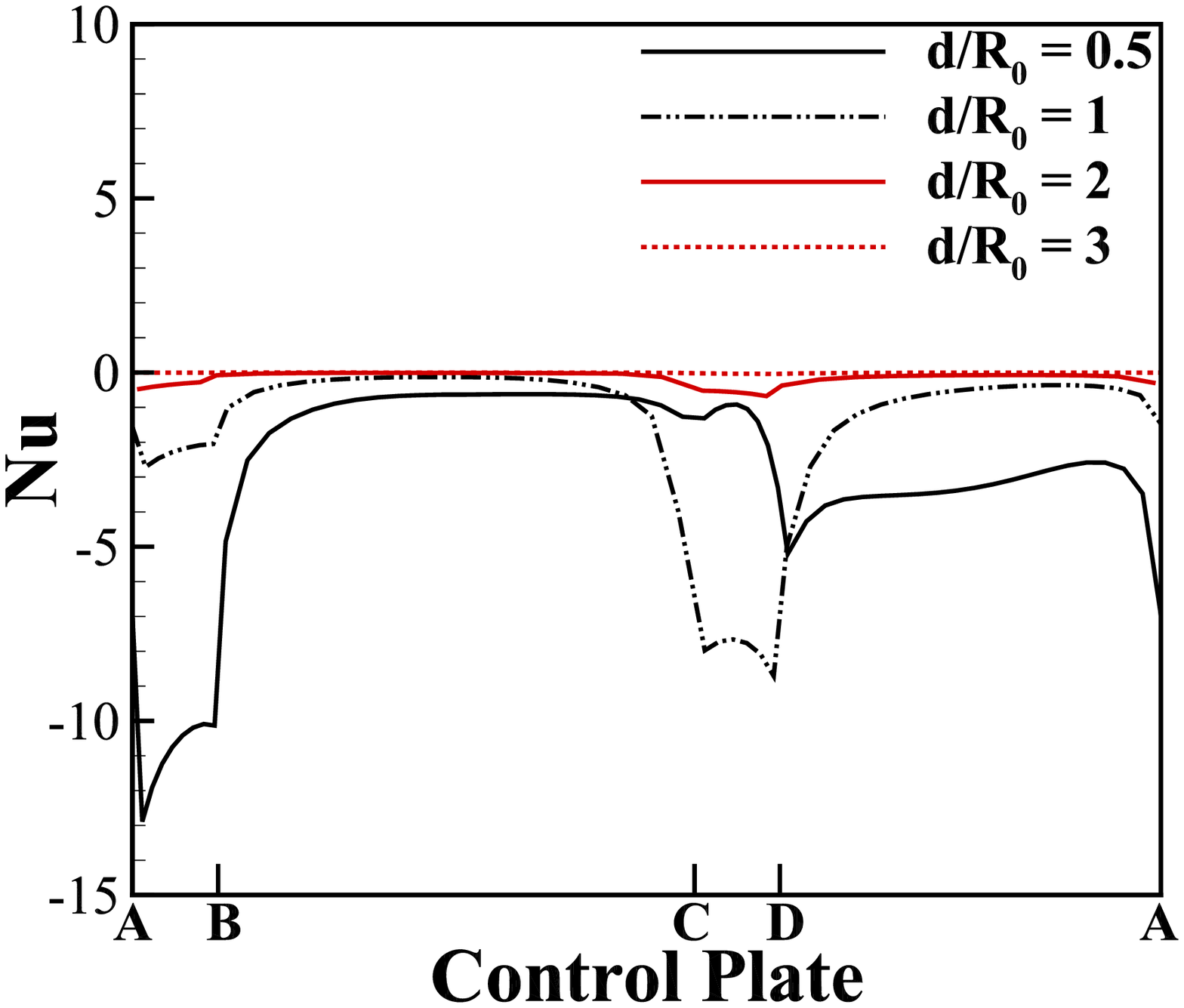}
\includegraphics[width=0.45\textwidth,trim={0.1cm 0.1cm 0.1cm 0.1cm},clip]{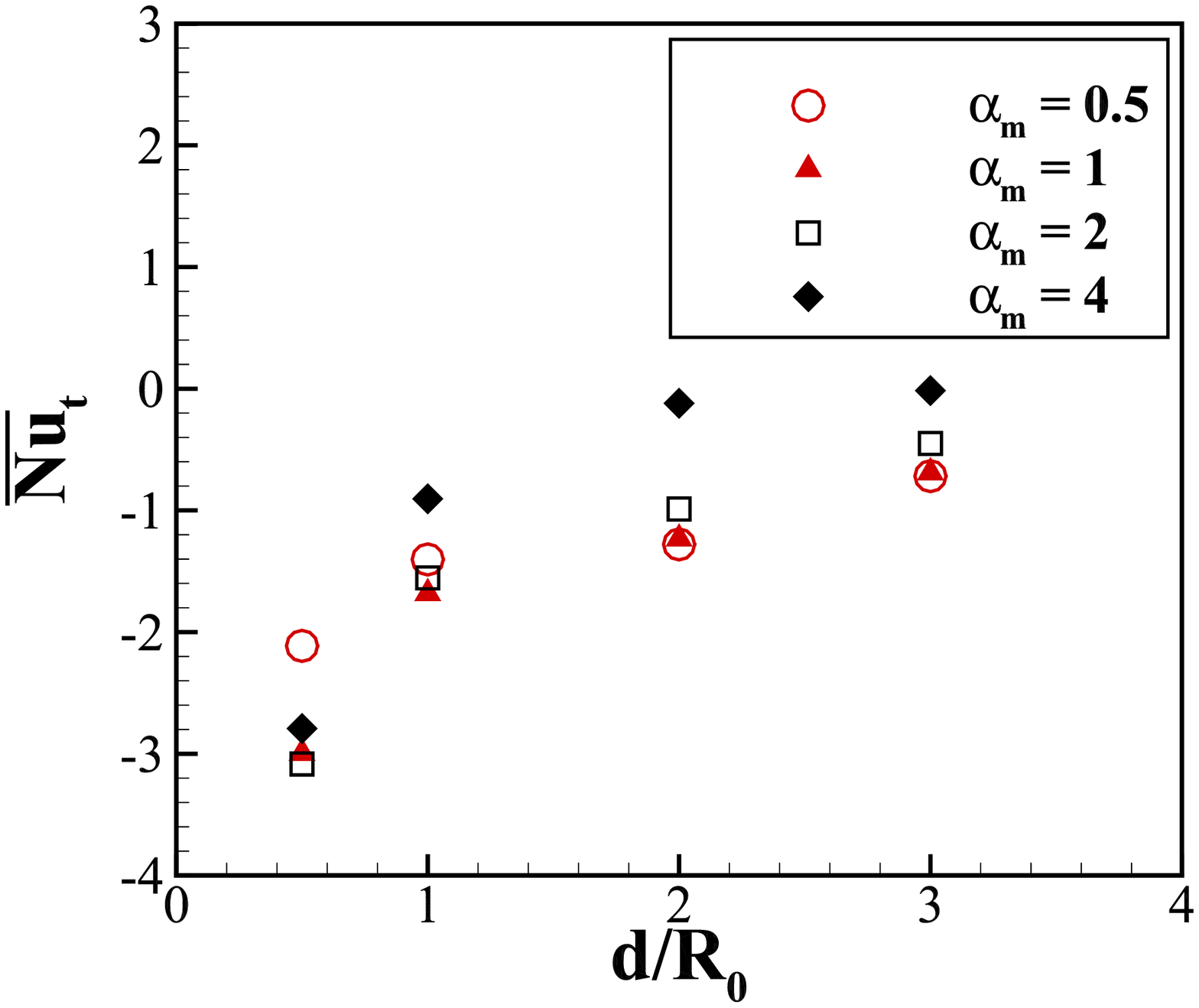}
\\
\hspace{2cm}(a) \hspace{5.5cm}(b)\hspace{2cm}
 \caption{The variation of (a) $Nu$ at $t=t_0+3T$ for various $d/R_0$ values with $\alpha_m=4$ and (b) the time-averaged Nusselt number $\overline{Nu_t}$ on the surface of the control plate in relation to $\alpha_m$ for $Re=150$, $Pr=0.7$, $f/f_0=0.5$.}
 \label{fig:total_Nu_f_0-5_plate}
\end{figure*}
%%%%%%%%%%%%%revise
\cref{fig:total_Nu_f_0-5_plate}(a) exhibits the instantaneous local Nusselt number distribution with varying gap ratios, $d/R_0$ for $\alpha_m=4$, $f/f_0=0.5$, $Re=150$ and $Pr=0.7$ at  $t=t_0+3T$. The highest amount of heat absorption occurs around the top ($CD$) and bottom ($AB$) surfaces of the control plate at this instant. However, the amount of heat absorption decreases with the increasing gap ratios of the control plate, from $d/R_0=0.5$ to $d/R_0=3$. It occurs due to the reduction in the density of the isotherm contours with increasing distances from the surface of the cylinder. The heat absorption is found to be negligible at a gap ratio $d/R_0=3$. It occurs as the density of the isotherm contours gradually decreases with increasing distance from the cylinder surface. \cref{fig:total_Nu_f_0-5_plate}(b) displays the time-averaged Nusselt number on the surface of the control plate with varying maximum angular velocities, $\alpha_m$ at different gap ratios of the control plate for $Re=150$, $Pr=0.7$ and $f/f_0=0.5$. It shows that the highest values of $\overline{Nu_t}$ are found for $\alpha_m=4$ at all the gap ratios except $d/R_0=0.5$. The thick thermal boundary layer around the cylinder at $\alpha_m=4$ decreases the overall heat transfer rate from the cylinder surface and eventually decreases the density of the isotherm contours downstream. As a result, the overall heat absorption on the surface of the control plate is lowest for $\alpha_m=4$ at all control plate gap ratios except $d/R_0=0.5$. The close proximity ($d/R_0=0.5$) of the control plate to the cylinder for $\alpha_m=2$ causes the most heat absorption on the surface of the control plate.
%%%%%%%%%%%
	
%\clearpage
\section{Conclusions\protect}\label{Conclusions}
The impact of a cold, vertical, arc-shaped control plate on the force convective heat transfer and fluid flow over an isothermally heated, rotary oscillating circular cylinder in a two-dimensional, unsteady, incompressible, laminar, and viscous flow of a Newtonian, constant property fluid is investigated numerically in this study. The simulations are performed with an in-house code. According to the findings of the study, the gap ratio of the control plate has a significant effect on both the fluid flow and heat transfer. Simulations are carried out using the parameters, $Re=150$, $Pr=0.7$, $0.5\leq\alpha_m\leq4$, $f/f_0=\{0.5,\ 3\}$, and $d/R_0\in[0,\ 3]$. For certain rotary oscillation parameters of the cylinder, the non-lock-on vortex shedding process can be locked-on by introducing the control plate into the system. The smallest gap ratio of $d/R_0=0.5$ is found to be more effective than other gap ratios in increasing the heat transfer rate for all values of $\alpha_m$ with $f/f_0=0.5$. Also, $\alpha_m=2$ exhibits highest amount of heat transmission for all gap ratios when $f/f_0=0.5$. The heat absorption on the surface of the control plate decreases to zero with increasing gap ratio when $\alpha_m=0.5$ and $f/f_0=0.5$ but never becomes zero when $\alpha_m=4$ and $f/f_0=3$. The highest peak of the drag coefficient is decreased by $9.87\%$ at a gap ratio of $d/R_0=3$ for $(\alpha_m,\ f/f_0)=(0.5,\ 0.5)$. However, at higher rotary oscillations like $(\alpha_m,\ f/f_0)=(4,\ 3)$, the maximum reduction in the peak value of drag coefficient is $1.92\%$ which is obtained at a gap ratio of $d/R_0=0.5$. Smallest gap ratio of $d/R_0=0.5$ is found to enhance the lift coefficient by $183\%$ for $(\alpha_m,\ f/f_0)=(4,\ 3)$.

\section*{Author Declarations}
%\subsection*{No conflicts of interest}
The authors have no conflicts to disclose.

\section*{Data Availability Statement}
The data that support the findings of this study are available from the corresponding author upon reasonable request.

 \bibliographystyle{elsarticle-num} 
 \bibliography{cas-refs}

\begin{thebibliography}{10}
\expandafter\ifx\csname url\endcsname\relax
  \def\url#1{\texttt{#1}}\fi
\expandafter\ifx\csname urlprefix\endcsname\relax\def\urlprefix{URL }\fi
\expandafter\ifx\csname href\endcsname\relax
  \def\href#1#2{#2} \def\path#1{#1}\fi

\bibitem{kalita2009transformation}
J.~C. Kalita, R.~K. Ray, {A transformation-free HOC scheme for incompressible
  viscous flows past an impulsively started circular cylinder}, Journal of
  Computational Physics 228~(14) (2009) 5207--5236.

\bibitem{ray2016higher}
R.~K. Ray, J.~C. Kalita, Higher-order-compact simulation of unsteady flow past
  a rotating cylinder at moderate reynolds numbers, Computational and Applied
  Mathematics 35~(1) (2016) 219--250.

\bibitem{mittal2017numerical}
H.~V.~R. Mittal, R.~K. Ray, Q.~M. Al-Mdallal, {A numerical study of initial
  flow past an impulsively started rotationally oscillating circular cylinder
  using a transformation-free HOC scheme}, Physics of Fluids 29~(9) (2017)
  93603.

\bibitem{mittal2017locked}
H.~V.~R. Mittal, Q.~M. Al-Mdallal, R.~K. Ray, {Locked-on vortex shedding modes
  from a rotationally oscillating circular cylinder}, Ocean Engineering 146
  (2017) 324--338.

\bibitem{gao2017experimental}
Y.-y. Gao, C.-s. Yin, K.~Yang, X.-z. Zhao, S.~K. Tan, Experimental study on
  flow past a rotationally oscillating cylinder, China Ocean Engineering 31~(4)
  (2017) 495--503.

\bibitem{yawar2019transient}
A.~Yawar, M.~Ebrahem, S.~Manzoor, N.~Sheikh, M.~Ali, Transient cross flow and
  heat transfer over a rotationally oscillating cylinder subjected to gust
  impulse, International Journal of Heat and Mass Transfer 137 (2019) 108--123.

\bibitem{ganta2019analysis}
N.~Ganta, B.~Mahato, Y.~G. Bhumkar, Analysis of sound generation by flow past a
  circular cylinder performing rotary oscillations using direct simulation
  approach, Physics of Fluids 31~(2) (2019) 026104.

\bibitem{strouhal1878besondere}
V.~Strouhal, {\"U}ber eine besondere Art der Tonerregung, Stahel, 1878.

\bibitem{kumar2013flow}
S.~Kumar, C.~Lopez, O.~Probst, G.~Francisco, D.~Askari, Y.~Yang, {Flow past a
  rotationally oscillating cylinder}, Journal of Fluid Mechanics 735 (2013)
  307--346.

\bibitem{lu1996numerical}
X.-Y. Lu, J.~Sato, {A numerical study of flow past a rotationally oscillating
  circular cylinder}, Journal of Fluids and Structures 10~(8) (1996) 829--849.

\bibitem{sellappan2014vortex}
P.~Sellappan, T.~Pottebaum, {Vortex shedding and heat transfer in rotationally
  oscillating cylinders}, Journal of Fluid Mechanics 748 (2014) 549--579.

\bibitem{Zebib1989}
A.~Zebib, Y.~K. Wo, {A Two-Dimensional Conjugate Heat Transfer Model for Forced
  Air Cooling of an Electronic Device}, Journal of Electronic Packaging 111~(1)
  (1989) 41--45.

\bibitem{yang2001thermal}
R.-J. Yang, L.-M. Fu, Thermal and flow analysis of a heated electronic
  component, International Journal of Heat and Mass Transfer 44~(12) (2001)
  2261--2275.

\bibitem{roshko1955wake}
A.~Roshko, On the wake and drag of bluff bodies, Journal of the Aeronautical
  Sciences 22~(2) (1955) 124--132.

\bibitem{Apelt1973}
C.~J. Apelt, G.~S. West, A.~A. Szewczyk, {The effects of wake splitter plates
  on the flow past a circular cylinder in the range
  $10^4${\textless}R{\textless}$5\times10^4$}, Journal of Fluid Mechanics
  61~(1) (1973) 187--198.

\bibitem{apelt1975effects}
C.~Apelt, G.~West, The effects of wake splitter plates on bluff-body flow in
  the range 104< r< 5$\times$ 104. part 2, Journal of Fluid Mechanics 71~(1)
  (1975) 145--160.

\bibitem{kwon1996}
K.~Kwon, H.~Choi, {Control of laminar vortex shedding behind a circular
  cylinder using splitter plates}, Physics of Fluids 8~(2) (1996) 479--486.

\bibitem{MITTAL2001291}
S.~Mittal, Control of flow past bluff bodies using rotating control cylinders,
  Journal of Fluids and Structures 15~(2) (2001) 291--326.

\bibitem{asadullah2018counter}
M.~Asadullah, S.~A. Khan, W.~Asrar, E.~Sulaeman, Counter clockwise rotation of
  cylinder with variable position to control base flows, in: IOP Conference
  Series: Materials Science and Engineering, Vol. 370, IOP Publishing, 2018, p.
  012058.

\bibitem{tokumaru1991rotary}
P.~T. Tokumaru, P.~E. Dimotakis, {Rotary oscillation control of a cylinder
  wake}, Journal of Fluid Mechanics 224 (1991) 77--90.

\bibitem{shiels2001investigation}
D.~Shiels, A.~Leonard, {Investigation of a drag reduction on a circular
  cylinder in rotary oscillation}, Journal of Fluid Mechanics 431 (2001)
  297--322.

\bibitem{he2000active}
J.-W. He, R.~Glowinski, R.~Metcalfe, A.~Nordlander, J.~Periaux, Active control
  and drag optimization for flow past a circular cylinder: I. oscillatory
  cylinder rotation, Journal of Computational Physics 163~(1) (2000) 83--117.

\bibitem{cheng2001}
M.~Cheng, G.~Liu, K.~Lam, Numerical simulation of flow past a rotationally
  oscillating cylinder, Computers \& Fluids 30~(3) (2001) 365--392.

\bibitem{cheng2001numerical}
M.~Cheng, Y.~T. Chew, S.~C. Luo, {Numerical investigation of a rotationally
  oscillating cylinder in mean flow}, Journal of Fluids and Structures 15~(7)
  (2001) 981--1007.

\bibitem{saxena1978heat}
U.~C. Saxena, A.~D.~K. Laird, {Heat transfer from a cylinder oscillating in a
  cross-flow}, Journal of Heat Transfer 100~(4) (1978) 684--689.

\bibitem{cheng1997experimental}
C.-H. Cheng, H.-N. Chen, W.~Aung, {Experimental study of the effect of
  transverse oscillation on convection heat transfer from a circular cylinder},
  Journal of Heat Transfer 119~(3) (1997) 474--482.

\bibitem{mahfouz2000forced}
F.~M. Mahfouz, H.~M. Badr, {Forced convection from a rotationally oscillating
  cylinder placed in a uniform stream}, International Journal of Heat and Mass
  Transfer 43~(17) (2000) 3093--3104.

\bibitem{FU20023033}
W.-S. Fu, B.-H. Tong, Numerical investigation of heat transfer from a heated
  oscillating cylinder in a cross flow, International Journal of Heat and Mass
  Transfer 45~(14) (2002) 3033--3043.

\bibitem{ghazanfarian2009numerical}
J.~Ghazanfarian, M.~R.~H. Nobari, {A numerical study of convective heat
  transfer from a rotating cylinder with cross-flow oscillation}, International
  Journal of Heat and Mass Transfer 52~(23-24) (2009) 5402--5411.

\bibitem{nobari2010convective}
M.~R.~H. Nobari, J.~Ghazanfarian, {Convective heat transfer from a rotating
  cylinder with inline oscillation}, International Journal of Thermal Sciences
  49~(10) (2010) 2026--2036.

\bibitem{al2017heat}
Q.~M. Al-Mdallal, F.~M. Mahfouz, {Heat transfer from a heated non-rotating
  cylinder performing circular motion in a uniform stream}, International
  Journal of Heat and Mass Transfer 112 (2017) 147--157.

\bibitem{celik2008flow}
B.~Celik, U.~Akdag, S.~Gunes, A.~Beskok, Flow past an oscillating circular
  cylinder in a channel with an upstream splitter plate, Physics of Fluids
  20~(10) (2008) 103603.

\bibitem{ghiasi2018numerical}
A.~Ghiasi, S.~E. Razavi, A.~Rouboa, O.~Mahian, Numerical study on flow over a
  confined oscillating cylinder with a splitter plate, International Journal of
  Numerical Methods for Heat \& Fluid Flow 29~(5) (2019) 1629--1646.

\bibitem{gerrard1966mechanics}
J.~Gerrard, The mechanics of the formation region of vortices behind bluff
  bodies, Journal of Fluid Mechanics 25~(2) (1966) 401--413.

\bibitem{bearman1965investigation}
P.~Bearman, Investigation of the flow behind a two-dimensional model with a
  blunt trailing edge and fitted with splitter plates, Journal of Fluid
  Mechanics 21~(2) (1965) 241--255.

\bibitem{razavi2008impact}
S.~E. Razavi, V.~Farhangmehr, F.~Barar, Impact of a splitter plate on flow and
  heat transfer around circular cylinder at low {Reynolds} numbers, Journal of
  Applied Sciences 8~(7) (2008) 1286--1292.

\bibitem{deep2022pod}
D.~Deep, A.~Sahasranaman, S.~Senthilkumar, {POD} analysis of the wake behind a
  circular cylinder with splitter plate, European Journal of Mechanics-B/Fluids
  93 (2022) 1--12.

\bibitem{liu2016experimental}
K.~Liu, J.~Deng, M.~Mei, {Experimental study on the confined flow over a
  circular cylinder with a splitter plate}, Flow Measurement and
  Instrumentation 51 (2016) 95--104.

\bibitem{bao2013passive}
Y.~Bao, J.~Tao, {The passive control of wake flow behind a circular cylinder by
  parallel dual plates}, Journal of Fluids and Structures 37 (2013) 201--219.

\bibitem{lu2014numerical}
L.~Lu, M.-m. Liu, B.~Teng, Z.-d. Cui, G.-q. Tang, M.~Zhao, L.~Cheng, Numerical
  investigation of fluid flow past circular cylinder with multiple control rods
  at low reynolds number, Journal of Fluids and Structures 48 (2014) 235--259.

\bibitem{eastman1985aerodynamics}
D.~Eastman, D.~Wenndt, Aerodynamics of maneuvering missiles with wrap-around
  fins, in: 3rd Applied Aerodynamics Conference, Colorado Springs, CO, U.S.A,
  1985, p. 4083.

\bibitem{martins2021influence}
D.~Martins, J.~Correia, A.~Silva, The influence of front wing pressure
  distribution on wheel wake aerodynamics of a f1 car, Energies 14~(15) (2021)
  4421.

\bibitem{mittal2016class}
H.~V.~R. Mittal, A class of higher order accurate schemes for fluid interface
  problems, Ph.D. thesis, Indian Institute of Technology Mandi (2016).

\bibitem{mittal2018numerical}
H.~V.~R. Mittal, Q.~M. Al-Mdallal, {A numerical study of forced convection from
  an isothermal cylinder performing rotational oscillations in a uniform
  stream}, International Journal of Heat and Mass Transfer 127 (2018) 357--374.

\bibitem{huang2015natural2}
Z.~Huang, W.~Zhang, G.~Xi, {Natural convection in square enclosure induced by
  inner circular cylinder with time-periodic pulsating temperature},
  International Journal of Heat and Mass Transfer 82 (2015) 16--25.

\bibitem{ray2022heat}
R.~K. Ray, A.~Haty, A.~Kumar, Heat transfer past rotationally oscillating
  circular cylinder heated with time-periodic pulsating temperature in a
  uniform flow, Heat Transfer 51~(3) (2022) 2808--2836.

\end{thebibliography}





% \end{thebibliography}
\end{document}